\documentclass[a4paper,11pt]{article}

\usepackage{adjustbox}
\usepackage{amsfonts}
\usepackage{amssymb, amscd}
\usepackage{graphicx}
\usepackage{empheq}
\usepackage{mathrsfs}
\usepackage[table,xcdraw]{xcolor}
\usepackage{colortbl}
\usepackage{slashed}
\usepackage{pdfpages}
\usepackage{braket}
\usepackage{sidecap}
\usepackage{arydshln}
\usepackage[font=small,labelsep=none]{caption}
\usepackage{mathtools}
\usepackage{cite}
\usepackage{hyperref}
\usepackage{tikz}
\usetikzlibrary{arrows,decorations.markings}
\usepackage{subcaption}
\usepackage[mode=buildnew]{standalone}
\usepackage[normalem]{ulem}
\usepackage[left=2cm,right=2cm,top=3.5cm,bottom=3.5cm]{geometry}

\hypersetup{
    colorlinks=true,
    linkcolor=ceruleanblue,
    filecolor=ceruleanblue,      
    urlcolor=ceruleanblue,
    citecolor=ceruleanblue,
}

\definecolor{myblue}{RGB}{174, 198, 219}
\definecolor{myred}{RGB}{157,31,68}
\definecolor{ceruleanblue}{rgb}{0.0, 0.2, 0.6}
\definecolor{grey}{rgb}{0.4,0.4,0.4}
\definecolor{dullmagenta}{rgb}{0.4,0,0.4}
\definecolor{darkblue}{rgb}{0,0,0.4}
\definecolor{midblue}{rgb}{0,0,0.5}
\definecolor{midred}{rgb}{0.5,0,0}
\definecolor{orange}{rgb}{1,0.5,0}
\definecolor{lightbrown}{rgb}{0.75,0.5,0.25}
\definecolor{tan}{cmyk}{0.14,0.42,0.56,0}
\definecolor{djunglegreen}{cmyk}{0.99,0,0.52,0}
\definecolor{lightgreen}{rgb}{0,1,0}
\definecolor{olivegreen}{cmyk}{0.64,0,0.95,0.40}
\definecolor{midgreen}{rgb}{0.0,0.675,0.0}
\definecolor{darkgreen}{rgb}{0,0.5,0}
\definecolor{ceruleanblue}{rgb}{0.0, 0.2, 0.7}
\definecolor{burgundy}{rgb}{0.5, 0.0, 0.13}
\definecolor{blue_light}{RGB}{0, 102, 204}

\newlength\mytemplen
\newsavebox\mytempbox

\makeatletter
\newcommand\mybox{%
    \@ifnextchar[
       {\@mybox}%
       {\@mybox[0pt]}}

\def\@mybox[#1]{%
    \@ifnextchar[
       {\@@mybox[#1]}%
       {\@@mybox[#1][0pt]}}

\def\@@mybox[#1][#2]#3{
    \sbox\mytempbox{#3}%
    \mytemplen\ht\mytempbox
    \advance\mytemplen #1\relax
    \ht\mytempbox\mytemplen
    \mytemplen\dp\mytempbox
    \advance\mytemplen #2\relax
    \dp\mytempbox\mytemplen
    \colorbox{black!10!white}{\hspace{1em}\usebox{\mytempbox}\hspace{1em}}}

\newcommand{\td}{\tilde{\delta}}

\def\MP{M_{\rm Pl}}
\newcommand*{\Cc}{\mathcal}%
\newcommand{\vect}[1]{\boldsymbol{#1}}
\newcommand{\de}{\, \mathrm{d}}	 
\newcommand{\nn}{\nonumber}
\def\b{\tilde{b}}
\def\bzeta{\bar{\zeta}}
\def\etaf{\eta_{\rm f}}

\newcommand{\be}{\begin{equation}}      
\newcommand{\ee}{\end{equation}}

\numberwithin{equation}{section}

\tikzset{cross/.style={cross out, draw=black, minimum size=2*(#1-\pgflinewidth), inner sep=0pt, outer sep=0pt},
cross/.default={4pt}}

\tikzset{
    master/.style={
        execute at end picture={
            \coordinate (lower right) at (current bounding box.south east);
            \coordinate (upper left) at (current bounding box.north west);
        }
    },
    slave/.style={
        execute at end picture={
            \pgfresetboundingbox
            \path (upper left) rectangle (lower right);
        }
    }
}

\date{\today}

\begin{document}
	\begin{flushright} {\footnotesize IPMU23-0040}  \end{flushright}
	
	\begin{center}
		\LARGE{\bf Non-perturbative Wavefunction of the Universe in Inflation with (Resonant) Features}
		\\[1cm] 
		
		\large{Paolo Creminelli$^{\,\rm a, \rm b}$, S\'ebastien Renaux-Petel$^{\,\rm c}$, Giovanni Tambalo$^{\,\rm d, e}$ \\ and Vicharit Yingcharoenrat$^{\,\rm f}$}
		\\[0.5cm]
		
		\small{
			\textit{$^{\rm a}$
				ICTP, International Centre for Theoretical Physics\\ Strada Costiera 11, 34151, Trieste, Italy}}
		\vspace{.2cm}
		
		\small{
			\textit{$^{\rm b}$
				IFPU - Institute for Fundamental Physics of the Universe,\\ Via Beirut 2, 34014, Trieste, Italy}}
		\vspace{.2cm}
		
		\small{
			\textit{$^{\rm c}$
				Institut d'Astrophysique de Paris, UMR 7095 du CNRS et de Sorbonne Universit\'e,\\ 98 bis bd Arago, 75014 Paris, France}}
		\vspace{.2cm}
		
		\small{
			\textit{$^{\rm d}$
				Max Planck Institute for Gravitational Physics (Albert Einstein Institute)\\ Am M\"uhlenberg 1, D-14476 Potsdam-Golm, Germany}}
		\vspace{.2cm}
		
		\small{
			\textit{$^{\rm e}$
				Institut f\"ur Theoretische Physik, ETH Z\"urich, 8093 Z\"urich, Switzerland}}
		\vspace{.2cm}
		
		\small{
			\textit{$^{\rm f}$
				Kavli Institute for the Physics and Mathematics of the Universe (WPI), The University of Tokyo Institutes for Advanced Study (UTIAS), The University of Tokyo, Kashiwa, Chiba 277-8583, Japan}}
		\vspace{.2cm}
				
	\end{center}
	
	\vspace{0.3cm} 
	
	\begin{abstract}\normalsize
	We study the statistics of scalar perturbations in models of inflation with small and rapid oscillations in the inflaton potential (resonant non-Gaussianity). We do so by deriving the wavefunction $\Psi[\zeta(\vect x)]$  non-perturbatively in $\zeta$, but at first order in the amplitude of the oscillations. The expression of the wavefunction of the universe (WFU) is explicit and does not require solving partial differential equations. One finds qualitative deviations from perturbation theory for $ |\zeta| \gtrsim \alpha^{-2}$, where $\alpha \gg 1$ is the number of oscillations per Hubble time. Notably, the WFU exhibits distinct behaviours for negative and positive values of $\zeta$ (troughs and peaks respectively). While corrections for $\zeta <0$ remain relatively small, of the order of the oscillation amplitude, positive $\zeta$ yields substantial effects, growing exponentially as $e^{\pi\alpha/2}$ in the limit of large $\zeta$.  This indicates that even minute oscillations give large effects on the tail of the distribution.		
	\end{abstract}
	
	\vspace{0.3cm} 
	
	\vspace{2cm}
	
	\newpage
	{
		\hypersetup{linkcolor=black}
		\tableofcontents
	}
	
	\flushbottom
	
	\vspace{1cm}

\newpage

\section{Introduction}\label{sec:intro}

It is widely believed that the initial conditions of our Universe were established during the period of inflation. Over the past two decades, significant progress has been made in understanding how to calculate the statistics of these initial perturbations. These calculations rely on in-in perturbation theory. The use of perturbation theory appears to be particularly justified, as experiments have placed strict constraints on non-Gaussianity, i.e.~departure from a free theory. This demonstrates that fluctuations during inflation are very weakly coupled. While {\em typical} fluctuations are indeed weakly coupled, it turns out that perturbation theory is not appropriate to describe large, unlikely fluctuations: non-linearities become sizeable when looking at large perturbations. To investigate such improbable perturbations, one can turn to non-perturbative semiclassical methods \cite{Celoria:2021vjw}.

Now, why should we concern ourselves with studying unlikely events that, by definition, are rarely observed?  One reason is conceptual. One is studying the most fundamental object of cosmology, the wavefunction of the Universe (WFU), which contains the complete information about the initial conditions. The possibility of studying it in a regime where perturbation theory fails is of the utmost importance, even if it were not relevant experimentally. The tail of the probability distribution also carries phenomenological significance: for instance the probability of generating a primordial black hole depends on the probability distribution far away from its typical values. Moreover, there are other inquiries where comprehensive knowledge of the WFU proves relevant, such as in the physics of eternal inflation, which may be sensitive to the tail of the probability distribution \cite{Creminelli:2008es,Dubovsky:2008rf,Dubovsky:2011uy}. Additionally, one might be interested in the probability of non-perturbative transitions to other vacua. We will also see that going beyond perturbation theory will also give new insights for the study of typical fluctuations.

In reference \cite{Celoria:2021vjw} it was shown that the WFU in the semiclassical limit can be expressed as
\be\label{eq:WFUsemi}
\Psi[\bar\zeta(\vect x)] \sim e^{i S[\zeta_{\rm cl}]/\hbar}\;.
\ee
The WFU is a functional of the scalar perturbation $\bar\zeta(\vect x)$ at late times. (In this paper we do not consider tensor perturbations and we focus on single-field models of inflation, so that $\zeta$ is the only variable.) The action on the right-hand side of the equation above is evaluated on-shell, i.e.~on the classical trajectory $\zeta_{\rm cl}$ that satisfies the boundary condition $\zeta_{\rm cl} = \bar\zeta(\vect x)$ at late times and the Bunch-Davies condition at early times. The semiclassical approximation is valid when the action is large compared with $\hbar$, and this occurs when the configuration $\bar\zeta(\vect x)$ is large, compared with a typical fluctuation. In this limit, loop corrections can be neglected and the functional integral that gives the WFU reduces to a single semiclassical configuration. Notice that we are keeping the full non-linear action and not expanding in perturbation theory: the semiclassical expression \eqref{eq:WFUsemi} resums all non-linearities that are enhanced by the large $\bar\zeta$. This method is general, but one still needs to solve a non-linear partial differential equation (PDE) with prescribed boundary conditions and evaluate the action on this classical solution. The solution of the PDE must be found numerically in general and this makes the process somewhat cumbersome and obscures the physics. (Other studies that emphasize the whole probability distribution of $\zeta$, beyond the usual expansion in correlation functions, include \cite{Chen:2018uul, Chen:2018brw, Panagopoulos:2019ail, Panagopoulos:2020sxp,Palma:2023idj}.)

In this paper we apply the method outlined above to a particular model of 
single-field inflation, where one has small and rapid oscillations of the inflaton potential superimposed to a standard slow-roll scenario. This model, sometimes dubbed resonant non-Gaussianity \cite{Chen:2008wn,Flauger:2009ab, Flauger:2010ja,Behbahani:2011it,DuasoPueyo:2023viy}, can be motivated by UV completions with monodromy \cite{McAllister:2008hb}. The amplitude of the oscillations will be taken to be a small parameter $\tilde b$. We will calculate the WFU non-perturbatively in $\bar\zeta$, but at first order in $\tilde b$. The important simplification is that at first order in $\tilde b$ one does not need to solve any PDE: it is enough to plug the solution in the absence of oscillations in the ${\cal O}(\tilde b)$ action.  Another reason why it is interesting to study the non-perturbative WFU in this class of models is related to the frequency of oscillations, which is usually taken to be mush faster than Hubble, $\omega \sim \alpha H$, $\alpha \gg 1$. The rapidity of the oscillations in the potential tells us that perturbation theory, which is based on expanding the potential in Taylor series around a given point, will break down soon: one cannot hope to study in perturbation theory a fluctuation that jumps from one minimum of the modulation to another. Indeed we will see that one needs the full non-perturbative WFU for $\alpha^2 |\bar\zeta| \gtrsim 1$. 
Eventually, notice that our approach takes into account all interactions, including here resonant interactions \textit{inside} the Hubble radius. This has to be contrasted with stochastic inflation, which only resums nonlinearities outside the Hubble radius, without taking into account non-Gaussianities at horizon crossing (see e.g.~\cite{Gorbenko:2019rza, Pinol:2020cdp, Vennin:2020kng, Cespedes:2023aal} and references therein).

In Sec.~\ref{sec:EFT_pi} we derive a particularly useful form of the effective field theory (EFT) of inflation \cite{Cheung:2007st} in the decoupling limit, when gravity perturbations can be neglected. This form, Eq.~\eqref{eq:action_pi_simple}, retains all non-linearities, it is valid for a generic potential and it has the advantage of making explicit the conservation of $\zeta$ in the long wavelength limit. (In App.~\ref{App:action_check} we verify that this form of the action is equivalent to others used in the literature, and in App.~\ref{App:mixing} we discuss corrections coming from the mixing with gravity.) In Sec.~\ref{sec:general} we focus on the case of periodic features in the potential and we derive a closed-form expression for the WFU at first order in $\tilde b$, Eq.~\eqref{eq:DeltaS1_E}. We also discuss the regime of validity of this formula, studying loop corrections (see also Apps.~\ref{App:bndry_loops} and \ref{app:1-loop_estimate}). It turns out that loop corrections are subdominant, in this particular model, even for typical fluctuations. The evaluation of the WFU still requires an integral over space and time. We do this both numerically and using a saddle-point approximation, valid in the regime $\alpha \gg 1$. These two methods are compared both in the case of a monochromatic profile for $\bar\zeta$, Sec.~\ref{sec:ODE_analysis}, and for a more general spherically symmetric configuration, Sec.~\ref{sec:PDE_analysis}. (Some details about the numerical analysis are deferred to App.~\ref{app:num_method}.)

The results show many new qualitative features that are absent in perturbation theory. First, the WFU has a large asymmetry between peaks and troughs of $\bar\zeta(\vect x)$: the effect of the oscillations in the potential is parametrically larger for peaks. Second, the modifications are not uniformly of order $\tilde b$: for $\alpha^2\bar\zeta \gtrsim 1$ one has very large effects, asymptotically going as $e^{\pi \alpha/2}$ for $\bar \zeta \gtrsim 1$. 
Third, the WFU exhibits oscillatory features as $\bar\zeta$ varies with frequency $\alpha$.
All these features may be understood in a quantum mechanical toy model (see Sec.~\ref{app:qm_example}), with a periodic perturbations of the Hamiltonian. The periodic modulation at high frequency induces transitions to excited states and these dominate completely the tail of the wavefunction compared to the ground state wavefunction.  

The study of the full WFU just began and many open questions remain as we discuss in the Conclusions, Sec.~\ref{sec:conclusions}.

\section{Decoupling limit of the EFT of Inflation}\label{sec:EFT_pi}

It is useful in many cases to study inflation in the limit $\epsilon \to 0$, keeping the power spectrum and the other slow-roll parameters fixed. Since the power spectrum $P_\zeta \sim H^2/(\MP^2 \epsilon)$ is fixed, this limit corresponds to inflation taking place at low energy, $H \to 0$, or equivalently this is the limit in which gravity decouples, $\MP \to \infty$. Data are progressively pushing towards this limit: the observation of a non-zero tilt together with the upper bounds on tensor modes imply a certain hierarchy $\epsilon \ll |\eta|$. There are various simplifications in this limit. First, the background geometry becomes exactly de Sitter. Second, perturbations can be studied in the decoupling limit, i.e.~without evaluating the perturbations of the geometry. Technically this means that the metric remains unperturbed and one does not need to solve the constraint equations for the lapse function $N$ and the shift vectors $N^i$. These variables are indeed $\epsilon$ suppressed: in the limit $\MP \to \infty$ one expects that the metric becomes non-dynamical. The solutions of the constraints thus give terms in the action suppressed compared with the action of the inflaton. The $\epsilon \to 0$ limit of slow-roll inflation is discussed in detail in \cite{Pajer:2016ieg}, while the decoupling limit is justified for typical fluctuations in the case of oscillatory features in \cite{Behbahani:2011it}. In App.~\ref{App:mixing}, we derive its regime of validity in a non-perturbative manner, as required for our analysis.

In this section we derive the action for the scalar perturbations $\pi$ in the decoupling limit, using the EFT of Inflation \cite{Cheung:2007st}. Throughout the paper, the background metric is assumed to be the flat Friedmann-Robertson-Walker (FRW) metric: $\de s^2 = -\de t^2 + a(t)^2 \de \vect{x}^2$, where $a(t)$ denotes the scale factor. The EFT action for a scalar field with a minimal kinetic term is \cite{Cheung:2007st}
\begin{align}\label{eq:EFT_action}
	S 
	= 
	\int \de^4x
	\,\sqrt{-g} 
	\bigg[
	\frac{\MP^2}{2} R  - \MP^2(3H(t)^2 + \dot{H}(t)) +  \MP^2 \dot{H}(t) g^{00} 
	\bigg] 
	\;, 
\end{align} 
where $R$ is the 4d Ricci scalar, $g^{00}$ is the $(00)$-component of $g^{\mu\nu}$ and $\MP$ is the Planck mass. We define the Hubble parameter as $H(t) \equiv \dot{a}/a$, where the dot is a time derivative. 
The action above is formulated in the \emph{unitary gauge}, having all the degrees of freedom inside the metric. The presence of the scalar degree of freedom can be made manifest by performing a space-time dependent time diffeomorphism and promoting the gauge parameter to a field $-\pi(t, \vect x)$.
This is nothing but the usual Stueckelberg trick: $t \rightarrow t + \pi(t,\vect{x})$, so that $g^{00}$ then transforms as
\begin{align}
	g^{00} \rightarrow (1+\dot\pi)^2 g^{00} + 2 (1+\dot\pi) \partial_i\pi g^{0i} + g^{ij}\partial_i\pi\partial_j\pi \simeq -1 - 2\dot{\pi} + (\partial_\mu \pi)^2\;.
\end{align}
In the last step we took the decoupling limit and neglected metric perturbations; notice that this can only be done after reintroducing $\pi$.
Therefore, the action (\ref{eq:EFT_action}) becomes
\begin{align}
	S 
	= 
	\int \de^4x\,\sqrt{-g} 
	\bigg[
	\frac{\MP^2}{2} R  
	- \MP^2(3H(t + \pi)^2 
	+ 2\dot{H}(t + \pi)) 
	+  \MP^2 \dot{H}(t + \pi)
	(-2\dot{\pi} + (\partial_\mu \pi)^2) 
	\bigg]
	\;.
\end{align}
The Einstein-Hilbert term, as expected, does not contain the field $\pi$ because it is invariant under 4d diffeomorphisms: since we are interested in the action for $\pi$ we can disregard this term from now on and write the action as
\begin{align}
	S 
	= 
	\int \de^4x \, a^3 \MP^2 
	\bigg[ 
	- 3H(t + \pi)^2 
	- 2 (1 + \dot{\pi}) \dot{H}(t + \pi) 
	+  \dot{H}(t + \pi) (\partial_\mu \pi)^2
	\bigg] 
	\;.
	\label{eq:action_pi_2} 
\end{align}
At linear order in $\pi$ the action above vanishes after performing an integration by parts: this is a consequence of the background equations of motion.

Let us now show that in the decoupling limit $\epsilon \to 0$, the field $\pi$ must be time-independent outside the horizon. Indeed, when all modes are well outside the horizon, the relation between $\pi$ and $\zeta$ reads \cite{Behbahani:2011it}
\be
	\zeta(t,\boldsymbol{x})
	=
	\int_{t}^{t+T(t,\boldsymbol{x})} 
	H(t') \de t'
	\,, 
	\quad \textrm{with} \quad \pi(t+T(t,\boldsymbol{x}),\boldsymbol{x})+T(t,\boldsymbol{x})
	=
	0 \;,
\label{change-gauge}
\ee
where the implicit equation defining $T$ is found by working out the time-diffeomorphism mapping the $\pi$ to the $\zeta$-gauge.~\footnote{The spatial diffeomorphism that is required beyond linear order in $\pi$ becomes negligible when all modes are well outside the horizon, see \cite{Behbahani:2011it} for an explicit proof in the context of resonant models.} In the decoupling limit, one can consider the Hubble rate constant, $H=H_\star$, hence $\zeta=H_\star T$. From the time-independence of $\zeta$ outside the horizon, one deduces that $T$ is also constant. Inspecting the relationship between $T$ and $\pi$ in \eqref{change-gauge}, one finds that this is realized only with $\pi$ constant, and hence with $\zeta=-H_\star \pi$. It is also instructive to go beyond the decoupling limit, in which case Eq.~\eqref{change-gauge} gives $\zeta=-H \pi+H \pi \dot{\pi}+ \dot{H}\pi^2/2+{\cal O}(\pi^3)$ up to quadratic order. This makes it manifest that the time-dependence of $H$ implies a time-dependence of $\pi$.

Although we are working in the decoupling limit, the time-independence of $\pi$ is not manifest in the action \eqref{eq:action_pi_2}: polynomial terms in $\pi$ naively induce an evolution outside the horizon. One can rewrite the action in a more transparent form. 
We rewrite the second term on the RHS as $-2\de [H(t + \pi)]/\de t$ and we integrate it by parts: up to terms that do not depend on $\pi$ the action \eqref{eq:action_pi_2} becomes
\begin{align}\label{eq:action_pi_recast}
	S 
	= 
	\int \de^4x\, a(t)^3 \MP^2
	\bigg[
	-3(H(t+\pi) - H(t))^2 
	+ \dot{H}(t + \pi) (\partial_\mu \pi)^2
	\bigg]
	\;.
\end{align}
The first term, $(H(t + \pi) - H(t))^2$, scales as $\sim \epsilon^2$ since it involves variations of $H$ during inflation. This is subdominant compared to the second term, $\dot{H}(t + \pi) (\partial_\mu \pi)^2$, which scales as $\sim \epsilon$.~\footnote{More precisely, $(H(t+\pi) - H(t))^2=(\int_{t}^{t+\pi} \dot{H}(t') \de t')^2 \leq \dot{H}^2_\textrm{max}  \pi^2 \leq \epsilon^2_\textrm{max} H^4_\textrm{ini} \pi^2$, where we used that $H$ is decreasing in the last step, shows that the first term is indeed negligible, inside the horizon, compared to the second one of order $\epsilon H^2 (\partial_\mu \pi)^2$. The impact outside the Hubble radius is discussed in App.~\ref{App:mixing}.} Actually, one is not allowed to retain the first term: as we discussed, solving the constraint equations for the lapse function and the shift vector would give extra terms in the action that scale as $\epsilon^2$, i.e.~of the same order as the first term, see App.~\ref{App:mixing}.
Therefore in the decoupling limit the action takes the simplified form: 
\begin{empheq}[box={\mybox[5pt][5pt]}]{equation}
\label{eq:action_pi_simple}
	S 
	= 
	\int \de^4x\, a(t)^3 \MP^2
	 \dot{H}(t + \pi) (\partial_\mu \pi)^2 \;,
\end{empheq}
where in $a(t)$, one should consider for consistency a de Sitter evolution $a(t) \propto e^{H_\star t}$. Since the only term in the action contains two derivatives, one can see explicitly that $\pi =$ const is a solution of the complete non-linear equation of motion. This action describes, in the limit $\epsilon \to 0$, a model of inflation with a minimal kinetic term and a generic potential that may include oscillations or features as we are going to discuss momentarily. One can add extra terms in the EFT, like $(g^{00}+1)^n$ or extrinsic curvature terms. All these terms contain at least two derivatives on $\pi$ and therefore do not affect the argument for the conservation of $\pi$ outside the horizon.

It is noteworthy that the action \eqref{eq:action_pi_simple} is not formulated perturbatively: the nonlinearities that it contains are expressed in a resummed manner, which is crucial when dealing with non-perturbative phenomena and rare large fluctuations. Eventually, our action can be used for any single-clock model of inflation, provided one is not interested in $\mathcal O (\epsilon^2)$ terms. If one is agnostic about the underlying dynamics driving inflation, the dynamics of $\pi$ can then be obtained simply by parametrizing the time evolution of the Hubble rate during inflation, from which one deduces
\begin{align}
\ddot{\pi}+\left[3 H_\star+\frac{\ddot{H}(t+\pi)}{\dot{H}(t+\pi)} \right] \dot{\pi}-\frac{\partial^2_i \pi}{a^2}=-\frac{\ddot{H}(t+\pi)}{2 \dot{H}(t+\pi)} \left[\dot \pi^2-\frac{(\partial_i \pi)^2}{a^2} \right]  \,.
\label{eq:full-eom}
\end{align}
Remarkably, this compact expression is the full non-linear equation of motion, encapsulating all non-linearities, in any model of inflation involving a canonical single scalar field, in the decoupling limit.

The classical solutions of this equation, with suitable boundary conditions, can be used, following \cite{Celoria:2021vjw}, to analyse the WFU in the large $\zeta$ limit. The corresponding PDE, however, can only be solved numerically. In this paper we concentrate on the case in which one has a feature, localised or periodic, superimposed to a smooth slow-roll potential. In this case one has another expansion parameter, the amplitude of the feature, and in this case we will be able to get analytic results. For concreteness we focus on periodic features, i.e.~the case of resonant non-Gaussianity.

\section{Wavefunction of the universe for resonant features}
\label{sec:general}

\subsection{Resonant features}\label{sec:potential}

In the following, we will compute the WFU when the time-dependence of the Hubble rate is assumed to verify
\be
\dot{H}(t)=\dot{H}_\star \left[1-\b \cos(\omega t+\delta) \right]\,,
\label{sec:model-EFT}
\ee
where all parameters $\dot{H}_\star,\b, \omega, \delta$ are constant, and when treating the oscillatory part as a perturbation, i.e.~at first order in the parameter $\b$. As we stressed, our method is readily applicable beyond these assumptions, but this simple form will enable us to derive analytical results.

While the form \eqref{sec:model-EFT} is a perfectly legitimate starting point from an EFT point of view, in this section we explain that it is indeed a good approximation to the dynamics of the Hubble rate in motivated models, and discuss its regime of validity in this context. Explicitly, let us consider models of inflation driven by a scalar field with canonical kinetic term and potential
\begin{equation}\label{eq:potential}
	V(\phi) 
	=
	V_{\rm sr}(\phi) 
	+ 
	\Lambda^{4} \cos\left(\phi / f\right)
	\;,
\end{equation}
where $V_{\rm sr}(\phi)$ is a generic slow-roll potential, $f$ is the analogue of the axion decay constant and $\Lambda$ is 
the scale that controls the amplitude of the oscillations of the potential. The specific model with $V_{\rm sr}(\phi) = \mu^3 \phi$ has been studied in detail in \cite{Flauger:2009ab} but we keep $V_{\rm sr}$ generic as in \cite{Flauger:2010ja,Leblond:2010yq}. We could also allow $\Lambda$ to depend on the scalar field in a slow-roll manner and results would equally hold, but we consider $\Lambda$ constant for simplicity.

The full equations governing the background dynamics are the standard ones:
\begin{align}
	&\ddot \phi + 3H \dot \phi + V'(\phi) = 0 \;, 	 \label{eq:eom_bkg_phi}\\ 
	& 3 H^2 \MP^2 = \frac{\dot \phi^2}{2} + V(\phi)\;,\label{eq:eom_bkg_friedman}
\end{align}
implying $-2\MP^2 \dot{H} = \dot \phi^2$. At zeroth order in the oscillatory component, 
$V_{\rm sr}$ is driving a standard phase of slow-roll inflation, whose corresponding quantities  $\phi_0(t),H_0(t)$ we denote with an index $0$. Up to first order in the oscillatory component (the precise expansion parameter will be made explicit below), the time derivative of the Hubble rate reads $-2\MP^2 \dot{H}=\dot{\phi}_0^2+2 \dot{\phi}_0 \dot{\phi}_1$, where quantities at first order are denoted with an index $1$, and one has
\be
\ddot{\phi}_1+3 H_0 \dot{\phi}_1+3 H_1 \dot{\phi}_0+V''_\textrm{sr}(\phi_0) \phi_1=\frac{\Lambda^4}{f} \sin(\phi_0/f)\,.
\label{eq-phi1}
\ee
We are interested in the regime where the frequency of variation of the oscillatory component $\dot \phi_0/f$ is large compared to the Hubble scale, i.e.~where $\alpha \equiv |\dot \phi_0|/(H_0 f) \gg 1$, a regime in which non-Gaussianities are resonantly enhanced \cite{Chen:2008wn,Flauger:2010ja,Leblond:2010yq}. In this regime, the left-hand side is dominated by the two-derivative term, with the approximate solution
\be
\phi_1=-\frac{\Lambda^4 f}{\dot{\phi}_0^2} \sin(\phi_0/f)\,.
\label{phi1-solution}
\ee
Note that all quantities like $H_0, \phi_0$ and $\alpha$ have a mild, slow-roll, time dependence with the usual successive Hubble slow-roll parameters $\epsilon_0=-\dot{H}_0/H_0^2=\dot{\phi}_0^2/(2 H_0^2 \MP^2), \eta_0=\dot{\epsilon}_0/(H_0 \epsilon_0), \ldots$ much smaller than unity.
Hence one can check that \eqref{phi1-solution} is indeed an approximate solution to \eqref{eq-phi1}: as the sine term varies much more rapidly than $\dot{\phi}_0$,  $\dot{\phi}_1$ scales like $\alpha H_0 \phi_1$, and hence it is immediate that the friction term $3 H_0 \dot{\phi}_1$ and the mass term $V''_\textrm{sr}(\phi_0) \phi_1$ are negligible compared to $\ddot{\phi}_1$. Let us show that $3 H_1 \dot{\phi}_0$ is also negligible. For this, note that \eqref{eq:eom_bkg_friedman} expanded at first order gives
\be
6 H_0 H_1 \MP^2=\dot{\phi}_0 \dot{\phi}_1+V'_\textrm{sr}(\phi_0) \phi_1+\Lambda^4 \cos(\phi_0/f)\,.
\label{H1}
\ee
The first and the last term on the right-hand side of \eqref{H1} contribute to $H_1 \dot{\phi}_0/\ddot{\phi}_1$ as $\epsilon_0/\alpha \ll 1$, and the second one is even further suppressed by $1/\alpha$. 

Now, using the solution \eqref{phi1-solution}, one finds the expression for $\dot{H}$ up to first order:
\be
	\dot{H}
	=
	-\epsilon_0 H_0^2 
	\bigg[
	1-\frac{2\Lambda^4}{\dot{\phi}_0^2} \cos(\phi_0/f) 
	\bigg]
	\,.
\label{Hdot-solution}
\ee
That is, in addition to the slow-roll dependence $\dot{H}_0=-\epsilon_0 H_0^2$ at zeroth order, $\dot{H}$ acquires a rapidly varying oscillatory component. This is similar to the form \eqref{sec:model-EFT} on which we will concentrate. It corresponds to the approximation in which the slow-varying quantities, both $\epsilon_0 H_0^2$ and the relative size of the oscillations $2 \Lambda^4/\dot{\phi}_0^2$, are considered as constant. More precisely, if one expands \eqref{Hdot-solution} around a pivot time $t_\star$, then on time scales $\Delta t=t-t_\star$ smaller than the scales of variation of the slow-roll part, i.e.~for $H_0 \Delta t \ll (1/\epsilon_0(t_\star),1/\eta_0(t_\star))$ and other combinations of inverse of slow-roll parameters at higher-order, one can consider all quantities as constant except for $\phi_0/f$ in the cosine term, which can be approximated by $(\phi_0(t_\star)+\dot{\phi}_0(t_\star)(t-t_\star))/f$. One then obtains the form \eqref{sec:model-EFT} with 
\begin{equation}\label{matching}
\begin{aligned}
	\dot{H}_\star &= \dot{H}_0(t_\star) \,, \quad \quad \quad 
	\omega = |\dot{\phi}_0(t_\star)|/f \,,
	 \\
	 \b &= 2 \Lambda^4/\dot{\phi}_0^2(t_\star)\,, \quad  \ \delta = \textrm{sign}(\dot{\phi}_0(t_\star))\phi_0(t_\star)/f-\omega t_\star
	 \,,
\end{aligned}
\end{equation}
and we chose $\omega >0$.~\footnote{From the approximate time-dependence of the scalar field we have derived, $(\phi(t)-\phi_0(t_\star))~\textrm{sign}(\dot{\phi}_0(t_\star)) \simeq \omega f (t-t_\star)  -\frac12 \b f \sin(\omega t+\delta)$, one obtains the link between the Goldstone boson $\pi$ and the fluctuation of the scalar field $\varphi(t)=\phi(t+\pi)-\phi(t)$, namely $\pm \varphi/f=\omega \pi-\frac12 \b [\sin(\omega (t+\pi)+\delta)-\sin(\omega t+\delta) ]$. With $\zeta=-H_\star \pi$ outside the horizon in the decoupling limit that we consider, this gives the fully nonlinear relationship between $\varphi$ and the observed curvature perturbation $\zeta$. The probability density function of $\zeta$ can then be deduced from the one of $\varphi$: $\mathcal{P}(\zeta)=\mathcal{P}(\varphi(\zeta)) \alpha f [1-\frac12 \b \cos(\omega t+\delta -\alpha \zeta)]$. However, this is of little practical use. Considering that $\varphi$ soon after Hubble crossing is Gaussian, as one would do in stochastic inflation in slow-roll models, would lead to completely wrong results. Instead, our approach takes into account the resonant interactions \textit{inside} the Hubble radius (actually all interactions), directly at the level of the Goldstone boson, which enables us to derive $\mathcal{P}(\zeta)$ straight away.} 
Note that another parameter $b=\frac{\Lambda^4}{f V'_\textrm{sr}(\phi_0(t_\star))}$ is used in references \cite{Flauger:2009ab,Flauger:2010ja,Leblond:2010yq}. It is simply related to our parameter $\b \simeq 6 |b| /\alpha(t_\star)$ upon using the slow-roll equations. We prefer to use $\b$ since, as pointed out in \cite{Behbahani:2011it}, $b$ does not have to be $\ll 1$. Instead, $\b < 1$ is a necessary requirement to satisfy the null energy condition, i.e.~to ensure that $\dot{H}$ in \eqref{sec:model-EFT} is always negative.

To summarize, the simple form \eqref{sec:model-EFT} for $\dot{H}(t)$ on which we will concentrate is a good local approximation of the dynamics, in models of the type \eqref{eq:potential} when $\b \ll 1$. As this is only a local approximation in time, it means that 
one cannot consider arbitrarily large values of $\pi$ in \eqref{eq:action_pi_simple} to study the WFU for rare large values of $\zeta=-H \pi$. As we work in the decoupling limit in which $\epsilon_0 \ll \eta_0$, one finds that one should restrict to values of $|\zeta| \ll 1/\eta_0$.~\footnote{More generally, one should also require $|\zeta| \ll \left[n!\dot{H}_0 H_0^n\left(\frac{\de^{n+1} H_0}{\de t^{n+1}}\right)^{-1}\right]^{1/n}$ ($n \geq 1$).} However, we stress that this is not a theoretical limitation of our method, which is valid for arbitrarily large values of $|\zeta|$ once the expansion history $H(t)$ is known: it simply illustrates that the extreme tail of the WFU is sensitive to the whole inflationary history.~\footnote{For completeness, we show in App.~\ref{App:mixing} that neglecting the mixing with gravity requires $|\zeta| \ll 1/\sqrt{\epsilon \alpha^3}$.}

\subsection{Wavefunction of the universe}\label{sec:finite-wfu}

In this section we are going to express the wavefunction of the curvature perturbation $\zeta$ up to first order in $\b$ in the semiclassical limit, i.e.~neglecting loops (we will discuss below the range of validity of this approximation) in models with resonant features described by the expansion history \eqref{sec:model-EFT}. We will pay attention to put it in a manifestly finite form so that it can be computed in the subsequent sections using both numerical and analytical methods.

Before delving into the analysis, it is useful to compare our action (\ref{eq:action_pi_simple}) to the one used in the literature to study resonant features. The ``derivative" form of the action we are using is in fact equivalent to the one that was used to calculate the $n$-point functions of $\zeta$ in these models (see Eq.~(24) of \cite{Behbahani:2011it} or Eq.~(4.1) of \cite{Leblond:2010yq}), up to integrations by parts. We explicitly show this in App.~\ref{App:action_check}. 
The advantage of using our action is that the conservation of $\pi$ is manifest, i.e.~the equation of motion admits a constant solution. On the other hand, in the form of the action used in the literature, the conservation of $\pi$ is not manifest because 
the action contains non-linear self-interactions $\pi^n$. This ``polynomial" form has also the disadvantage that boundary terms must be kept in order to compute correlation functions, see App.~\ref{App:action_check} for more details.

Specifying Eq.~\eqref{eq:action_pi_simple} to the time dependence \eqref{sec:model-EFT}, our action of interest reads
\begin{align} 
	S
	= 
	\int \de^4x
	\,a(t)^3 
	\MP^2 \dot{H}_\star
	\bigg[
	1 - \b \cos(\omega t + \omega \pi+\delta) 
	\bigg] 
	(\partial_\mu \pi)^2 
	\;,
	\label{eq:Lag_pi} 
\end{align}
where remember that one should write $a(t) \propto e^{H_\star t}$ for consistency.
Let us now proceed with the calculation of the WFU. Starting from the action (\ref{eq:Lag_pi}) one can write down the classical non-linear equation of motion of $\pi$, Eq.~\eqref{eq:full-eom}, and solve such a differential equation with boundary conditions at $\eta \rightarrow -\infty$ and $\eta \rightarrow 0$, where $\eta$ is the conformal time such that  $a(\eta)=-1/(H_\star \eta)$. Then, one can compute the WFU in the semiclassical limit, which is essentially the exponential of the on-shell action \cite{Celoria:2021vjw}. 
This procedure is quite involved since it requires solving a non-linear partial differential equation for $\pi$.
Since we are interested in computing the WFU at linear order in $\b$, we will not need to perform this complicated task. Indeed, it is sufficient to calculate the action \eqref{eq:Lag_pi} on the ``free'' solution of $\pi$ ---the one with $\b=0$. 
This comes from the fact that
\begin{equation}
	S[\pi=\pi_0+\b \pi_1]
	=
	S[\pi_0]
	+ \b \int \de^4 x\, \pi_1\, 
	\left(
	\frac{\delta {\cal L}}{\delta \pi}
	\right)\bigg|_{\pi=\pi_0,\b=0}
	+{\cal O}(\b^2) \;, 
\end{equation}
where we explicitly factored out $\b$ in the expansion $\pi = \pi_0 + \b \pi_1+\ldots$ and where, by definition, $\pi_0$ satisfies the free equation of motion, 
i.e.~$(\delta {\cal L} / \delta \pi)\big|_{\pi=\pi_0,\b=0}=0$. 
As we see, the equality $S[\pi=\pi_0+\b \pi_1]=S[\pi_0]+{\cal O}(\b^2)$ is a generic fact that can be used in any model with a small expansion parameter, and is not tied to the specific form \eqref{eq:Lag_pi}.~\footnote{For the sake of completeness, let us verify it in this case. The action \eqref{eq:Lag_pi} up to first order in $\b$ reads
\begin{align} 
	S[\pi_0 + b\pi_1] 
	\simeq S_0[\pi_0] 
	&
	- 2 \b \int \de\eta \de^3 \vect x\,
	\frac{\MP^2 \dot{H}_\star}{H_\star^2\eta^2}
	\bigg[
	\pi_0' \pi_1' - \partial_i \pi_0 \partial_i \pi_1
	\bigg]  
	\nonumber \\ 
	&
	+\b \int \de\eta \de^3 \vect x\,
	\frac{\MP^2 \dot{H}_\star}{H_\star^2\eta^2}
	\bigg[
	\pi_0'^2 - (\partial_i \pi_0)^2
	\bigg]
	\cos(\omega t + \omega \pi_0+\delta) 
	\;,
	\label{eq:Lag_pi_2} 
\end{align}
where a prime denotes a derivative with respect to the conformal time $\eta$, the first term is defined by $S_0 [\pi_0] 
	\equiv 
	- \int \de\eta \de^3 \vect x \, 
	\frac{\MP^2 \dot{H}_\star}{H_\star^2\eta^2} 
	\big[
	\pi'^2_0 - (\partial_i \pi_0)^2
	\big]$, 
and $\pi_0$ verifies the linear equation of motion $\pi''_0 - (2/\eta) \pi'_0 - \partial_i^2 \pi_0 = 0$. Performing an integration by parts in the second term on the RHS of \eqref{eq:Lag_pi_2} thus gives rise to the equation of motion of $\pi_0$ (plus boundary terms, which vanish since we are imposing $\pi_1 = 0$ at early and late times). Thus, the terms that contain $\pi_1$ vanish using the free equation of motion of $\pi_0$, and only the first and last term remain, whose sum is nothing else than the full action, evaluated on the free solution $\pi_0$. Hence, the on-shell action at first order in $\b$ can indeed be evaluated by using the free solution $\pi_0$.}

Let us now define $\zeta \equiv -H_\star \pi_0$. The on-shell action as a function of the late-time value $\bar\zeta(\vect x)$ of the curvature perturbation becomes
\begin{align}
	S[\bzeta]
	=  
	\int \de\eta \de^3 \vect x~
	\frac{1}{2 \eta^2 P_\zeta} 
	\bigg[1 -  \b
	\cos\left(\alpha \left(\log(\eta/\eta_\star)+ \zeta \right)  -\td \right)
	\bigg] 
	\bigg[\zeta'^2 - (\partial_i \zeta)^2\bigg]
	\;, 
	\label{eq:action_Zeta}
	{}
\end{align}
where we have defined $P_\zeta \equiv H_\star^4/(2 \MP^2 |\dot{H}_\star|)$, $\td \equiv \delta+\omega t_\star$ is simply $\textrm{sign}(\dot{\phi}_0(t_\star)) \phi_0(t_\star)/f$ when \eqref{sec:model-EFT} comes from the scalar field model \eqref{eq:potential}, and $\alpha \equiv \omega/H_\star$ (this naturally coincides with the previously defined $\alpha$ in the scalar field model, simply evaluated at $t_\star$). Note one thing: $\zeta$ inside the integrand does not coincide with the curvature perturbation at all times, but it is simply a rescaled version of the free $\pi_0$. However, as we have seen in Sec.~\ref{sec:EFT_pi}, $-H_\star \pi$ does agree with the curvature perturbation at late times, and we also impose that $\pi_1$ vanishes at the end of inflation, so that $\zeta \equiv -H_\star \pi_0$ approaches the curvature perturbation field $\bzeta$. Hence, Eq.~\eqref{eq:action_Zeta} gives the on-shell action as a function of the late-time $\bar\zeta(\vect x)$.

~
\newline
\textbf{Late-time divergences and Euclidean action:}
Let us study the divergences of the action \eqref{eq:action_Zeta} at late times, in order to express its physical part in a manifestly finite form.
Note that $\zeta(\eta,\vect x)$ is defined as the unique solution of the partial differential equation $\zeta'' - (2/\eta) \zeta' - \partial_i^2 \zeta= 0$ that asymptotes to the configuration $\bzeta(\vect x)$ at the end of inflation, and with suitable behaviour at past infinity. Its Fourier transform $\zeta(\eta, \vect k) =
		\int \de^3 \vect x\,
		\zeta(\eta, \vect x) e^{-i \vect k \cdot \vect x}$ reads
\begin{equation}
		\zeta(\eta, \vect k) 
		= 
		\bzeta(\vect k) 
		\frac{(1-i k \eta) e^{i k \eta}}{(1 - i k \etaf) e^{i k \etaf}} 
		\;, 
		\label{eq:zeta_free_fourier}
	\end{equation}
where $\bzeta(\vect k)$ denotes the Fourier transform of the late-time configuration, $k = |\vect{k}|$, $\etaf$ is an arbitrary late-time regulator (which will be sent to $0$). We have selected only the $e^{i k \eta}$ solution corresponding to the usual Bunch-Davies vacuum. It corresponds to deforming the contour of the time integration $\eta \to \eta(1-i \epsilon)$ in \eqref{eq:action_Zeta}. Notice that $\zeta^*(\eta, \vect k) \neq \zeta(\eta, -\vect k)$, where the star denotes complex conjugation: $\zeta(\eta,\vect x)$ is not real, i.e.~it does not correspond to any physical configuration of the field, but it simply gives a configuration that dominates the path integral in the semiclassical limit.

Let us take $\etaf = 0$ and consider the limit $\eta \to 0$
\begin{equation}
	\zeta(\eta, \vect k) 
	\simeq
	\bzeta(\vect k) 
	\left[ 
	1 
	+\frac{1}{2}k^2\eta^2
	+\frac{i}{3}k^3\eta^3
	+ \mathcal O(\eta^4)
	\right]
	\,.
\label{eq:zeta_late_times}
\end{equation}
Then, using the inverse Fourier transform $\zeta(\eta, \vect x) =
		\int \frac{\de^3 \vect k}{(2 \pi)^3}
		\zeta(\eta, \vect k) e^{i \vect k \cdot \vect x}$, 
we arrive at
\begin{equation}
		\zeta(\eta, \vect x) 
		\simeq
		\left[ 
		\bzeta(\vect x)  
		-\frac{1}{2}\eta^2 \nabla^2 \bzeta(\vect{x})
		+\frac{i}{3}\eta^3 \nabla^3 \bzeta(\vect{x})
		+ \mathcal O(\eta^4)
		\right]
		\;,
\label{eq:zeta_late0}
\end{equation}
where we defined the inverse Fourier transform of $k^3  \bzeta(\vect k)$ as the non-local operator $\nabla^3 \bzeta(\vect{x})$. From the expression above, the only term in the free action that diverges at late time is $(\partial_i \bzeta)^2/\eta^2$. This divergent term, as pointed out in \cite{Maldacena:2002vr} (see also \cite{Celoria:2021vjw}), is real and thus it gives a pure phase in the WFU that depends on the late-time regulator $\eta_{\rm f}$. This phase does not contribute to the modulus squared of the WFU and so it does not matter if we are interested in observables related to $\zeta$. (The phase of the WFU is relevant if one is interested in the momentum conjugate to $\zeta$, which however decays exponentially at late times.) Since it is irrelevant for late-time observables, in order to deal with finite quantities, we can subtract this divergence from the action.

Apart from the free term, in \eqref{eq:action_Zeta} there is a divergent contribution at order $\b$ as well. Using \eqref{eq:zeta_late0} one obtains 
\begin{align}\label{eq:cos_late}
	\cos\left(\alpha \left(\log(\eta/\eta_\star)+ \zeta \right)  -\td \right)
	=  
	\ & \cos\left(\alpha \left(\log(\eta/\eta_\star)+ \bzeta \right)  -\td \right)  \nonumber \\
	&+\frac{1}{2}\alpha \eta^2 
	\sin\left(\alpha \left(\log(\eta/\eta_\star)+ \bzeta \right)  -\td \right) 
	\nabla^2 \bzeta 
	+ \ldots 
\end{align}
The second term on the right-hand side gives a finite contribution in the action since the $\eta^2$ factor cancels the same factor at the denominator of Eq.~\eqref{eq:action_Zeta}.
The divergent term in the action is thus 
$\cos\left(\alpha \left(\log(\eta/\eta_\star)+ \bzeta \right)  -\td \right) (\partial_i \bzeta)^2/\eta^2.$ 
This again contributes to a pure phase in the WFU that can be dropped. Therefore, the finite action at first order in $\b$ is
\begin{align}
	\Delta S
	= 
	\Delta S_0
	+ 
	\tilde{b} \, \Delta S_1
	\;. 
\label{eq:action_zeta_finite}
\end{align}
Here we define
\begin{align}
	\Delta S_0 
	\equiv 
	\int \de\eta \de^3 \vect x
	\,
	\frac{1}{2\eta^2 P_\zeta} 
	\big[
	\zeta'^2 - (\partial_i \zeta)^2 +  (\partial_i \bzeta)^2
	\big] 
	\;, 
	\label{eq:action_b0}
\end{align}
and
\begin{align}
	\Delta S_1 
	\equiv 
	-  
	\int \de\eta \de^3 \vect x \,
	\frac{1}{2\eta^2 P_\zeta} 
	\bigg\{
	&
	\big[
	\zeta'^2 - (\partial_i \zeta)^2 
	\big]  
	\cos\left(\alpha \left(\log(\eta/\eta_\star)+ \zeta \right)  -\td \right) 
	\nonumber \\
	&+ (\partial_i \bzeta)^2 
	\cos\left(\alpha \left(\log(\eta/\eta_\star)+ \bzeta \right)  -\td \right)
	\bigg\}
	\,, 
	\label{eq:action_zeta_b1}
\end{align}
where, as above, the subscripts $0$ and $1$ refer to the actions at zeroth and first order in $\b$ respectively. $\Delta S_0$ gives the Gaussian wavefunction, while $\Delta S_1$ gives all the deviations from Gaussianity of the probability distribution of $\bar{\zeta}$.

It is convenient to rotate to Euclidean time, where $\zeta$ is real and exponentially decaying at early times, instead of oscillating. Since the action \eqref{eq:action_zeta_finite} is analytic everywhere in the upper-left quadrant of the complex $\eta$ plane, and the integrand decays sufficiently fast at infinity, one can indeed perform a rotation to Euclidean space: 
$\eta \to -i \tau$ where $\tau$ denotes the Euclidean time. Notice that it was necessary to make the integrals in \eqref{eq:action_b0}--\eqref{eq:action_zeta_b1} convergent at $\eta \to 0$ before doing the rotation.~\footnote{Some more details about the rotation. $\zeta$ is an analytic function of $\eta$ and also the logarithm is analytic in the quadrant of interest, except at the origin. Since also the cosine is analytic, one has only to worry about the arc at infinity and the origin. Regarding the arc at infinity, notice that the imaginary part of the logarithm is bounded in the quadrant of interest: this makes the modulus of the cosine bounded and the convergence is guaranteed by the $1/\eta^2$ term (notice that $\zeta$, but not $\bar\zeta$, is exponentially decaying at infinity). Regarding the origin: one can neglect the integration along the infinitesimal quarter of the circle close to the origin since the modulus of the integrand is bounded.}
We write the exponent of the WFU as $i \Delta S_1 = - \Delta S_{\rm{E}, 1}$, where we define $\Delta S_{\rm{E}, 1}$ as the Euclidean action.
After the analytical continuation to Euclidean time, \eqref{eq:action_zeta_b1} leads to
\begin{empheq}[box={\mybox[5pt][5pt]}]{equation}
	\begin{split}
	\Delta S_{\rm{E},1}[\bzeta] 
	=
	\int_{-\infty}^0 \de\tau \int \de^3 \vect x\,
	\frac{1}{2\tau^2 P_{\zeta}} 
	\bigg\{	 
	\big[
	 \zeta'^2 + (\partial_i \zeta)^2 
	 \big]
	\cos\left( \alpha \left(\log(\tau/\eta_\star)+ \zeta \right)  -\td-i \alpha \pi/2\right)
	 \\ \label{eq:DeltaS1_E}
	-(\partial_i \bzeta )^2 
	\cos\left(\alpha \left(\log(\tau/\eta_\star)+ \bzeta \right)  -\td-i \alpha \pi/2\right) 
	\bigg\} \;,
	\end{split}
\end{empheq}
where a prime now refers to a derivative with respect to $\tau$, and explicitly the real variable $\zeta(\tau,\vect x)=\int \frac{\de^3 \vect k}{(2 \pi)^3}
		\bzeta(\vect k) e^{i \vect k \cdot \vect x}(1-k \tau)e^{k \tau}$. Note that $\zeta(\tau,\vect x)$ is real, but in the action one has an imaginary constant inside the cosine as a consequence of the Euclidean rotation. Therefore $\Delta S_{\rm{E},1}$ is complex. In the rest of the paper we will evaluate the action \eqref{eq:DeltaS1_E} using analytical and numerical methods.
		
To understand the behaviour of the WFU, it is useful to fix a certain ``shape" for $\bzeta(\vect x)$, say a spherically symmetric Gaussian, and study the WFU as a function of the overall size: in this way one has a function of a single variable that we can call $\bzeta$ with an abuse of notation. One expects that the result for small $\bzeta$ is the one of perturbation theory, while things get non-perturbative for large $\bzeta$. It is interesting to notice that the WFU as a function of this single variable $\bzeta$ has no singularity in the whole complex $\bzeta$ plane. This is evident from the explicit expression Eq.~\eqref{eq:DeltaS1_E}, since the cosine is an entire function. This means that the series in $\bzeta$, which is the perturbation-theory series, has infinite radius of convergence. The non-perturbative results we are going to present are the sum of the perturbative series.

Our results can be applied to the general case of bounded features in $H(t)$.
In situations where $\dot H(t) = \dot H_0 + \tilde b \dot H_1(t)$, with the feature assumed to be controlled by the small parameter $\tilde b$, most of the steps of the previous section still hold, so that we can write down the leading correction to the on-shell action.
If we define the dimensionless function $h(t) \equiv \dot H_1(t) / \dot H_0$, then the correction to the action is
\begin{empheq}[box={\mybox[5pt][5pt]}]{equation}
\label{eq:DS_1_general}
	\Delta S_{1}[\bzeta] =
	\int \de \eta \de^3 \vect x\,
	\frac{1}{2 \eta^2 P_{\zeta}} 
	\bigg\{
	\big[
	 \zeta'^2 - (\partial_i \zeta)^2 
	 \big]
	h\left(t - \zeta / H_0 \right) 
	+(\partial_i \bzeta )^2 
	h
	\left( t - \bzeta / H_0 \right) 
	\bigg\}
	\;,
\end{empheq}
where $t$ must be understood as function of the conformal time $\eta$, we subtracted the divergent contribution at late times, assuming that $h$ is bounded in this limit, and we replaced $\pi$ with $\zeta$ as before.
Formula \eqref{eq:DS_1_general} represents one of the main results of our paper. It can be used to compute the WFU in models with a small feature in $H(t)$, at first order in the amplitude of the feature, but non-perturbatively in $\bar{\zeta}$. Note that the rotation to Euclidean time $\tau$ can however be more subtle and needs to be studied case-by-case: the function $h$ could feature singularities in the complex plane.

\subsection{Regime of validity and relationship with perturbation theory}
Let us study the regime of validity of the semiclassical result of Eq.~\eqref{eq:DeltaS1_E}. In the methodology employed in \cite{Celoria:2021vjw}, the semiclassical approximation proves reliable for the tails of $\Psi$, specifically when $|\bar\zeta| \gg P_{\zeta}^{1/2}$. On the tails, tree-level diagrams are enhanced relative to loops: at a given order in perturbation theory, tree-level diagrams are enhanced compared to loops by the amplitude $|\bar\zeta|$ of the external legs.
This conclusion remains true in the presence of features. However, in this case loops are negligible compared with tree-diagrams even for typical values of $\bar \zeta$, as we will show momentarily. Unless specified, the whole discussion is a first order in $\tilde b$.

\begin{figure}
\centering
  \includestandalone[width=1\textwidth]{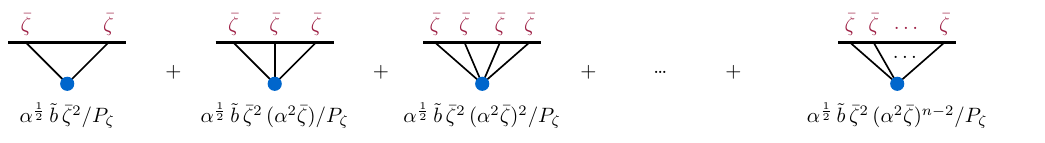}
  \newline
  \vspace*{0.1cm}
  \includestandalone[width=1\textwidth]{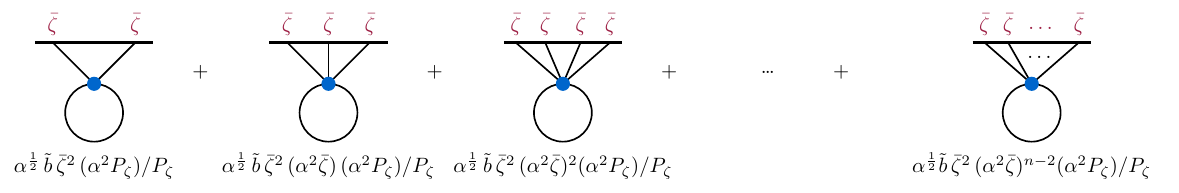}
  \caption{~Witten diagrams contributing to first order in $\tilde b$, with corresponding scalings with $\alpha$, $\bzeta$ and $P_{\zeta}$, including the effect of the resonance. 
  \textbf{Top:} Tree-level diagrams. The first diagram on the left corresponds to the correction to the power spectrum. The second is the contribution from the cubic part of the action and so on. 
  \textbf{Bottom:} One-loop diagrams at first order in $\tilde b$. 
  At any given order in $\bzeta$, one-loop diagrams are suppressed by $\alpha^2 P_{\zeta}$.}
  \label{fig:Witten_diags} 
\end{figure}
~
\newline
\textbf{Scaling of tree-level diagrams:} 
Let us study the scaling of the tree-level Witten diagrams. These correspond to the expansion in powers of $\zeta$ of the on-shell action of Eq.~\eqref{eq:action_Zeta}. We first determine the scaling of the vertex and then the $\alpha$ dependence of the time integral.
To obtain the scaling of the vertex, we need to Taylor-expand the action in powers of $\zeta$. 
The vertex with $n$ powers of $\zeta$ is obtained by expanding the cosine function at order $n-2$, see Eq.~\eqref{eq:vertex_n+2}: this gives $\alpha^{n-2}$. The time integral involves an oscillating function of $\eta$ that goes as either $\cos(\alpha \log(-\eta))$ or $\sin(\alpha \log(-\eta))$, depending on whether $n$ is even or odd.
At order $n$ the structure of the integral is
\begin{equation}\label{eq:estimate_tree_scaling}
	\frac{1}{n!} \int_{-\infty}^{0}\frac{\de \eta}{\eta^2}
	\cos(\alpha \log(-\eta)) 
	\, 
	e^{i k_{\rm t} \eta}
	\prod_{i = 1}^{n}(1-i k_i \eta)
	\underset{\alpha \gg 1}{\sim}
	\frac{k_{\rm t}}{n^{2n}} \sqrt{\alpha} \, \alpha^{n-2}\,,
\end{equation}
with $k_{\rm t} \equiv \sum_{i = 1}^{n} k_i$, and where we considered all $k_i$ to be comparable. The integral is estimated in saddle-point approximation with the saddle at $-k_{\rm t}\eta_s \sim \alpha$ (\footnote{The saddle point $\eta_s$ moves to later and later times when increasing $n$, since $k_{\rm t}$ scales with $n$ if all the external momenta are of the same order. We will come back to this point in the next sections, when evaluating the time integral in the WFU using a saddle-point approximation at large $\alpha$ and $\bar \zeta$.}).  (The estimate is the same for $n$ odd, when cosine is replaced by sine.)  Including the scaling of the vertex discussed above, the tree-level Witten diagrams go as $\sqrt\alpha\,\alpha^{2(n-2)}$. The Gaussian action is of order $\bzeta^2$: compared to this the $n^{\rm th}$ term in the action contains an additional $\bzeta^{n-2}$. Putting all together the scaling of the tree-level Witten diagram is $\b \sqrt{\alpha}\,\bzeta^2 (\alpha^2 \bzeta)^{n-2}/P_\zeta$ as shown in the first line of Fig.~\ref{fig:Witten_diags}. This estimate just reproduces what obtained from the explicit calculations of the $n$-point correlation functions \cite{Leblond:2010yq}.
From this estimate, the expansion in powers of $\bar \zeta$  needs to be resummed when $ \alpha^2 |\bar\zeta| \gtrsim 1$: in this case the expansion in powers does not make any sense and one has to rely on the full non-perturbative result. 

It is quite straightforward to analyse the structure of tree-level diagrams at higher order in $\tilde b$. Tree level diagrams will give for the WFU
\be\label{eq:wfu_scaling_b}
\Psi \sim e^{-\frac{1}{P_\zeta} \left[ \bzeta^2+ \tilde{b} \Delta S_{\rm E,1}(\bzeta, \alpha)+\tilde{b} ^2 \Delta S_{\rm E,2}(\bzeta, \alpha)  +{\cal O}(\tilde{b}^3)\right]}.
\ee
(For tree diagrams, each new vertex gives $1/P_\zeta$ from the normalization of the action, which is compensated by an extra propagator $\propto P_\zeta$.)
Since we are expanding in $\tilde b$, the $\Delta S_{\rm E,1}(\bzeta, \alpha)$ term must be small compared with the Gaussian part. However notice that this condition is compatible with $\tilde{b} \Delta S_{\rm E, 1}(\bzeta, \alpha) \gtrsim 1$: in this case one cannot expand the term $\Delta S_1$ in the exponential.

~
\newline
\textbf{Scaling of loop diagrams:}
Let us now discuss loop diagrams at $\mathcal{O}(\b)$, see Fig.~\ref{fig:Witten_diags}.
First, we need to make a distinction between loops at the level of the WFU and of correlators. Once the WFU is obtained at one loop, in order to obtain correlation functions one still needs to perform an integration over $\bzeta(\vect x)$  (i.e.~apply the Born rule). 
This average over late-time field configurations gives extra contributions, which we dub ``boundary" loops. 
Of course the final result must coincide with the direct evaluation of the correlator using the in-in calculation. We verify this in App.~\ref{App:bndry_loops} 
for the simple case of $\lambda \phi^4$ in de Sitter for the equal-time two-point function at one loop. The effect of the boundary loop combines with the  WFU loop in such a way to give the loop of the in-in propagator.

Let us focus, for the time being, on the WFU and estimate the size of loop corrections in the resonant scenario. In App.~\ref{app:1-loop_estimate} we show that one-loop diagrams are suppressed by $\alpha^2 P_{\zeta}$ compared to tree-level contributions with the same number of external legs, see Fig.~\ref{fig:Witten_diags}.~\footnote{There also exists a one-loop diagram with one external leg. This tadpole contribution is a spacetime constant whose effect is to redefine the background solution.}  
Let us now try to understand the suppression of the one-loop graphs. 
First notice that, because of the loop, the vertex has two more $\zeta$'s than in the tree-level diagram with the same number of external legs. 
(For instance, the one-loop graph with two external legs, contains the four-point vertex of Eq.~{\eqref{eq:vertex_4}}.). This gives two extra powers of $\alpha$ times $P_\zeta$. There are no extra powers of $\alpha$ from the loop: the bulk-to-bulk propagator is zero at late times, which physically says loop particles are not resonantly produced.~\footnote{In terms of the canonical scalar $\phi$ of \eqref{eq:potential}, fluctuations inside the loop have a size $\delta \phi \sim H$, while external particles have $\delta \phi \sim \alpha H$, since they are produced well inside the horizon when their frequency is $\sim \alpha H$.} Indeed,  the integral over the physical momentum $k_{\rm p}$ of the bulk-to-bulk propagator that describes the loop does not depend on $\eta$ (in the limit $\eta_{\rm f} \rightarrow 0$, see Eq.~\eqref{eq:M_1-loop_2}), so that the integration over $\eta$ remains the same as at tree level, Eq.~\eqref{eq:estimate_tree_scaling}. We conclude that one-loop diagrams at order $\tilde{b}$ scale as the tree-level diagrams with an additional factor $\alpha^2 P_\zeta$ (the exception is the one-point correlation function, which has no tree-level analogue). It is straightforward to see that with $\ell$ loops at first order in $\tilde{b}$ one gets a factor $(\alpha^2 P_{\zeta})^{\ell}$.  Loops are thus small if $\alpha^2 P_\zeta \ll 1$. As we will discuss below, the validity of the EFT will give a bound which is more stringent than this, so that loops are automatically tiny.

We saw above that the expansion parameter of the tree-level diagrams is $\alpha^2\bar\zeta$. This implies that even for typical fluctuations, $\bar\zeta \sim P_{\zeta}^{1/2}$, tree-level diagrams, which are suppressed by powers of $\alpha^2 P_\zeta^{1/2}$, are more important than loop corrections, which are suppressed by powers of $\alpha^2 P_\zeta$. This is at variance with the general case of \cite{Celoria:2021vjw}, where loop diagrams can be neglected compared to the tree-level ones only for unlikely fluctuations, on the tail of the probability distribution. 

What happens at the level of correlators? In this case the effect of loops, at first order in $\b$, {\em vanishes exactly} (see the related discussion in \cite{Lee:2023jby}).  This is reminiscent of what happens for loops of massless particles in S-matrix calculations, when the loop does not depend on the external momenta. Indeed one can check that the in-in loops vanish exactly in dimensional regularization. This  does not happen for the Witten loops in the WFU as discussed in App.~\ref{app:1-loop_estimate}: one remains with physical logarithms that cannot be removed by counterterms. The difference between the in-in and wavefunction calculation is particularly evident in Minkowski, as explained in \cite{Lee:2023jby}. For the in-in case the loop reduces to evaluating the in-in propagator at coincident points: this gives a divergent constant which can be reabsorbed by local counter-terms. For the WFU calculation the propagator has a different boundary condition: it vanishes on the late-time boundary. Therefore, its contribution cannot be a space-time constant as the boundary conditions break translational invariance. In momentum space this corresponds to a momentum inflow inside the loop, and ultimately leads to $\log$ contributions that cannot be removed by counter-terms. (We thank Enrico Pajer for enlightening discussions about this.) 

In conclusion, loops in the WFU at order $\b$ are suppressed by $\alpha^2 P_{\zeta} \ll 1$, while they are exactly zero at the level of correlation functions. 

~
\newline
\textbf{Regime of validity of the EFT:} We saw above that loop corrections are small if $\alpha^2 P_{\zeta} \ll 1$ and that the tree-level action needs to be resummed if $\alpha^2 \bar\zeta \gtrsim 1$. One may wonder whether it is possible to be in a regime in which one needs resummation even for typical fluctuations $\alpha^2 P_\zeta^{1/2} \gtrsim 1$ (notice that this is compatible with small loop corrections). In this regime one would not be allowed to describe the non-Gaussian corrections to the typical fluctuations in terms of bispectrum, trispectrum and so on: all the $n$-point functions should be considered at once.  

Using the definitions of $\alpha$ and $P_{\zeta}$, the condition for requiring resummation for typical fluctuations can be written in terms of $\omega$ and $f$ (see Sec.~\ref{sec:potential}) and is equivalent to $\omega / (4 \pi f) \gtrsim 1$. (Here, we are also taking into account the factors of $\pi$ originating from the momentum integrals in the correlators, see \cite{Behbahani:2011it}.) One would naively think that the unitarity cutoff of the theory is $\Lambda_{\rm cutoff} \sim 4\pi f$, so that the regime $\omega / (4 \pi f) \gtrsim 1$ lies beyond the regime of validity of the EFT. However, the conclusion is too quick: for $\tilde b =0$ the theory is free, so there must be some $\tilde b$ dependence in the calculation of the unitarity cut-off. This calculation has been done recently in \cite{Hook:2023pba} and the result is 
 \begin{equation}\label{unitarity}
	\Lambda_{\rm cutoff} = 4 \pi f \log^{1/2}\left(\frac{f^4}{\Lambda^4}\right) \sim 4 \pi f \log^{1/2}\left(\frac{1}{\tilde b P_\zeta \alpha^4}\right) \sim 4 \pi f \log^{1/2}\left({\tilde b^{-1}}\right) \;.	
\end{equation}
In the same reference \cite{Hook:2023pba} an explicit UV completion with new states entering at a scale parametrically larger than $4 \pi f$ is studied.
It is therefore possible to have a situation where the resummation of the tree-level diagrams is necessary for typical fluctuations: this contrasts the usual expectation that deviations from Gaussianity should be encoded only in the three- and four-point functions. We defer a detailed study of this interesting case to a future publication.

\section{Single Fourier mode analysis}\label{sec:ODE_analysis}

The WFU as a function of the boundary value $\bzeta(\vect{x})$ is given by $\Delta S_{\rm{E},1}$, Eq.~(\ref{eq:DeltaS1_E}). To answer a specific physical question, one should integrate over the configurations of $\bzeta(\vect{x})$ with a weight given by their probability, i.e.~the modulus squared of the WFU. In order to understand how the WFU behaves for large fluctuations, when non-linearities become relevant, one can fix a configuration of $\bzeta(\vect{x})$ up to the overall amplitude and study the WFU as a function of this amplitude. In this way one can study a function of a single variable, instead of a functional of the whole $\bzeta(\vect{x})$. In this section we take $\bzeta$ to be a single Fourier mode and postpone to Sec. \ref{sec:PDE_analysis} the study of more general configurations. This Single Fourier Mode approximation (SFM) is similar to the ordinary-differential equation analysis done in Sec.~4.3 of \cite{Celoria:2021vjw}.

In Sec.~\ref{subsec:saddle_pt_ode} we explain that for $\alpha \gg 1$ the integral over $\tau$ is amenable of the saddle-point approximation. In Sec.~\ref{subsec:int_saddle} we study the regime when non-linearities become important $\alpha^2|\bzeta| \gtrsim 1$, while in Sec.~\ref{subsec:late_saddle} we focus on the extreme tail of the distribution $|\bar \zeta| \gg 1$, when the saddle-point analysis can be completed analytically. All analytical results are checked numerically in both Secs.~\ref{subsec:int_saddle} and \ref{subsec:late_saddle}.

\subsection{Saddle-point approximation}\label{subsec:saddle_pt_ode}
Let us consider the spherically symmetric profile for $\zeta$ at late times given by $\bar{\zeta}(r) = \bar \zeta \sin(k r) / (kr)$, where $\bar{\zeta}$ is the amplitude at late times and $k = |\vect{k}|$ is a fixed momentum scale. This profile, in momentum space, has support only for momenta equal to $k$, hence it can be considered as a single Fourier mode. 
Due to this property, its free time evolution is given by
\begin{equation}\label{eq:zeta_free_ode}
	\zeta(\tau, r) 
	= 
	 \bigg[\frac{\bar{\zeta} \sin(kr)}{kr} \bigg](1 - k \tau ) e^{k \tau}
	\;.
\end{equation}
Notice that the mode function above is not the same as the one used in Sec.~4.3 of \cite{Celoria:2021vjw} because of the radial dependence of the amplitude. 
At late times, when $k \tau \rightarrow 0$, one can see that the mode function exhibits a different behaviours between positive and negative values of $\bar{\zeta}$ (in the absence of the denominator $k r$ a change of sign in $\bar \zeta$ can be compensated by a radial shift).

With the reason above, we consider the function \eqref{eq:zeta_free_ode} as a radial single Fourier mode in position space, and refer to it as the SFM simplification. Notice that the profile we are considering is in real space, so that $\zeta$ remains dimensionless. 

The fact that in Eq.~(\ref{eq:zeta_free_ode}) the amplitude depends on $r$ makes our computation of the action \eqref{eq:DeltaS1_E} more complicated and less illuminating since the radial dependence cannot be separated from the time dependence, e.g. the cosine function contains both $\tau$ and $r$ dependences. 
However, since our main focus here is to capture the main features of the WFU of the full radial profile case (Sec.~\ref{sec:PDE_analysis}) as a function of $\bar{\zeta}$, we are going to neglect the radial dependence of the profile (\ref{eq:zeta_free_ode}), i.e. $\zeta(\tau, r) \simeq \bar{\zeta} (1 - k\tau) e^{k \tau}$, throughout this section.
Therefore, we do not need to perform the integral over $r$ to obtain the on-shell action and we are able to capture the main behaviour of the WFU encoded in the amplitude $\bar{\zeta}$.

Let us now proceed with the SFM simplification, using the mode function $\zeta(\tau, r) \simeq \bar{\zeta} (1 - k\tau) e^{k \tau}$. We define the variable $X(\tau)$ as
\begin{equation}
	X(\tau)
	\equiv 
	\zeta'^2 + k^2 \zeta^2
	= 
	\bar{\zeta}^2 k^2 
	\left[
	1 - 2 k \tau (1 - k \tau) 
	\right]
	e^{2k \tau}
	\;, \label{eq:X_tau_ODE}
\end{equation}
so that $\bar{X} \equiv X(0) = k^2 \bar{\zeta}^2$. Using \eqref{eq:zeta_free_ode} and \eqref{eq:X_tau_ODE}, the action \eqref{eq:DeltaS1_E} then becomes
\begin{equation}
\begin{aligned}
\label{eq:DeltaS1_E_ODE}
	\Delta S_{\rm{E},1}
	&
	=
	\frac{4 \pi}{P_\zeta k^3} 
	\int_{-\infty}^{0} \de\tau \,
	\frac{1}{2\tau^2} 
	\bigg\{	 
	X(\tau)
	\cos\left(\alpha \log\left(\tau/\eta_\star\right) -\tilde\delta - i\alpha \pi/2 + \alpha \zeta \right)
	\\
	& 
	\hspace*{6cm}
	- 
	\bar{X}
	\cos\left(\alpha \log\left(\tau/\eta_\star\right) -\tilde\delta- i \alpha \pi/2 + \alpha \bar{\zeta} \right) 
	\bigg\}
	\;,
	\end{aligned} 
\end{equation}
where we have replaced $(\partial_i \zeta)^2$ with $k^2 \zeta^2$ and the spatial-volume integral with $4 \pi k^{-3}$. Then, we rewrite the action above in an exponential form, which is convenient for one to use the saddle-point approximation. We thus obtain
\begin{align}
\Delta S_{\rm{E},1 } &=
	\frac{\pi}{P_\zeta} 
	\sum_{\sigma = \pm 1}
	\int_{-\infty}^{0} \de \tau \,
	\frac{e^{\sigma \alpha \pi/2}}{\tau^2} 
	\bigg\{	 
	X(\tau)
	\exp\left(i \sigma \alpha \log\left(-\tau\right)  + i \sigma \alpha \zeta \right)
	\nn \\
	& 
	\hspace*{5.5cm}	
	- 
	\bar{X}
	\exp\left(i \sigma\alpha \log\left(- \tau\right) + i \sigma\alpha \bar{\zeta} \right) 
	\bigg\}
	e^{- i \sigma (\td  + \alpha\log(-k\eta_\star))}
	\;, \label{eq:DeltaS1_E_ODE1}
\end{align}
where we have changed variable to $\tau \rightarrow k \tau $ and re-defined $X \rightarrow X/k^2$, so that $\tau$ and $X(\tau)$ that appear now in the integral are dimensionless. Naively, one expects that the integral above is dominated by the contributions with $\sigma = 1$, compared to the ones with $\sigma = - 1$. This is only the case for $\bar{\zeta} > 0$. However, we will find that in the case where $\bar{\zeta} < 0$ the two contributions are of the same order, so that one needs to take into account the terms with $\sigma = - 1$ as well.

Let us pause to comment on the dependence on the momentum scale $k$ of the late-time profile.   If we rescale $k \rightarrow k/\lambda$, then the free solution in real space goes to $\zeta(\vect x, \tau) \rightarrow \zeta(\lambda \vect x, \lambda \tau)$. 
One can then rescale the coordinates $\tau$ and $\vect x$ to $\tau' = \lambda \tau$, $\vect x' = \lambda \vect x$ in Eq.~\eqref{eq:DeltaS1_E}. 
In doing so, the only change in the action is due to the explicit $\tau$ dependence inside the cosine.  Hence, a change in $k$ is degenerate with a change in $\eta_\star$. This is explicit in Eq.~(\ref{eq:DeltaS1_E_ODE1}), since $k$ appears only in the very last term, in the combination $k \eta_\star$. The action is periodic in $\alpha\log(-\eta_\star)$ (the original scale-invariance of de Sitter is broken by the oscillations to a discrete subgroup) and therefore the result will be periodic in $\alpha\log k$. This is all we can say in terms of symmetries. Notice, however, that when one of the two terms $\sigma = \pm 1$ dominates in Eq.~(\ref{eq:DeltaS1_E_ODE1}), then the $k$ (or $\eta_\star$) dependence reduces to a sinusoidal modulation of the (real part of the) action.

We can write \eqref{eq:DeltaS1_E_ODE1} in a more a compact form as 
\begin{align}
	\label{eq:DeltaS1_E_ODE2}
	\Delta S_{\rm{E},1}
	=
	\frac{\pi}{P_\zeta} 
	\sum_{\sigma = \pm 1}
	e^{i \sigma \psi}
	e^{\sigma \frac{\alpha \pi}{2}}
	\,
	\Cc I_{\sigma} 
	\;,
\end{align}
where $\psi \equiv  -\td - \alpha \log(-k\eta_\star)$ and the integral $\Cc I_{\sigma} $ is defined by
\begin{align}\label{eq:ode_integrals_I}
	\Cc I_{\sigma}
	&
	\equiv
	\int_{-\infty}^{0} \de \tau \,
	\left( 
	e^{\Phi_{\sigma}}
	- 
	e^{\Psi_{\sigma}}
	\right) \;, 
\end{align}
with the exponents $\Phi_\sigma$ and $\Psi_\sigma$ being
\begin{equation}
\begin{aligned}
	\Phi_{\sigma}
	&
	\equiv
	-(2 - i \sigma \alpha)\log(-\tau)
	+ \log\left(X(\tau)\right)
	+ i \sigma \alpha \zeta \;,
	\\
	\Psi_{\sigma}
	&
	\equiv
	-(2 - i \sigma \alpha)\log(-\tau)
	+ \log\left(\bar{X}\right)
	+ i \sigma \alpha \bar{\zeta}
	\;.
\end{aligned}
\end{equation}
The form \eqref{eq:ode_integrals_I} is particularly useful for an asymptotic expansion when $\alpha$ is much larger than unity.  
It should be noted that the integral $\Cc I_{\sigma}$ depends on both $\alpha$ and the late-time value $\bar{\zeta}$. Below, we are going to evaluate the integral \eqref{eq:ode_integrals_I} using the saddle-point approximation, which is valid for $\alpha \gg 1$.

~
\newline
\textbf{Saddle-point equation:}
Here we are going to find the relevant saddle points for the evaluation of the integral \eqref{eq:ode_integrals_I}. 
Before proceeding, we need to be careful in applying the saddle-point approximation. Indeed, Eq.~\eqref{eq:ode_integrals_I} contains two exponential terms, each of which diverges at late times. 
Notice that the first exponential term, $e^{\Phi_\sigma}$, contains the relevant physical information about the dynamics of $\zeta$, whereas the second term, $e^{\Psi_\sigma}$, has the only purpose of making the integral finite. Additionally, one cannot evaluate the two integrals in saddle point separately, as the second term has no saddle point solutions (solutions of $\partial_\tau \Psi_{\sigma} = 0$) since it only depends on $\log(-\tau)$. 
This would suggest that a proper treatment to find the saddle point when both terms contribute equally to the integral is required.
However, we will argue now that the contribution from $e^{\Psi_\sigma}$ can be neglected and one can just focus on the saddle point of $e^{\Phi_\sigma}$.

The argument is the following. Suppose the saddle point of $\Phi_\sigma$ is away from the origin $\tau = 0$. This means that the integral will be accumulating its value around this saddle. 
On the other hand, the contribution of $\Psi_{\sigma}$ will not grow around this specific point, as there is no saddle for $\Psi_{\sigma}$,
but only at later times, where however there will be a cancellation with $\Phi_{\sigma}$ as to make the integral finite. This suggests that indeed $\Psi_{\sigma}$ can be neglected.~\footnote{A similar situation happens when computing a single WFU coefficient at tree level. Although the integrands diverge at late times, with the divergence being just a phase, in the saddle-point estimate the divergent piece is lost.}

We thus disregard the contribution of $\Psi_{\sigma}$: we will see later that this is indeed in agreement with the numerical analysis.
Let us proceed with the saddle-point analysis for the term $\Phi_\sigma$. The saddle-point equation for $\Phi_{\sigma}$ is given by
\begin{equation}
\label{eq:saddle_pt_eq_ode}
	\partial_\tau \Phi_{\sigma} 
	= 
	-\frac{2-i\sigma \alpha}{\tau} 
	+ \frac{X'(\tau)}{X(\tau)} 
	+ i \sigma \alpha \zeta' 
	= 
	0
	\;.
\end{equation}
Generally, the solutions of the above equation lie in the complex plane. Here we want to look for the solutions with $\rm{Re}\, \tau < 0$, so that the wavemode \eqref{eq:zeta_free_ode} decays at large $|\tau|$.~\footnote{
More properly, we require that the path of integration can be smoothly deformed from the initial domain to meet the saddle point. In this sense, we can allow for saddles with $\rm{Re}\, \tau > 0$ as long as the overall integral remains convergent.
} 
Unfortunately, Eq.~\eqref{eq:saddle_pt_eq_ode} does not admit a closed-form solution; therefore, one needs to either solve the equation numerically or perform some additional expansion, e.g.~$|\bar{\zeta}| \gg 1$, such that an analytic solution can be found. Indeed, as we will show below, in the limit $|\bar{\zeta}| \gg 1$ we find an analytical late-time saddle point. 

Let us comment on the possibility of having a saddle point at late times ($k|\tau| \ll 1$). The motivation comes from the perturbative calculation. In order to compute the connected $n$-point correlation functions $\langle \pi^n \rangle$ in the perturbative regime, one encounters the integral of the product of the mode functions over the conformal time $\eta$. (See App.~\ref{bispectrum} for a computation of $\langle \pi^3 \rangle$.) Such an integral can be done analytically and the result is given in terms of  incomplete Gamma functions. 
However, in the large $\alpha$ limit, one can use the saddle-point approximation to evaluate the integral over $\eta$. Indeed, it was explicitly shown in \cite{Leblond:2010yq} that the saddle-point is located at $k_{\rm t}\eta = -\alpha$, where $k_{\rm t}$ is the total external momentum. For small $n$, this saddle is at early times. However, this is not the case when the number of external legs is much larger than unity and becomes comparable with $\alpha$. 
Indeed, if all the $n$ external momenta are of the same order, then $k_{\rm t} \sim n k$ and the saddle point schematically becomes $k\eta \sim -\alpha/n$.
This observation therefore motivates us to look for a late-time saddle point of the integral \eqref{eq:ode_integrals_I} at least for sufficiently large $\bzeta$.~\footnote{Although it is not clear how the limits of large $n$ and large $\bar{\zeta}$ are related, it is reasonable to expect that the saddle points for large $\bar{\zeta}$ case follow the one of large $n$ limit.}

\subsection{$\alpha^2|\bzeta| \gtrsim 1$: intermediate saddle point} \label{subsec:int_saddle}
Let us recall the scaling of the tree-level action at first order in $\tilde{b}$.
From Fig.~\ref{fig:Witten_diags} and the discussion below Eq.~\eqref{eq:estimate_tree_scaling}, we concluded that when $\alpha^2 \bar{\zeta} \sim 1$ all the tree-level $n$-point functions are equally important and need to be resummed.
We will distinguish this case from the far-tail of the distribution, $|\bar \zeta| \gg 1$.
The solution to Eq.~\eqref{eq:saddle_pt_eq_ode} cannot be found analytically in the regime where $\alpha^2 \bar{\zeta} \sim 1$.
We then solve Eq.~\eqref{eq:saddle_pt_eq_ode} numerically in two cases: $\bar{\zeta} > 0$ and $\bar{\zeta} < 0$.

At fixed $\alpha$ and $\bar \zeta$ the saddle-point equation has always multiple solutions in the $\tau$ complex plane.~\footnote{We can intuitively see how these branches arise by inspecting the saddle-point equation \eqref{eq:saddle_pt_eq_ode} at large $\tau$. In this limit, the equation is solved by the Lambert-$W$ function, which is known to have infinite branches labelled by a positive integer.} However, only some are relevant: when we deform the contour of integration, we only reach a sub-set of all the saddles.
Moreover, depending on the value of $\alpha$, we can have different saddles being relevant. In particular, we find discrete values of $\alpha$, at approximatively $\alpha = 4 \pi N$ with $N = 1,2,\ldots$, at which the behaviour of the saddles changes. 

\begin{figure*}[t!]
 \includegraphics[width=0.49\textwidth]{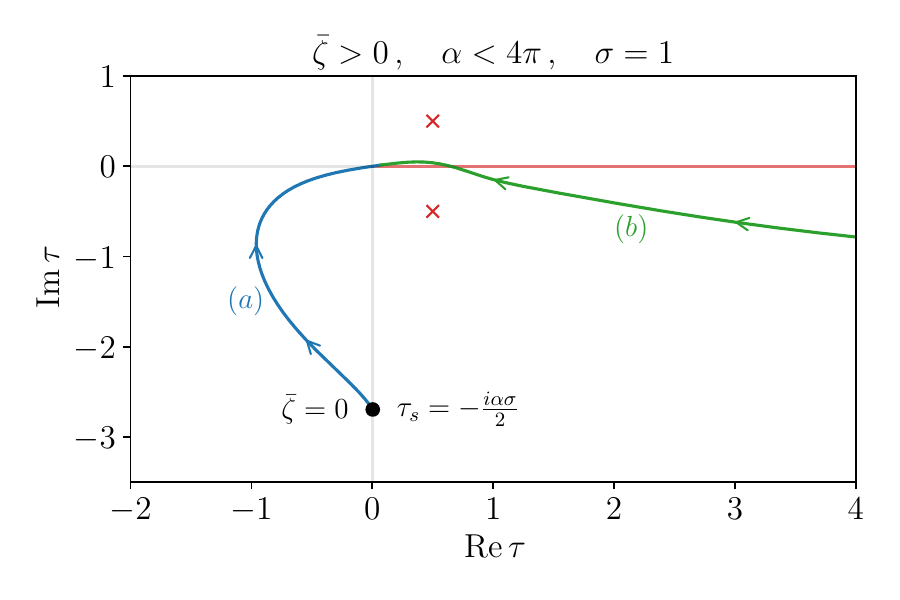}
 \includegraphics[width=0.49\textwidth]{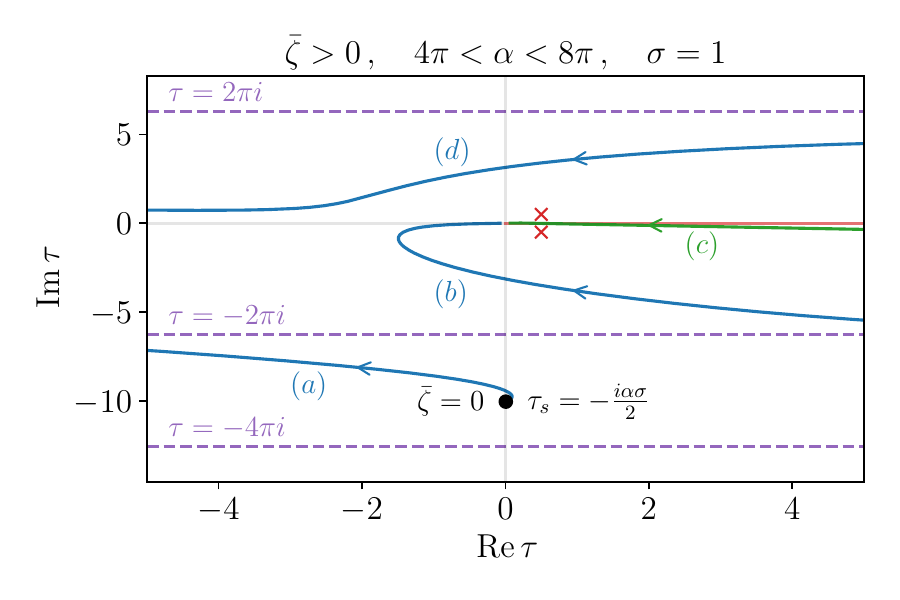}
 \\
 \center
 \includegraphics[width=0.49\textwidth]{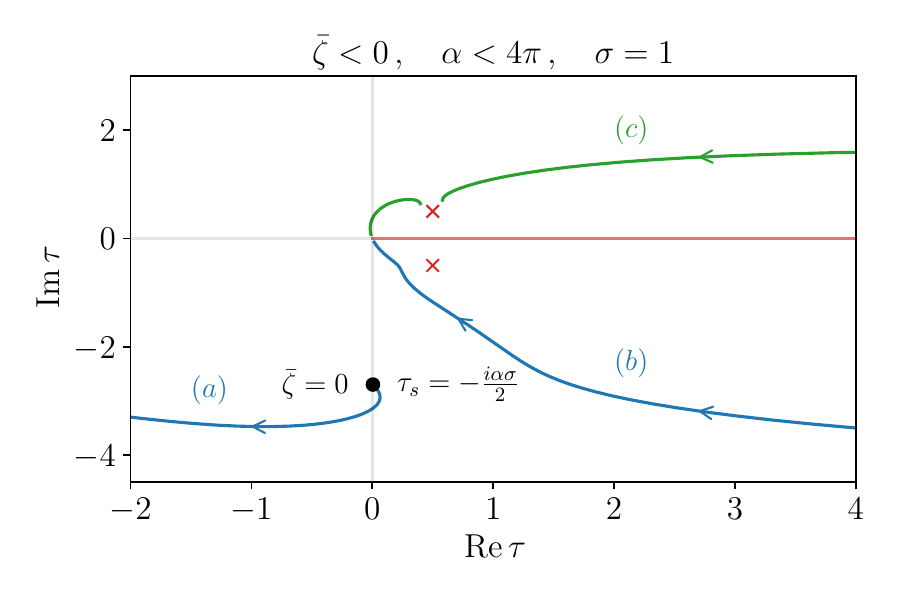}
 \caption{~Location of the solutions of the saddle-point equation \eqref{eq:saddle_pt_eq_ode} in the complex-$\tau$ plane when varying $\bar \zeta$. The arrows in the curves point to the direction of increasing $|\bar \zeta|$. The light-red line represents the branch cut of $\Phi_{\sigma}$ due to the $\log(-\tau)$ term. The zeros of $X(\tau)$, located at $\tau = e^{\pm i \pi / 4} / \sqrt{2}$, are indicated by the two red crosses.
 \label{fig:ode_saddles}
 \textbf{Top Left:} saddles for $\bar \zeta > 0$ and $\alpha < 4 \pi$. The perturbative saddle $(a)$ starts at $\tau_s = - i \alpha \sigma / 2$ 
when $\bar\zeta = 0$ and moves towards late times (the origin). The branch $(b)$ instead moves from infinity towards the origin, becoming the complex-conjugate of $(a)$.
 \textbf{Top Right:} saddles for $\bar \zeta > 0$ and $4 \pi < \alpha < 8 \pi$. The perturbative saddle $(a)$ moves to infinity whilst the branch $(b)$ moves towards the origin, dominating the action at large $\bar \zeta$. The branches $(c)$ and $(d)$ are instead irrelevant.
 \textbf{Bottom centre:} saddles for $\bar \zeta < 0$ and $\alpha < 4 \pi$. The perturbative branch $(a)$ moves towards infinity whilst the branch $(b)$ moves towards late times and becomes dominant. The branch $(c)$ is irrelevant. In its evolution it merges with one of the zeros of $X(\tau)$ before moving to the origin.
 }
\end{figure*}

Notice that when $\bar \zeta = 0$ we have a unique saddle, located at $\tau_s = - i \alpha \sigma / 2$. This saddle corresponds with the perturbative saddle point used to evaluate the correction to the power spectrum (this comes from the fact that we are setting $\bar \zeta = 0$ inside the cosine term in the action). We therefore interpret this saddle as the perturbative one.
We can then analyse how the relevant saddles evolve as $|\bar \zeta|$ increases.

First, we start with the case $\bar \zeta > 0$. When $\alpha < 4 \pi$ (but larger than $1$ so that we can apply the saddle-point approximation), the perturbative saddle point evolves towards late times. Moreover, we notice the presence of an additional saddle which starts from $\infty$ when $\bar \zeta = 0$ and that at large $\bar \zeta$ moves to the origin (its imaginary part is equal and opposite to the one of the $(a)$ branch). The evolution of the saddles is shown in the top-left panel of Fig.~\ref{fig:ode_saddles}. Additionally, one finds more solutions at larger values of $|\tau|$, both with positive and negative imaginary parts (again, these are related to the Lambert-$W$ function). In this case, we find that only the evolved perturbative saddle is relevant in the integral. 

The situation becomes more intricate when we move to larger $\alpha$'s. For instance, in the approximate window $4 \pi < \alpha < 8 \pi$, we notice that the perturbative saddle does not move to the origin anymore but instead moves towards $k\tau_s \to -2 \pi i - \infty$. Instead, a different saddle point becomes relevant and moves towards the origin. This is the branch $(b)$ in the top-right panel of Fig.~\ref{fig:ode_saddles}. Notice that again, this branch at late times has a ``conjugate'' saddle, the branch $(c)$. The figure also shows an additional branch $(d)$, that is however not relevant.
For larger values of $\alpha$ we find that this periodic behaviour continues. For example, when $8 \pi < \alpha < 12 \pi$ we still have a perturbative branch moving to infinity and a branch moving to the origin, with an additional branch in between. For larger $\alpha$s the number of branches in between increases.

Finally, we can briefly mention what happens when $\bar \zeta < 0$ in the case $\alpha < 4 \pi$. As opposed to the positive-$\bar \zeta$ case, even for these values of $\alpha$ the perturbative saddle moves to infinity. Also, an additional branch moves towards late times as $|\bar \zeta|$ increases (together with its complex-conjugate, as before). This is the branch $(b)$ in the bottom panel of Fig.~\ref{fig:ode_saddles}. Notice that in this case the branch $(b)$ approaches the origin from a different angle of the complex plane compared to the previous cases. We will see this more in detail when evaluating this late-times saddle analytically later on. 
For larger values of $\alpha$ we find a similar pattern as in the positive case, with new branches interposing between the perturbative and the late-times branches.

Despite the complex structure of the saddles, we will show by direct comparison with the numerical integration that for small $\alpha^2 |\bar \zeta|$ the integral is well approximated by the perturbative saddle. On the other hand, for $\alpha^2 |\bar \zeta| \gg 1$ the branch that is moving towards late times dominates.

Once the relevant saddle-point solutions $\tau_s$ are identified numerically, we can evaluate the action \eqref{eq:DeltaS1_E_ODE1} on saddle. Specifically, the integral $\Cc I_{\sigma}$ of Eq.~\eqref{eq:ode_integrals_I} is evaluated as (here we neglect $\Psi_{\sigma}$)
\begin{equation}
	\Cc I_\sigma
	\simeq 
	\sqrt{\frac{2 \pi}{- \partial_{\tau}^2\, \Phi_{\sigma}}} 
	\,
	\left.
	e^{\Phi_{\sigma}}
	\right\vert_{\tau_s}
	\;.
\end{equation}
Finally, we can evaluate the action (\ref{eq:DeltaS1_E_ODE2}) using the integral we found above.
Note that if we have more than one relevant saddle, we also need to sum over them. 

\begin{figure*}[t!]
 \includegraphics[width=0.49\textwidth]{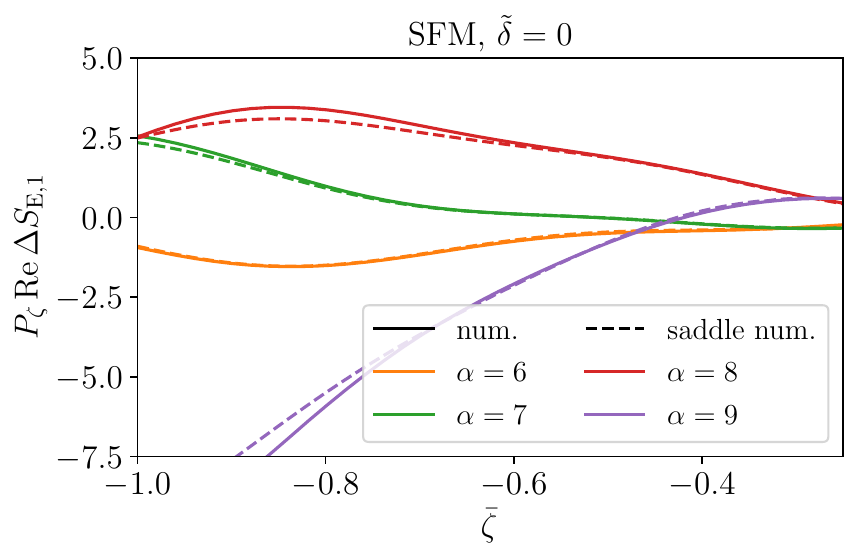}
 \includegraphics[width=0.49\textwidth]{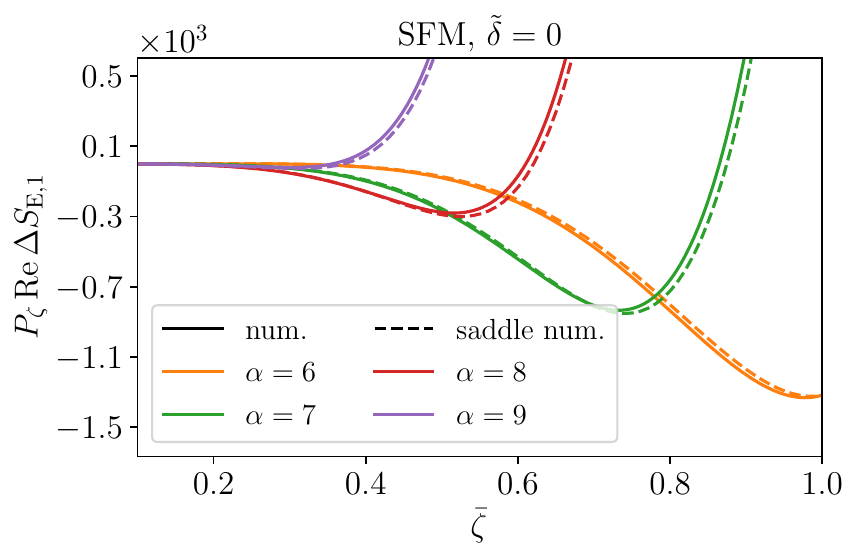}
 \caption{~Action with the SFM simplification as a function of $\bar \zeta$ in the regime of $\alpha^2 \bar \zeta \gtrsim 1$ for different values of $\alpha$. Solid lines are obtained by numerically integrating Eq.~\eqref{eq:DeltaS1_E_ODE}. Dashed lines are obtained using the saddle-point approximation, solving numerically Eq.~\eqref{eq:saddle_pt_eq_ode} for the perturbative branch. 
 \label{fig:ode_alpha2z_large}
 \textbf{Left:} case of negative $\bar \zeta$. \textbf{Right:} case of positive $\bar \zeta$. 
 }
 \label{fig:SFM-intermediate}
\end{figure*}
Apart from the saddle-point approximation presented above, we now compute the action (\ref{eq:DeltaS1_E_ODE}) numerically. 
First, we limit the range of integration up to $k \tau_{\rm f} = -10^{-6}$ (the lower limit $k \tau_{\rm i}$ is chosen large enough in modulus so to reach numerical convergence).
Then, we use the free solution of $\zeta(\tau)$ given by Eq.~\eqref{eq:zeta_free_ode} and plug it back into the action (\ref{eq:DeltaS1_E_ODE}).
We fix $\alpha = 6$, $7$, $8$ and $9$, and we compute the integral over $\tau$ numerically, varying the values of $\bar{\zeta}$ from $-1$ to $1$.

\begin{figure*}[t!]
\centering
 \includegraphics[width=0.49\textwidth]{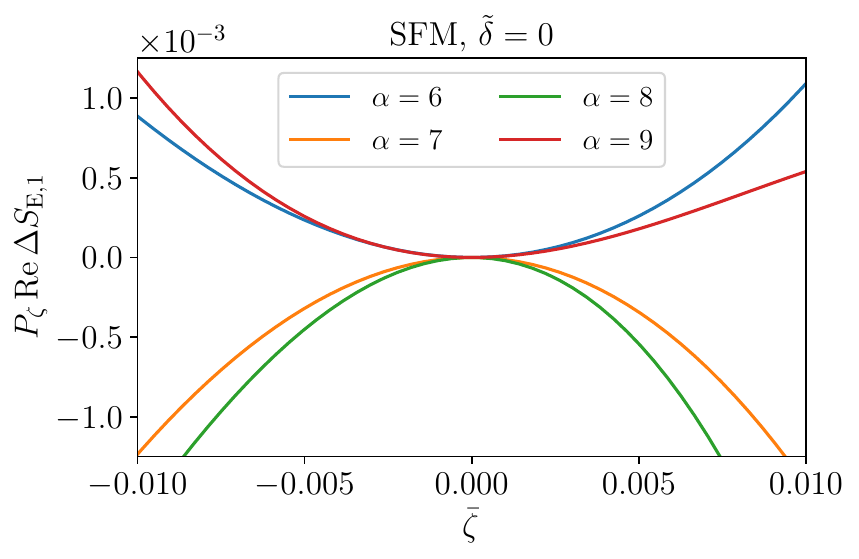}
 \caption{~Action with the SFM simplification as a function of $\bar \zeta$ in the regime of small $\alpha^2 \bar \zeta$. In this limit we recover the perturbation-theory expansion: the curves are well approximated by a polynomial in $\bar \zeta$.
 \label{fig:ode_small_zeta}
 }
\end{figure*}

The results for the cases with $\alpha = 6$, $7$, $8$ and $9$ are shown in Fig.~\ref{fig:ode_alpha2z_large} (both the full numerical integration of Eq.~\eqref{eq:DeltaS1_E_ODE} and the saddle-point approximation). The range of $\bar \zeta$ is chosen to highlight the region where $\alpha^2 \bar \zeta \sim 1$: here perturbation theory stops being reliable.
To obtain the saddle-point approximation in this range for $\bar \zeta$ we only included the branch $(a)$ of Fig.~\ref{fig:ode_saddles} (the perturbative branch) for both signs of $\bar \zeta$, while other branches are found to be subdominant here.
As one expects, when $\alpha^2 |\bar{\zeta}| \ll 1$ the saddle point is similar to the one of perturbation theory and the action $\Delta S_{\rm{E}, 1}$ is dominated by the quadratic term (the correction at order $\tilde b$ to the power-spectrum). In fact, this can be explicitly seen in Fig.~\ref{fig:ode_small_zeta} where the action $\Delta S_{\rm{E}, 1}$ is symmetric around the vertical axis for $\alpha^2 |\bar{\zeta}| \ll 1$. 
For consistency, in the regime where $\alpha^2 |\bar{\zeta}| \ll 1$ we check that summing the perturbative tree-level diagrams at $\mathcal{O}(\tilde{b})$ (see Eq.~(4.16) of \cite{Leblond:2010yq}) reproduces our results in Fig.~\ref{fig:ode_small_zeta}.~\footnote{It should be noted that the comparison of the two results in the perturbative regime was carried out using the SFM simplification, otherwise one needs to perform the integrals over all momenta or fix the late-time profile $\bar{\zeta}(\vect{x})$ as we will study in detail in Sec.~\ref{sec:PDE_analysis}.}
As $|\bar{\zeta}|$ increases, the action becomes asymmetric due to the presence of odd contributions in $\bar \zeta$, starting from cubic terms. 
As we approach the non-perturbative regime, the asymmetry is magnified as it can be noted from the different scales in the two plots of Fig.~\ref{fig:ode_alpha2z_large}.
It is, in fact, interesting to point out that even in the regime where $\alpha^2 |\bar{\zeta}| \gtrsim 1$ our results in Fig.~\ref{fig:ode_alpha2z_large} match with the summation of perturbative results, which indicates the fact that our non-perturbative WFU resums the perturbative tree-level graphs at first order in $\tilde{b}$, as discussed below Eq.~(\ref{eq:DeltaS1_E}).
Additionally, we find that the exponential growth for $\bar{\zeta} > 0$ in Fig.~\ref{fig:ode_alpha2z_large} can also be realized in the summation of perturbative series, see Sec~\ref{sec_crit_remark} for more detailed discussions. 
Finally, as a check, we confirm that the full numerical integration and the saddle-point approximation are in remarkably good agreement even for these moderate values of $\alpha$.
We did not choose larger values of $\alpha$ since the numerical integration becomes more challenging, while the qualitative features are unaffected.

\subsection{$|\bar{\zeta}| \gg 1$: late-time saddle point}\label{subsec:late_saddle}
Given the observation above, let us focus on the saddle point at late times. We will confirm that the contributions from the early-time saddle point are subdominant. Expanding $\zeta$ and $X(\tau)$ for $\tau \ll 1$ gives
\begin{equation}
\begin{aligned}
	&
 	\zeta 
 	\simeq 
 	\bar{\zeta} 
 	\left( 1 
 	- \frac{\tau^2}{2} 
 	- \frac{\tau^3}{3} 
 	+ \Cc O(\tau^4)
 	\right)
 	\;,
 	\\
 	&
 	X(\tau)
  	\simeq 
 	 \bar{\zeta}^2
 	\left( 
 	1 
	+ \frac{4}{3}\tau^3 
 	+ \Cc O(\tau^4)
 	\right)
 	\;.
\end{aligned}
\end{equation}
The leading behaviour of the saddle point can be captured by keeping the terms up to first order in $\tau$ in Eq.~(\ref{eq:saddle_pt_eq_ode}). We obtain
\begin{equation}\label{eq:dPhi_dt_ode}
 	\partial_\tau \Phi_{\sigma} 
 	\simeq
	-\frac{2-i\sigma \alpha}{\tau} 
	- i \sigma \alpha \bar{\zeta} \tau
	= 
	0
	\;.
\end{equation}
Note that the term $X'(\tau)/X(\tau)$ starts at second order in $\tau$; therefore, it is negligible compared to the ones we kept. The solution $\tau_s$ to Eq.~(\ref{eq:dPhi_dt_ode}) is given by
\begin{equation}
\label{eq:late_saddle_ode}
	\tau_s^2 \simeq \frac{1}{\bar{\zeta}} \left(1 + \frac{2 i \sigma}{\alpha}\right)
	\;,
\end{equation}
where the second term in the parenthesis is assumed to be small, since we are taking $\alpha$ large (it is anyway useful to keep it). Let us assume, for the moment, that the subleading term is negligible, so that $\tau^2_s \simeq 1/\bar{\zeta}$. For $\bar{\zeta} > 0$ we have $\tau_s \simeq - \bar{\zeta}^{-1/2}$, while the second root has $\rm{Re}\, \tau > 0$. The positive root will not be encountered as we deform the contour of integration (it corresponds to the ``conjugate'' branch showed in green in Fig.~\ref{fig:ode_saddles}). On the other hand, for $\bar{\zeta} < 0$ both solutions are close to the imaginary axis. 
We find that, depending on $\sigma$, the relevant saddle is approximatively $\tau_s \simeq - i \sigma |\bar{\zeta}|^{-1/2}$.
This is a striking result, which indicates that when $\bar{\zeta} < 0$ the saddle point moves from Euclidean to Lorentzian time. This finding is consistent with the discussion of the previous subsection and with the numerical saddles of Fig.~\ref{fig:ode_saddles}.
We will come back to this point later in Sec.~\ref{subsub:neg_zeta_ODE}. Let us now evaluate the integral \eqref{eq:ode_integrals_I}. 

\subsubsection{$\bar{\zeta} > 0$ case}

From Eq.~(\ref{eq:ode_integrals_I}), we compute the exponent $\Phi_\sigma$ on the saddle point (\ref{eq:late_saddle_ode}). Recall that we are neglecting the contribution from $\Psi_\sigma$. Since we are dealing with a late-time saddle point, then the relevant terms in the exponent $\Phi_\sigma$ are just $\sim \alpha \zeta$ and $\alpha \log(-\tau)$, while the rest can be evaluated at late times, e.g.~$X(\tau) \sim \bar{X}$. Therefore, in this case we obtain
\begin{equation}
	e^{\Phi_{\sigma}}\rvert_{\tau_s}
	\simeq 
	e \, \bar{\zeta}^3 
	\left(
	1
	+\frac{2i\sigma}{\alpha}
	\right) 
	^{i \sigma \alpha /2 -1}
	e^{- i \alpha \sigma / 2}
	e^{i \sigma \alpha (\bar{\zeta} - \log \sqrt{\bar{\zeta}})}
\;. \label{eq:exponent_Phi_saddle}
\end{equation} 
where we have kept the terms in $\Phi_\sigma$ up to $\mathcal{O}(\tau^2)$. We see that the presence of the amplitude $\bar{\zeta}$ affects both the overall scaling ($\sim \bar{\zeta}^3$) and the oscillating behaviour through a phase $\sim \sigma \alpha (\bar{\zeta} - \log \sqrt{\bar{\zeta}})$. This non-trivial dependence on $\bar{\zeta}$ can be checked against the full numerical integral. 

Apart from the exponent, we also need to evaluate the prefactor, which contains the second derivative of $\Phi_\sigma$ with respect to $\tau$, evaluated at the saddle point. 
Straightforwardly, taking an additional derivative on Eq.~\eqref{eq:dPhi_dt_ode} with respect to $\tau$ and evaluating such an expression on the saddle point (\ref{eq:late_saddle_ode}) we find 
\begin{align}\label{eq:prefactor_ODE_zeta_0}
\partial_\tau^2 \Phi_\sigma\rvert_{\tau_s} \simeq -2 i \sigma \alpha \bar{\zeta} \;.
\end{align}
Note that this result holds for both $\bar{\zeta} > 0 $ and $\bar{\zeta} < 0$. 

Using (\ref{eq:exponent_Phi_saddle}) and (\ref{eq:prefactor_ODE_zeta_0}) we therefore obtain
\begin{align}
\label{eq:I_s_zo_gr_0}
	\Cc I_{\sigma} 
	&
	\simeq
	\sqrt{\frac{2 \pi}{- \partial_\tau^2 \Phi_{\sigma}}}
	\left.
	e^{\Phi_{\sigma}}
	\right\vert_{\tau_s}
	=
	\sqrt{\frac{\pi \bar{\zeta}^5}{\alpha}}
	\left(
	1
	+\frac{2i\sigma}{\alpha}
	\right) 
	^{i \sigma \alpha /2 -1}
	e^{1 - i \alpha \sigma / 2 - i \pi \sigma/4}
	e^{i \sigma \alpha (\bar{\zeta} - \log \sqrt{\bar{\zeta}})}
	\;.
\end{align}
From the result above we see that the two signs for $\sigma$ do not affect the overall magnitude of $\Cc I_{\sigma}$. On the other hand, looking at Eq.~\eqref{eq:DeltaS1_E_ODE2} we then see that the dominant contribution corresponds to $\sigma = 1$, which is due to the exponential factor $\sim e^{\pi\alpha / 2}$, for large $\alpha$. Therefore, the dominant contribution ($\sigma = 1$) to the action $\Delta S_{\rm{E},1}$ is given by 
\begin{equation}
	\Delta S_{\rm{E},1}
	\simeq 
	\frac{\pi}{P_{\zeta}}
	\sqrt{\frac{\pi \bar{\zeta}^5}{\alpha}}\,
	e^{\pi\alpha/2}
	e^{i \alpha (\bar{\zeta} - \log \sqrt{\bar{\zeta}})}
	e^{i \tilde \psi}
	\;,
	\quad
	(\bar{\zeta} > 0)
	\;,
	\label{eq:ode_saddle_z0pos}
\end{equation}
where we have defined $\tilde \psi \equiv \psi -\alpha/2 - \pi /4$. Note that to obtain (\ref{eq:ode_saddle_z0pos}) we approximated the parenthesis in Eq.~\eqref{eq:I_s_zo_gr_0} with $e^{-1}$. Moreover, it should be noted that in principle, there are terms proportional to additional inverse powers of $\alpha$ which are generated by subleading terms in the saddle-point expansion, so we cannot trust them at this level. 

Before moving to the negative $\bar{\zeta}$ case, let us point out that one can obtain a better approximation for $\Delta S_{\rm E, 1}$ at intermediate values of $\alpha$ and $|\bar \zeta|$, by solving the saddle-point equation \eqref{eq:saddle_pt_eq_ode} numerically. This requires choosing the relevant saddle for the integral, as discussed previously. For large $\bar \zeta$ we find that this saddle is approximately the late-times one, Eq.~\eqref{eq:late_saddle_ode}.
We checked that by doing so, the saddle-point approximation matches the full numerical result with a good precision, even at moderate values for $\alpha$ and $|\bar \zeta|$.

\subsubsection{$\bar{\zeta} < 0$ case}\label{subsub:neg_zeta_ODE}
This case is parametrically different from the previous case. As we already obtained, the late-times saddle in this case is imaginary, see Eq.~\eqref{eq:late_saddle_ode}.
Among the two saddles (obtained when taking the square root in Eq.~\eqref{eq:late_saddle_ode}), we identify the relevant one to be~\footnote{To check this, we performed a Thimble decomposition of the original contour of integration (see e.g.~\cite{Tambalo:2022plm,Serone:2017nmd} for more details on this procedure). Moreover, this choice of saddles is in agreement with the numerical results.}
\begin{equation}
	\tau_s
	= 
	\frac{-i\sigma}{\sqrt{|\bar{\zeta}|}} 
	\left(
	1 
	+ \frac{2i\sigma}{\alpha}
	\right)^{1/2}
	\;.
	\label{eq:late_saddle_ode_neg}
\end{equation}
There are two cases one needs to consider, depending on $\sigma = \pm 1$. When $\sigma = 1$, the saddle point \eqref{eq:late_saddle_ode_neg} corresponds to a Lorentzian saddle point: 
$\tau_s \sim -i/\sqrt{|\bar{\zeta}|}$ for large $\alpha$. Therefore, in this case the exponential factor $e^{\pi\alpha/2}$ in Eq.~\eqref{eq:DeltaS1_E_ODE2} gets cancelled. Essentially, this cancellation occurs because one rotates back to the Lorentzian time.  
Instead, when $\sigma = -1$ we have 
$\tau_s \sim i / \sqrt{|\bar{\zeta}|}$
for large $\alpha$. This saddle point is still imaginary, but it now lies on the positive imaginary axis of the complex $\tau$-plane. In this case we obtain an additional contribution proportional to $e^{-\pi \alpha/2}$ to the action. Therefore, the contribution from $\sigma = -1$ is negligible for large $\alpha$.
Repeating the same steps as in the $\bar{\zeta} > 0$ case and using the saddle point \eqref{eq:late_saddle_ode_neg}, we obtain
\begin{equation}
	\Delta S_{\rm E, 1}
	\simeq 
	-\frac{\pi}{P_{\zeta}}
	\sqrt{\frac{\pi |\bar{\zeta}|^{5}}{\alpha}}
	e^{i \alpha (\bar{\zeta} - \log \sqrt{|\bar{\zeta}|})}
	e^{i \tilde \varphi}
	\;,
	\quad
	(\bar{\zeta} < 0)
	\;,
	\label{eq:ode_saddle_z0neg}
\end{equation}
where we defined $\tilde \varphi \equiv \psi - \alpha / 2 + \pi / 4$.
The main difference between Eqs.~\eqref{eq:ode_saddle_z0pos} and \eqref{eq:ode_saddle_z0neg} is the factor $e^{\pi\alpha/2}$, which enhances the positive case.
As already mentioned in the $\bar{\zeta} > 0$ case, one can improve the matching between the results of the saddle-point approximation and the numerical method, by solving the saddle-point equation numerically. 

\begin{figure}[t!]
\centering
  \includegraphics[width=1\textwidth]{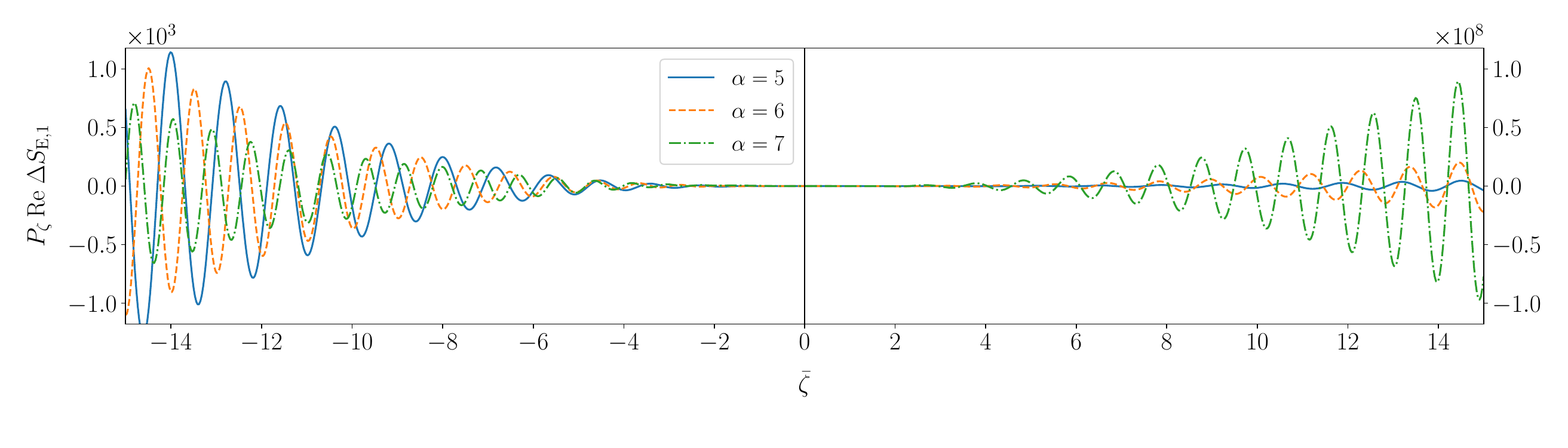}
  \caption{\quad Numerical results for the SFM action, Eq.~\eqref{eq:DeltaS1_E_ODE}, for different values of $\alpha$. Here we are using $\td = 0$ and $k \eta_\star = -1$. Note the difference between the sizes of the action for positive and negative values of $\bar{\zeta}$.}
  \label{fig:ode_action} 
\end{figure}

In order to explicitly see the behaviour of the results (\ref{eq:ode_saddle_z0pos}) and (\ref{eq:ode_saddle_z0neg}), we compute the action (\ref{eq:DeltaS1_E_ODE}) numerically in the limit $|\bar{\zeta}| \gg 1$. 
Following the same procedure described in Sec.~\ref{subsec:int_saddle}, in Fig.~\ref{fig:ode_action} we show the real part of $\Delta S_{\rm E, 1}$ as a function of $\bar{\zeta}$ for $\alpha = 5$, $6$, $7$ respectively. 
Such numerical results are obtained in the range $\bar{\zeta} \in [-15,15]$.
Note that it is straightforward to verify that the imaginary part of $\Delta S_{\rm E, 1}$ behaves in the same way as the real part. 
In the plot we have multiplied the action by $P_{\zeta}$ and we have set $\td= 0$ and $k \eta_\star = -1$, giving $\psi=0$. 
Let us comment on the features of our numerical results and their similarities to the saddle-point results (\ref{eq:ode_saddle_z0pos}) and (\ref{eq:ode_saddle_z0neg}) below. We leave the actual comparison between the results of these two methods to Sec.~\ref{sec:pde_num}.

As we already mentioned in the saddle-point calculation, the most striking feature of these results is the stark asymmetry between positive and negative $\bar{\zeta}$. Positive values of $\bar{\zeta}$ lead to larger contributions to the action, exponentially enhanced by $\alpha$. On the other hand, for negative $\bar{\zeta}$ the action simply scales as $\tilde b |\bar{\zeta}|^{5/2} / \sqrt{\alpha}$: larger $\alpha$ reduces the value of the action.
We note however that $\Delta S_{\rm E, 1}$ is oscillatory, hence the enhancement might not directly lead to large asymmetries when computing specific observables.

Another crucial feature is the presence of oscillations in $\bar{\zeta}$. As we were expecting from the saddle calculation, these oscillations have indeed frequency $\alpha$. In addition to this main frequency, there is also a modulation $\sim e^{i \alpha \log \sqrt{\bar{\zeta}}}$, see Eqs.~\eqref{eq:ode_saddle_z0pos} and \eqref{eq:ode_saddle_z0neg}. Although it is difficult to notice by eye, we checked that this feature leads to a better agreement between numerical and analytical results.

We will see in the next section that these main features also apply beyond the SFM simplification, for different late-time profiles for $\bar{\zeta}$.

\subsection{Comments on the results}\label{sec_crit_remark}
In this subsection we comment on the properties of the WFU we computed using the SFM approximation in Secs.~\ref{subsec:int_saddle} and \ref{subsec:late_saddle}.

As already explained in the previous section, the asymmetry of the WFU between positive and negative values of $\bar \zeta$ is manifest in the regime $\alpha^2 |\bar \zeta| \gg 1$.
More explicitly, for positive $\bar{\zeta}$, Eq.~\eqref{eq:ode_saddle_z0pos} contains the exponential factor $e^{\pi \alpha/2}$, unlike the WFU for negative $\bar{\zeta}$, Eq.~\eqref{eq:ode_saddle_z0neg}. 
We notice that the combination $\tilde b \, e^{\pi \alpha / 2}$ can exceed unity, unless the parameter $\tilde b$ is chosen to be catastrophically small.
One expects that perturbation theory in $\tilde b$ breaks down when the ${\cal O}(\tilde b)$ action becomes of order of the Gaussian one; see also the discussion below Eq.~(\ref{eq:wfu_scaling_b}). In order to be more quantitative one should look at the action at ${\cal O}(\tilde b^2)$, which requires to evaluate the solution at ${\cal O}(\tilde b)$.

Another interesting point to stress is the onset of the asymmetry, which starts to appear in the intermediate regime, $\alpha^2 |\bar \zeta| \sim 1$. Here, there is an approximate exponential growth on the positive-$\bar{\zeta}$ side (see right panel of Fig.~\ref{fig:ode_alpha2z_large}), while on the negative side the results remain small.
One way to understand this phenomenon is that, for $\bar{\zeta} > 0$, the saddle point moves away from the imaginary axis (see the top-left panel of Fig.~\ref{fig:ode_saddles}), so that the factor $e^{\pi \alpha/2}$ tends to dominate the integral $\mathcal{I}_\sigma$.
One can give a very rough estimate of the growth of the action considering the exponent of the WFU as an infinite sum over tree-level Witten diagrams (or equivalently over wavefunction coefficients, which are related to correlators).
The terms of order $n$ in the series scale as $(\alpha^2 \bar \zeta)^n / (n!)^2$, 
where we have used Eq.~(\ref{eq:estimate_tree_scaling}) and we have kept only terms that contain the power $n$.
Given this result, we naively expect that by maximizing such a series over $n$ we obtain $n_{\rm max} \sim \alpha \bzeta^{1/2}$, which yields  
$\tilde{b}\,e^{\alpha \bar{\zeta}^{1/2}}$ on the positive-$\bar{\zeta}$ side. Note that we assume no cancellations among different terms in the series.
Notice that for $\alpha^2 |\bar \zeta| \gg 1$,  the action is dominated by contributions around $n_{\rm max}$, with negligible contributions from the two- and three-point functions.
On the negative side instead, we expect large cancellations among the terms of the series so to keep the overall sum small (similarly to what happens in alternating series, such as the cosine function, as opposed to the series for a real exponential).

We will expand on these considerations in an analogue quantum mechanical model in Sec.~\ref{app:qm_example}, where a similar behaviour of the wavefunction can be realized and the series coefficients can be investigated directly.

\section{Spherical profile analysis}\label{sec:PDE_analysis}

In the previous section we analysed the WFU using a single Fourier mode. Here we are going to analyse the WFU assuming that the late-time profile $\bar{\zeta}(\vect{x})$, and therefore the whole solution $\zeta(\tau, \vect x)$ is spherically symmetric, {i.e.}~function of the single variable $r$.  
We also assume that the profile $\bar{\zeta}(r)$ is localised in space, as motivated for instance  by primordial black hole formation, with an extremum at the center of coordinates $r=0$. We denote this extremum value as $\bar{\zeta} \equiv \bar{\zeta}(r = 0)$. The choice of spherical symmetry is motivated by simplicity and also by the expectation that non-spherical profile are less likely: this is indeed the case in the Gaussian case.
As in Sec.~\ref{subsec:saddle_pt_ode}, we will use the saddle-point approximation to evaluate the action $\Delta S_{\rm{E}, 1}$. Then, we will numerically compute the same action and compare the two approaches.

\subsection{Saddle-point approximation}\label{subsec:PDE_saddle}
Here we are going to approximate the action (\ref{eq:DeltaS1_E}) using the saddle-point calculation for large $\alpha$ and large $|\bar{\zeta}|$. In fact, the main obstacle in performing analytic and numerical estimates of the action is that we do not have an explicit analytic form for $\zeta(\tau, r)$ for a generic late-time profile.~\footnote{For a Gaussian profile, $\zeta(\tau, r)$ can be written in terms of exponential integrals but manipulations of such expressions are cumbersome.}
However, thanks to the intuition gained in Sec.~\ref{subsec:saddle_pt_ode}, we expect the final result to be dominated by a late-time saddle where we will be able to write down approximate expressions for $\zeta(\tau, r)$ when $|\tau| \ll 1$.

Following the same procedure as done in Sec.~\ref{subsec:saddle_pt_ode}, we rescale the spacetime coordinates with the typical spatial momentum $k_0$ of the late-time profile $\bar{\zeta}(r)$. From the action (\ref{eq:DeltaS1_E}) we obtain
\begin{equation}
\begin{aligned}
\label{eq:DeltaS1_E_PDE}
	\Delta S_{\rm{E},1}
	&
	= 
	\frac{\pi}{P_\zeta} 
	\sum_{\sigma = \pm 1}
	\int_{-\infty}^{0} \frac{\de \tau}{\tau^2} 
	\int_{0}^{\infty} \de r \, r^2 \, 
	e^{\sigma \alpha \pi/2}
	\bigg\{	 
	X(\tau, r)
	\exp\left(i \sigma \alpha \log\left(-\tau\right)  + i \sigma \alpha \zeta(\tau, r) \right)
	\nn \\
	& 
	\hspace*{6cm}	
	- 
	\bar{X}(r)
	\exp\left(i \sigma\alpha \log\left(- \tau\right) + i \sigma\alpha \bar{\zeta}(r) \right) 
	\bigg\}
	e^{-i \sigma (\tilde{\delta} + \alpha  \log(-k_0\eta_\star) )} \;,
\end{aligned}
\end{equation}
where we have defined $X(\tau,r) \equiv (\partial_\tau \zeta(\tau,r))^2+(\partial_r \zeta(\tau,r))^2 $ and  $\bar{X}(r) \equiv X(0,r) = (\partial_r \bar{\zeta})^2$.
As in Eq.~(\ref{eq:DeltaS1_E_ODE2}), we write down the action above in a more compact form:
\begin{align}
	\label{eq:DeltaS1_E_PDE2}
	\Delta S_{\rm{E},1}
	=
	\frac{\pi}{P_\zeta} 
	\sum_{\sigma = \pm 1}
	e^{i \sigma \psi}
	e^{\sigma \frac{\alpha \pi}{2}}
	\,
	\Cc I_{\sigma} 
	\;,
\end{align}
where $\psi \equiv  -\tilde{\delta}  -  \alpha\log(-k_0 \eta_\star)$ and the integral $\Cc I_{\sigma}$ is defined by
\begin{align}\label{eq:pde_integrals_I}
	\Cc I_{\sigma}
	\equiv
	\int_{-\infty}^{0} \de \tau 
	\int_{0}^{\infty}\de r \,
	\left( 
	e^{\Phi_{\sigma}}
	- 
	e^{\Psi_{\sigma}}
	\right)
	\;,
\end{align}
with the exponents $\Phi_{\sigma}$ and $\Psi_{\sigma}$ being
\begin{equation}\label{eq:saddle_pde_Phi_Psi}
\begin{aligned}
	\Phi_{\sigma}
	&
	\equiv
	-(2 - i \sigma \alpha)\log(-\tau)
	+ 2 \log(r)
	+ \log\left(X(\tau, r)\right)
	+ i \sigma \alpha \zeta(\tau, r)
	\;,
	\\
	\Psi_{\sigma}
	&
	\equiv
	-(2 - i \sigma \alpha)\log(-\tau)
	+ 2 \log(r)
	+ \log\left(\bar{X}(r)\right)
	+ i \sigma \alpha \bar{\zeta}(r)
	\;.
\end{aligned}
\end{equation}
As in Sec.~\ref{sec:ODE_analysis}, the term with $\Psi_\sigma$ can be neglected in the saddle-point expansion. This is due to the fact that $\Psi_\sigma$ gives non-negligible contributions only at $\tau \rightarrow 0$ in which the cancellation between $\Phi_\sigma$ and $\Psi_\sigma$ happens. Therefore, around the saddle point the integral (\ref{eq:pde_integrals_I}) is dominated by the expansion of $\Phi_\sigma$.  In what follows, we will focus on the $\Phi_{\sigma}$ term.  Note that, as we are going to show, the saddle point over the integral in $r$ is close to the peak of the late-time profile.

We now look for the saddle both in $\tau$ and in $r$: the saddle-point equations are 
\begin{equation}
\begin{aligned}
	\partial_\tau \Phi_{\sigma} 
	&= 
	-\frac{2-i\sigma \alpha}{\tau} 
	+ \frac{X'(\tau, r)}{X(\tau, r)} 
	+ i \sigma \alpha \zeta' (\tau, r)
	= 
	0
	\;,
	\\
	\partial_r \Phi_{\sigma} 
	&= 
	\frac{2}{r}
	+ \frac{\partial_r X(\tau, r)}{X(\tau, r)} 
	+ i \sigma \alpha \, \partial_r\zeta(\tau, r)
	= 
	0
	\;.
\label{eq:saddle_pt_eq_pde}
\end{aligned}
\end{equation}
This system of equations can be solved numerically for a generic late-time profile $\bar{\zeta}(r)$ with amplitude $\bar{\zeta}$. However, for large $|\bar{\zeta}|$ we can find an analytic solution, corresponding to a late-time location of the saddle point. For simplicity, we concentrate on this in what follows, although there would be no obstacle to study the intermediate regime $\alpha^2 \bar{\zeta} \gtrsim 1$ with a numerical solution to the saddle-point equations, as we did in Sec.~\ref{subsec:saddle_pt_ode} for a single Fourier mode.

~
\newline
\textbf{Late-time saddle:} We assume the profile to be such that $\partial_r \bar{\zeta}(r)|_{r = 0} = 0$ and $\partial^2_r \bar{\zeta}(r)|_{r = 0} \neq 0$, i.e.~the origin can be either a maximum or a minimum.
Then, using the late-time expansion \eqref{eq:zeta_late0}, in Euclidean time, we obtain
\begin{equation}
	\zeta(\tau, r) 
	\simeq 
	\bar{\zeta} 
	+\left( \frac{r^2}{6} + \frac{\tau^2}{2}\right) \nabla^2 \bar{\zeta} \bigg|_{r = 0}
	+ \ldots
	\;, \label{eq:exp_saddle_PDE}
\end{equation}
where we have used the fact that around $r = 0$ the late-time profile $\bar{\zeta}(r)$ can be expanded as
\begin{align}
\bar{\zeta}(r) = \bar{\zeta} + \frac{1}{6} r^2 \nabla^2 \bar{\zeta} \bigg|_{r = 0} + \ldots \;.
\end{align}
Note that for a spherically symmetric profile we have $\nabla^2 \bar{\zeta} = 3 \partial_r^2 \bar{\zeta}$. ~\footnote{
For a profile with $\nabla^2 \bar{\zeta} |_{r = 0} = 0$, one should go to next order in the expansion.}
For later convenience, we will drop the evaluation symbol, $|_{r = 0}$, and denote as $\nabla^2 \bar{\zeta}$ the Laplacian of $\bar{\zeta}(r)$ computed at $r = 0$.

Let us now determine the saddle point for the integrals over $\tau$ and $r$. Using the expansion (\ref{eq:exp_saddle_PDE}) in \eqref{eq:saddle_pt_eq_pde} the saddle point of the $\tau$-integral is 
\begin{equation}
	\tau_s^2 
	\simeq 
	\frac{1}{- \nabla^2 \bar{\zeta}}
	\left( 1 + \frac{2i \sigma}{\alpha} \right)
	\;,
	\label{eq:sol_tau_saddle}
\end{equation}
and the saddle point of the $r$-integral is
\begin{equation}
	r_s^2 
	\simeq 
	\frac{-6 i \sigma}{\alpha(- \nabla^2 \bar{\zeta})}
	\;,
	\label{eq:sol_r_saddle}
\end{equation}
where we have self-consistently assumed that $|r_s| \ll |\tau_s| \ll 1$ and we have neglected the term $\partial_r X(\tau,r) /X(\tau, r)$ in Eq.~(\ref{eq:saddle_pt_eq_pde}), which we can check in retrospect to be consistent with the assumption $\alpha^2 \bar{\zeta} \gg 1$.
The saddle-point location \eqref{eq:sol_tau_saddle} reduces to the one of the previous section, Eq.~\eqref{eq:late_saddle_ode}, when $\bar{\zeta}(r)$ is treated as a single Fourier mode. Below, we are going to evaluate the action $\Delta S_{\rm{E}, 1}$ on the saddle point (\ref{eq:sol_tau_saddle})--(\ref{eq:sol_r_saddle}) in the two cases $-\nabla^2\bar{\zeta} > 0$ and $-\nabla^2\bar{\zeta} < 0$, corresponding respectively to a local maximum and a local minimum of the profile.

\subsubsection{Local maximum}
Here we are interested in the case $-\nabla^2\bar{\zeta} > 0$. 
Let us compute the exponent $\Phi_\sigma$ evaluated on the saddle point (\ref{eq:sol_tau_saddle})--(\ref{eq:sol_r_saddle}). There are four contributions in Eq.~(\ref{eq:saddle_pde_Phi_Psi}) giving
\begin{equation}\label{eq:each_Phi_sigma_PDE}
\begin{aligned}
-(2 - i \sigma \alpha) \log(-\tau_s) &\simeq -1 + \bigg(1 - \frac{i \sigma \alpha}{2}\bigg) \log(|\nabla^2 \bar{\zeta}|) \;, \quad
2 \log(r_s) = \log\bigg[\frac{-6 i \sigma}{\alpha |\nabla^2 \bar{\zeta}|}\bigg] \;, \\ 
\log(X(\tau_s, r_s)) &\simeq \log(|\nabla^2 \bar{\zeta}|) \;, \quad 
i\sigma \zeta(\tau_s, r_s) \simeq i \sigma \alpha \bar{\zeta} - \frac{i \sigma \alpha}{2} \;,
\end{aligned}
\end{equation}
where we have taken the limit $\alpha \gg 1$ and we have used 
\begin{align}
(\partial_\tau \zeta)^2 \big\rvert_{\tau_s, r_s} \simeq \bigg(1 + \frac{2 i \sigma}{\alpha}\bigg) |\nabla^2 \bar{\zeta}| \;, 
\quad (\partial_r \zeta)^2\rvert_{\tau_s, r_s} \simeq \frac{2i \sigma}{3 \alpha} |\nabla^2 \bar{\zeta}| \;. 
\end{align}
Combining all the pieces we get
\begin{align}
e^{\Phi_\sigma} \rvert_{\tau_s, r_s} \simeq -\frac{6 i \sigma }{e \alpha}  |\nabla^2 \bar{\zeta}| \, e^{i  \sigma \alpha  (\bar{\zeta}- \log(\sqrt{|\nabla^2 \bar{\zeta}|}))}
	e^{- i \sigma \alpha /2} \;. \label{eq:expo_Phi_sigma_PDE}
\end{align}
Moreover, in order to evaluate the integral (\ref{eq:pde_integrals_I}), it is necessary to compute the Hessian matrix of $\Phi_\sigma$ evaluated at the saddle point. We find
\begin{equation}
	\partial_\tau^2 \Phi_\sigma\big\rvert_{\tau_s, r_s}
	\simeq 
	2 i \sigma \alpha \nabla^2 \bar{\zeta}
	\;, \quad
	\partial_r^2 \Phi_\sigma \big\rvert_{\tau_s, r_s} 
	\simeq 
	\frac{2}{3}i \sigma \alpha \nabla^2 \bar{\zeta} \;, \label{eq:Hessian_pde}
\end{equation}
while the off-diagonal term is negligible at this order. 
Note that the expression (\ref{eq:Hessian_pde}) is valid for both $-\nabla^2 \bar{\zeta} > 0$ and $-\nabla^2 \bar{\zeta} < 0$.
Therefore, using Eqs.~(\ref{eq:expo_Phi_sigma_PDE}) and (\ref{eq:Hessian_pde}), the integral (\ref{eq:pde_integrals_I}) can be approximated as 
\begin{align}
	\Cc I_{\sigma} 
	&\simeq
	\frac{2\pi}{\sqrt{\det\left(- \partial_{\mu}\partial_\nu\Phi_\sigma\right)\big\rvert_{\tau_s, r_s}}}
	\, e^{\Phi_{\sigma}}\big\rvert_{\tau_s, r_s} \label{eq:saddle_pt_general} \\
	&\simeq
	-
	\frac{6\pi \sigma \sqrt{3} }{\alpha^2 e}\,
	e^{i  \sigma \alpha  (\bar{\zeta}- \log(\sqrt{|\nabla^2 \bar{\zeta}|}))}
	e^{- i \sigma \alpha /2}
	\;,
	\label{eq:I_s_zo_gr_0_pde}
\end{align}
where we have kept the leading order for large $\alpha$ and $\bar{\zeta}$. 
We note that the factor $\nabla^2 \bar{\zeta}$ in the determinant of the Hessian matrix cancels with the one in (\ref{eq:expo_Phi_sigma_PDE}).
Looking at Eq.~(\ref{eq:DeltaS1_E_PDE2}), we see that the action is dominated by the terms with $\sigma = 1$, since there is no parametric difference at the level of $\Cc I_{\sigma}$, similarly to the case of the SFM analysis.
We therefore obtain the action at first order in $\tilde{b}$, 
\begin{empheq}[box={\mybox[5pt][5pt]}]{equation}
\begin{aligned}
	\Delta S_{\rm E, 1} 
 	\simeq 
 	-
 	\frac{6 \pi^{2}}{P_{\zeta}}
	\frac{\sqrt{3}}{\alpha^2 e}
	\,
	e^{\pi \alpha/2} 
	e^{i  \sigma \alpha  (\bar{\zeta}- \log(\sqrt{|\nabla^2 \bar{\zeta}|}))}
	e^{i \chi}
	\;,
	\quad
	(\textrm{local maximum})
	\;,
	\label{eq:pde_saddle_z0pos}
\end{aligned}
\end{empheq}
\noindent where $\chi \equiv \psi - \alpha / 2$. 
Let us comment on the features of the result above. 
First, we have a different overall scaling with $\nabla^2 \bar{\zeta}$, compared with Eq.~\eqref{eq:ode_saddle_z0pos}. 
This is simply due to the fact that in this analysis we are dealing with the two dimensional integral, instead of one dimensional integral as in the SFM simplification.  
Thus, it leads to the fact that $\Delta S_{\rm E,1}$ above does not grow as $\bar{\zeta}$ increases.
Apart from this, we also have a different scaling in $\alpha$, compared with Eq.~\eqref{eq:ode_saddle_z0pos}.
Moreover, it is useful to point out that our result (\ref{eq:pde_saddle_z0pos}) cannot be obtained by simple rescaling of $\zeta$, as in \cite{Celoria:2021vjw} where the on-shell action is simply proportional to an arbitrary function of the expansion parameter.   
Essentially, this is due to the resonant effect, resulting in a non-trivial dependence of $\alpha$ in the action $\Delta S_{\rm E, 1}$, e.g. the enhancement factor $e^{\pi \alpha/2}$.  
Finally, we see that the action (\ref{eq:pde_saddle_z0pos}) behaves as an oscillating function in both $\bar{\zeta}$ and 
$\log(\sqrt{|\nabla^2 \bar{\zeta}|})$, with frequency $\alpha$, which cannot be captured by any order in perturbation theory.

\subsubsection{Local minimum}
\label{sec:pde_nagative_zeta}
Here we are interested in the $-\nabla^2 \bar{\zeta} < 0$ case, which is similar to the SFM simplication with $\bar{\zeta} < 0$.
As explained in Sec.~\ref{subsub:neg_zeta_ODE}, the role of $\tau_s$ changes depending on the sign of $\nabla^2\bar{\zeta}$. We find that in this case the relevant saddle point of the $\tau$-integral is 
\begin{equation}
	\tau_s
	= 
	\frac{-i\sigma}{\sqrt{\nabla^2 \bar{\zeta}}} 
	\left(
	1 
	+ \frac{2i\sigma}{\alpha}
	\right)^{1/2}
	\;.
	\label{eq:late_saddle_pde_neg}
\end{equation}
We see that for $\sigma = 1$ this saddle point lies on the negative imaginary axis in the limit $\alpha \gg 1$. This implies that the factor $e^{\pi \alpha/2}$ in (\ref{eq:DeltaS1_E_PDE2}) will disappear (essentially one rotates back to the Lorentzian time). On the other hand, for $\sigma = -1$ the saddle point then lies on positive imaginary axis, which gives another factor $e^{-\pi \alpha/2}$ to the action. Therefore, the dominant piece in $\Delta S_{\rm{E}, 1}$ in this case is given by the term with $\sigma = 1$. Following the same procedure as before, we obtain
\begin{empheq}[box={\mybox[5pt][5pt]}]{equation}
\begin{aligned}
	\Delta S_{\rm E, 1} 
 	\simeq 
 	\frac{6 \pi^2}{P_{\zeta}}
	\frac{\sqrt{3}}{\alpha^2 e}
	\, 
	e^{i  \sigma \alpha  (\bar{\zeta}- \log(\sqrt{|\nabla^2 \bar{\zeta}|}))}
	e^{i \chi}
	\;,
	\quad
	(\textrm{local minimum})
	\;.
	\label{eq:pde_saddle_z0neg}
\end{aligned}
\end{empheq}
\noindent It is important to point out that there is no issue of selecting the right saddle point of the $r$-integral because at leading order the dependence on this variable is only through $r^2$. 
The action above indicates that the overall scalings in $\alpha$ and $\bar{\zeta}$ are different from the one of the single-mode simplification, Eq.~(\ref{eq:ode_saddle_z0neg}). 
We also see that the result (\ref{eq:pde_saddle_z0neg}) is similar to (\ref{eq:pde_saddle_z0pos}) in the sense that they share the same oscillatory behaviour as a function of $\bar{\zeta}- \log(\sqrt{|\nabla^2 \bar{\zeta}|})$, a behaviour of the WFU that cannot be captured by perturbative computations.
Moreover, the fact that this result (\ref{eq:pde_saddle_z0neg}) does not contain the enhancement factor $e^{\pi\alpha /2}$ implies that the amplitude of the WFU is much smaller for a local minimum than for a local maximum, resulting in an interesting asymmetry between the two situations.
 
In the next subsection we will compute the action $\Delta S_{\rm{E}, 1}$ numerically and compare it to the analytic results, Eqs.~(\ref{eq:pde_saddle_z0pos}) and (\ref{eq:pde_saddle_z0neg}).

\subsection{Numerical results}\label{sec:pde_num}
In this section we evaluate the WFU using the numerical integration of Eq.~\eqref{eq:DeltaS1_E_PDE}. For concreteness, we focus on a Gaussian profile at late times:
\begin{equation}
	\bar \zeta(r)
	= 
	\bar \zeta \, e^{-(k_0 r)^2}
	\;,
	\label{eq:gaussian_prof}
\end{equation}
where $k_0$ is a given momentum scale and $\bar{\zeta}$ denotes the peak value of this Gaussian profile.
Note that, for this profile we have $\nabla^2 \bar \zeta = -6 k_0^2 \bar \zeta$  at $r = 0$. Therefore, 
we have $\nabla^2 \bar{\zeta} < 0$ for positive $\bar \zeta$, while $\nabla^2 \bar{\zeta} > 0$ for negative $\bar{\zeta}$.
We refer the reader to App.~\ref{app:num_method} where we explain in detail our numerical method used to compute the action (\ref{eq:DeltaS1_E_PDE}) with the late-time configuration (\ref{eq:gaussian_prof}).
Below, we report our numerical results and compare them with the results obtained from the saddle-point approximation in Sec.~\ref{subsec:PDE_saddle}.

\begin{figure}[t!]
\centering
  \includegraphics[width=0.7\textwidth]{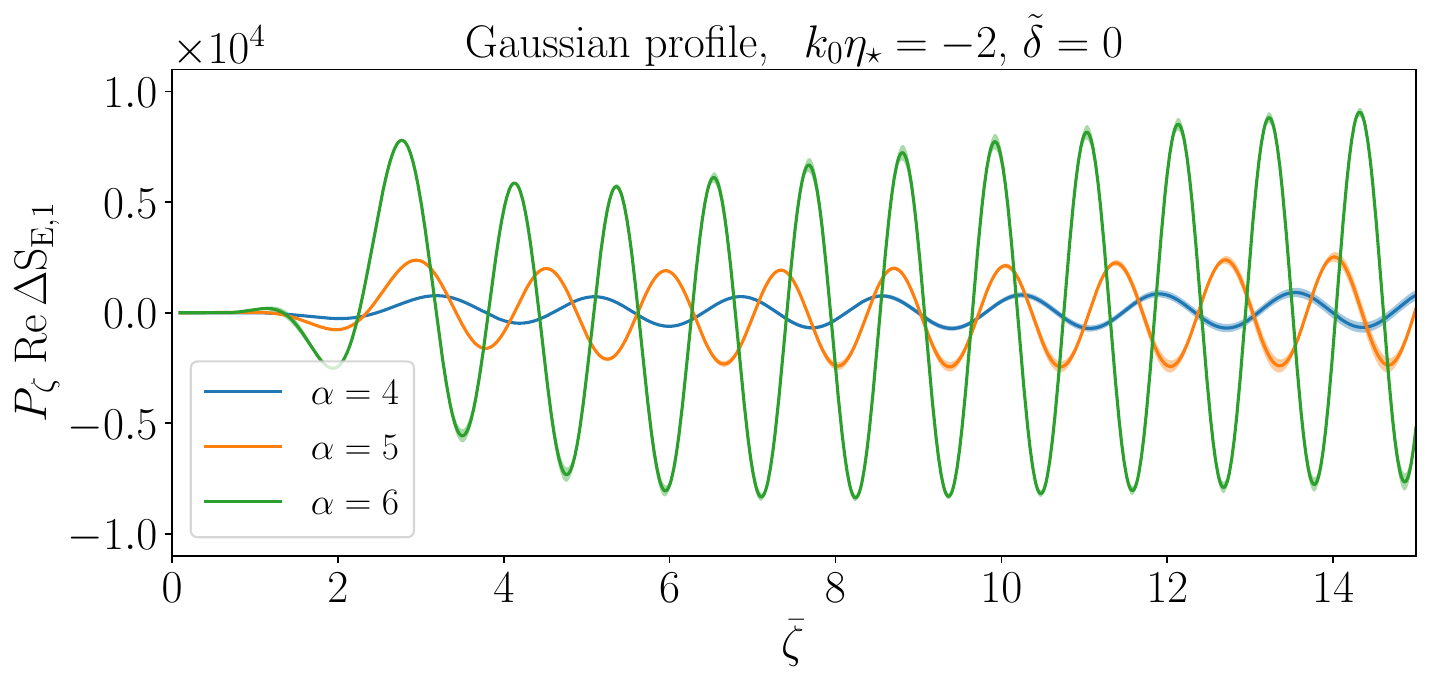}
  \caption{\quad Numerical results for the real part of the on-shell action with Gaussian profile \eqref{eq:gaussian_prof} for different values of $\alpha$ as a function of $\bar \zeta$ (positive). We use $\tilde{\delta}= 0$ and $k_0 \eta_\star = -2$. The numerical integration is performed using the FFT method. For each curve the shaded area represents the estimated numerical error, obtained by comparing the results from the FFT and the PDE methods (see App.~\ref{app:num_method} for details).}
\label{fig:pde_pos} 
\end{figure}
\begin{figure}[t]
\centering
  \includegraphics[width=0.7\textwidth]{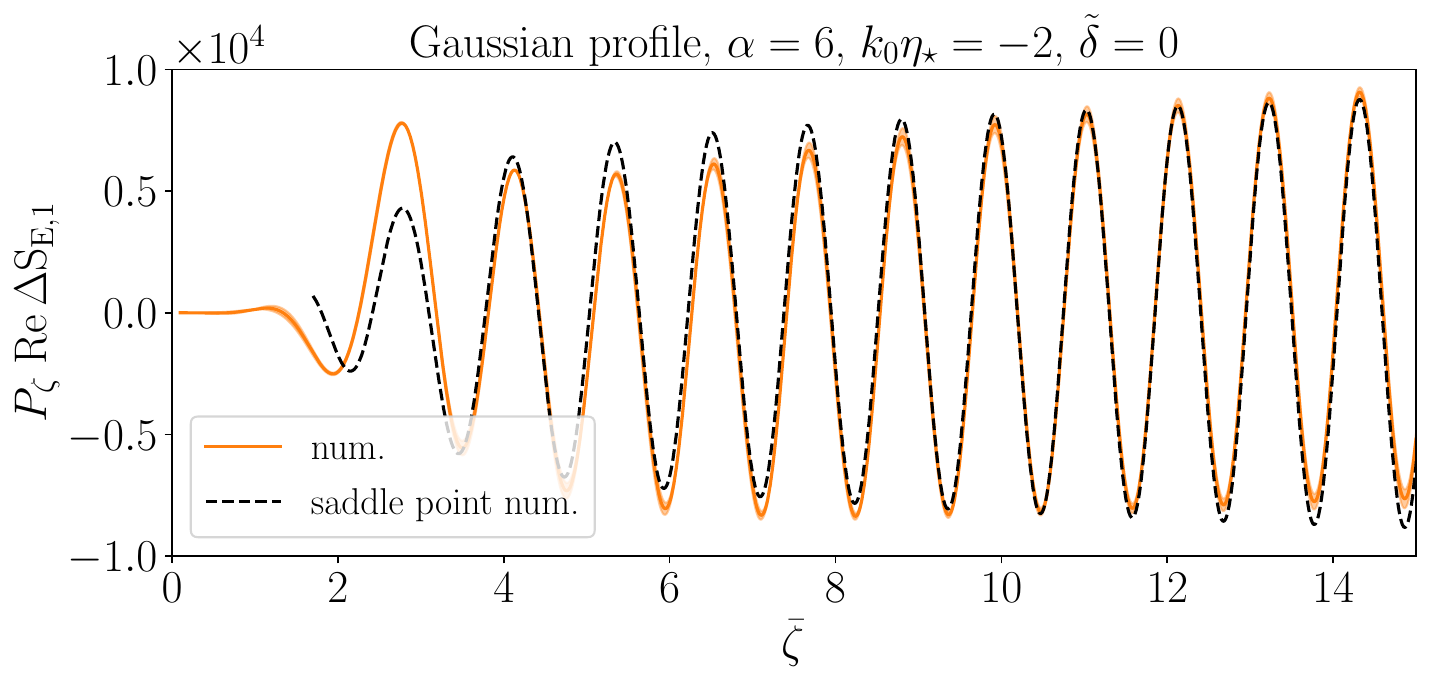}
  \caption{\quad Comparison between the numerical integration (orange) of the action and the numerical saddle-point approximation for the integral (black dashed) as a function of $\bar \zeta$ for $\alpha = 6$. The other parameters are fixed as in Fig.~\ref{fig:pde_pos}.
  The saddle-point curve obtained from a late-time expansion is truncated at small $\bar \zeta$, where the expansion stops being reliable. The shaded orange area represents the difference between the results obtained from the FFT and the PDE methods.}
\label{fig:pde_pos_saddle_vs_num} 
\end{figure}

The results for $\Delta S_{\rm E, 1}$ are obtained by combining $\Delta S_{1, \rm{early}}$ (Eq.~(\ref{Delta_S_1early})) and $\Delta S_{1, \rm{late}}$ (Eq.~(\ref{Delta_S_1late})) with the numerical integral performed in the interval $\{\tau_{\rm min}, r_{\rm min}\}$ to $\{\tau_{\rm max}, r_{\rm max}\}$, as explained in App.~\ref{app:num_method}. 
Following the analytical result in saddle point obtained in Sec.~\ref{subsec:PDE_saddle}, we discuss separately the cases $\bar \zeta < 0$ and $\bar \zeta > 0$ (corresponding to $\nabla^2 \bar \zeta > 0$ and $\nabla^2 \bar \zeta < 0$, respectively). In particular, we focus on evaluating the real part of the resonant action $\Delta S_{\rm E, 1}$ as a function of $\bar \zeta$ for several values of $\alpha$ (it is straightforward to verify that the imaginary part of $\Delta S_{\rm E, 1}$ qualitatively behaves in the same way as the real part). To have a better control on the numerical errors, we focus on moderate values of $\alpha$ and $\bar \zeta$: larger values require higher spatial resolution in the integration. 

Fig.~\ref{fig:pde_pos} shows our numerical results for $\bar \zeta > 0$ ($\nabla^2 \bar{\zeta} < 0$) for several values of $\alpha$, with $k_0 \eta_\star = -2$ and $\td$. 
In the plot, the blue, orange and green lines correspond to $\alpha = 4$, $5$ and $6$ respectively. 
We can see that the real part of the action exhibits oscillatory features with frequency $\alpha$. 
Moreover, for $\bar \zeta > 0$ the amplitude of the action grows exponentially with $\alpha$. 
To assess the convergence of our numerical implementations, we compare the results obtained using the FFT and the PDE methods with similar resolutions (see App.~\ref{app:num_method} for details). The difference between the two methods gives an estimate of the numerical error, which is represented by the shaded area around the numerical curves in Fig.~\ref{fig:pde_pos}.
We find that the results from the two methods coincide with small numerical errors. 
Furthermore, we also compare the numerical integration with the saddle-point approximation in Fig.~\ref{fig:pde_pos_saddle_vs_num}. 
Actually, it is important to note that the comparison is carried out with the \emph{numerical} solution of the saddle-point equations \eqref{eq:saddle_pt_eq_pde}.~\footnote{To obtain this numerical solution, we include very high-order terms in the late-time expansion \eqref{eq:zeta_late0} and use the analytic solutions Eqs.~\eqref{eq:sol_tau_saddle} and \eqref{eq:sol_r_saddle} as initial guesses. The action is then obtained by evaluating Eq.~\eqref{eq:saddle_pt_general} on the numerical saddle point. 
For $\bar \zeta \sim \mathcal O(10)$ the approximate analytic solution \eqref{eq:pde_saddle_z0pos} differs from the numerically-evaluated saddle by $\sim 30\,\%$.
We checked that the difference decreases for higher $\bar \zeta$. This is in agreement with the expectation that corrections to Eq.~\eqref{eq:pde_saddle_z0pos} scale as $\sim 1 / |\bar \zeta|^{1/2}$.} This is due to the fact that the analytic formula \eqref{eq:pde_saddle_z0pos} is not expected to be very accurate since we are dealing with moderate values for $\bar \zeta$. In Fig.~\ref{fig:pde_pos_saddle_vs_num} we see that the agreement improves as $\bar \zeta$ increases, where the saddle-point approximation is expected to become more accurate. The same behaviour is expected to happen for higher values of $\alpha$'s.

\begin{figure}
\centering
  \includegraphics[width=0.7\textwidth]{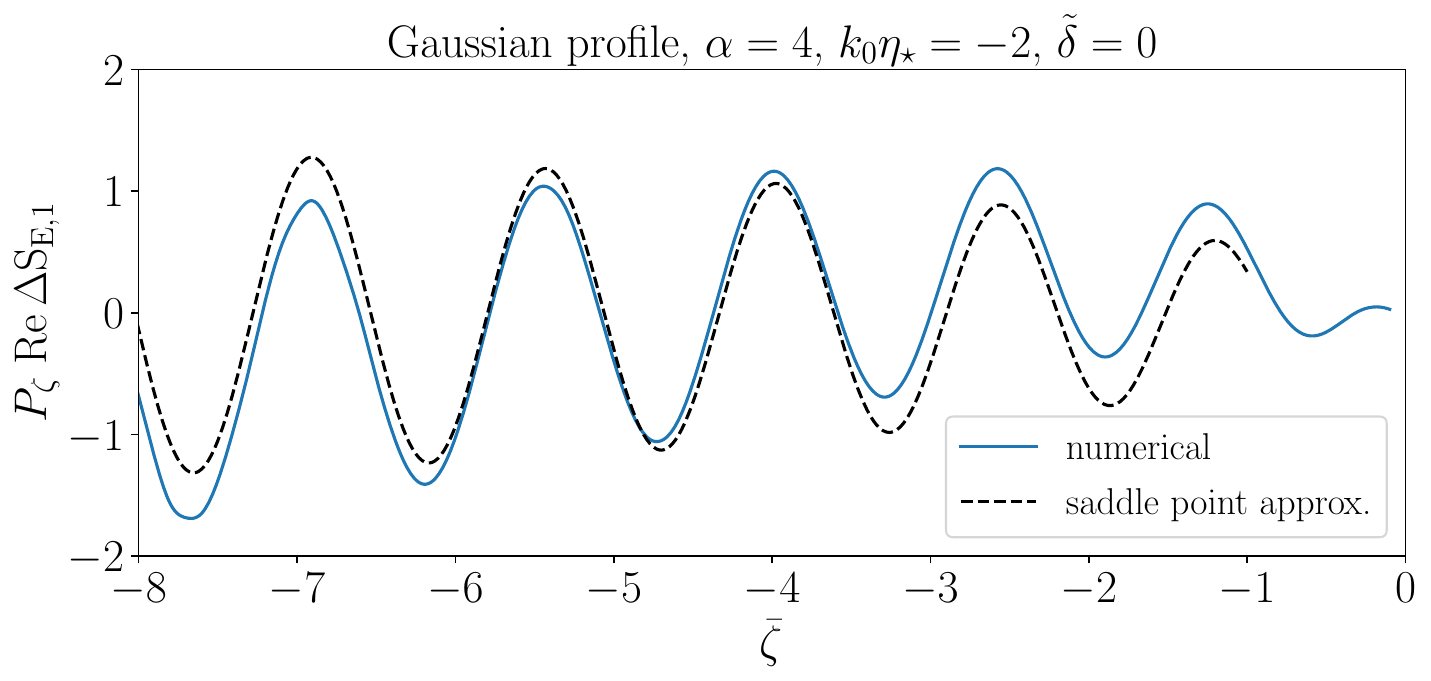}
  \caption{\quad The real part of the on-shell action with Gaussian profile for $\alpha = 4$ as a function of $\bar \zeta$ (negative). 
  We compare the numerical integration (blue) with the numerical saddle-point approximation (black dashed).
  The numerical curve is obtained using the analytic formula for $\zeta(\tau, r)$, Eq.~\eqref{eq:zeta_analytic}.
  The difference at larger $|\bar \zeta|$ is expected to be mainly due to numerical errors.
  We use $\td= 0$ and $k_0 \eta_\star = -2$. 
  }
\label{fig:pde_neg} 
\end{figure}

Let us move to the case where $\bar \zeta < 0$ ($\nabla^2 \bar{\zeta} > 0$). In this case the numerical integration becomes considerably more challenging. The reason can be explained as follows. As already discussed in Sec.~\ref{sec:pde_nagative_zeta}, the resonant action is not exponentially enhanced in $\alpha$ since the saddle point $\tau_s$ becomes Lorentzian. Then, the fact that the integrand in \eqref{eq:DeltaS1_E} contains exponentially large factors, $e^{\pi \alpha/ 2}$, implies that there should be cancellations of such a large numerical value in the integration. Therefore, to resolve these cancellations and obtain numerically convergent results, we then need an accuracy of order $\sim e^{-\pi\alpha / 2}$.~\footnote{Performing the numerical integration along the Lorentzian contour does not improve the situation: the exponential in $\alpha$ is removed but $\zeta(\eta, r)$ is now complex, leading to exponential terms of order $\sim e^{\alpha \bar \zeta}$.} Given the complication in this case, we only perform the numerical integration using the analytic expression \eqref{eq:zeta_analytic} of $\zeta(\tau,r)$. In Fig.~\ref{fig:pde_neg} we plot the numerical result (blue solid line) with $\alpha = 4$, $k_0 \eta_\star = -2$ and $\td= 0$, and the numerical saddle-point approximation (black dashed line). The two results agree and illustrate the fact that the resonant action oscillates as a function of $\bar{\zeta}$ with frequency $\alpha$. 
Finally, we confirm the asymmetric property of the WFU, i.e.~the amplitude of $\Delta S_{\rm E,1}$ for negative $\bar{\zeta}$ is much smaller than the one for positive $\bar{\zeta}$, see the different vertical axes of Figs.~\ref{fig:pde_pos_saddle_vs_num} and \ref{fig:pde_neg}.

\section{Quantum mechanics example}\label{app:qm_example}

In this section we will study a simple quantum-mechanical model that shares many features with the inflationary scenario we have been discussing. This will help in understanding the asymmetry between negative and positive values of $\bzeta$ and the size of the effect of oscillations.

Let us consider a quantum harmonic oscillator perturbed by a small time-dependent interaction. The Hamiltonian is taken to be
\begin{align}\label{eq:H_QM}
	{\rm{H}} 
	&= 
	{\rm{H}}_0
	+
	\tilde b \, \mathcal{E}_0 W(t) \cos(\alpha \log(-t / t_0) + \alpha x / \ell)
	\nonumber \\
	&= 
	{\rm H}_0 + \tilde b \, \Delta {\rm H}(t)\;,
	\qquad
	{\rm H}_0 \equiv \frac{p^2}{2 m} + \frac{m\omega^2}{2} x^2\;.
\end{align}
Here $\tilde b$ is our small expansion parameter, $W(t)$ is a window function that turns off the interaction at early times ($t\to -\infty$) and late times ($t \to 0$): the time $t$ here plays the role of conformal time in the inflationary case.
For convenience we take $W(t) = -\omega t  \, e^{\varepsilon \cdot \omega t }$ with $\varepsilon > 0$ and small. This toy model is chosen to mimic some characteristic features of our inflationary setup, of action \eqref{eq:Lag_pi}. The window function turns off the effect of the forcing at late times, effectively replacing the Hubble friction, while at early times one has an analogue of the usual $i \epsilon$ prescription.
Note that with this choice, resonance effects between the forcing term and the 
$n^\textrm{th}$ energy eigenstate of the harmonic oscillator happen around the time $\omega t \simeq - \alpha / n$.
The coefficient $\ell$ is a classical length scale (i.e.~it is finite as $\hbar \to 0$). 
As in the inflationary case, we want to be in a regime where quantum fluctuations of $x$ (which are of order $d \equiv \sqrt{\hbar / (m \omega)}$) do not jump to other minima of the cosine: thus we need $\alpha d/\ell \ll 1$. 
Finally, $\mathcal{E}_0$ is a classical energy scale.

Let us assume to be in the vacuum $\ket{0}$ of ${\rm H}_0$ at early times $t \to -\infty$ and seek the evolved wavefunction when the interaction drops to zero at $t = 0$. 
At linear order in $\tilde b$ the wavefunction, at all orders in $\hbar$, can be obtained using time-dependent perturbation theory. Indeed, if we write $\ket{\Psi(t = 0)} = e^{-i E_0 T / \hbar}\ket{0} + \sum_{n} c_n e^{-i E_n T / \hbar}\ket{n}$ with $n = 0, 1, \ldots$ (and $T$ being the total time of the evolution) then the coefficients $c_n$ are given by
\begin{equation}
	c_n 
	=
	-\frac{i \tilde b}{\hbar} 
	\int_{-\infty}^{0} \de t \braket{n|\Delta {\rm H}(t)|0} e^{-i (E_0 - E_n) t / \hbar} 
	e^{i (E_n - E_0) T / \hbar}
	\;.
	\label{eq:cn_integr_dt}
\end{equation}
Notice that the final exponential phase factor, containing $T$, cancels at the level of the wavefunction. Therefore, we are going to drop $T$ in all the expressions. To compute the matrix element we split the cosine into a sum of exponentials. Then, we notice that the operator ${\rm D}(A) = e^{\pm i \alpha x / \ell}$ is nothing but the boost operator and that when applied to the vacuum it generates a coherent state $\ket{A}$, with $A \equiv \pm i \alpha d / (\sqrt{2} \ell)$. Therefore, $\braket{n|{\rm D}(A)|0} = A^n e^{-|A|^2 / 2} / \sqrt{n!}$.
Using this relation, it is then straightforward to obtain the $c_n$ as
\begin{equation}\label{eq:cn_QM}
	c_n 
	=
	-\frac{i \tilde b}{2 \sqrt{2^n n!}}
	\frac{\mathcal{E}_0}{\hbar \omega}
	e^{-\alpha ^2 d^2/(4 \ell^2)}
	\sum_{\sigma = \pm 1}
	\left(\frac{i \alpha d \sigma }{\ell}\right)^n
	\frac{\Gamma (2 + i \alpha  \sigma)}{(\varepsilon+i n)^{2 + i \alpha \sigma}}
	\;.
\end{equation}
(The time integral can also be performed in saddle-point approximation for large $\alpha$, equivalently, one can expand the Gamma functions in the last expression.) Notice that, for $n > 0$, the denominator $(\varepsilon+i n)^{2 + i \alpha \sigma} \simeq (i n)^{2 + i \alpha \sigma}$ is exponentially suppressed for $\sigma = 1$ compared to $\sigma = -1$. Hence, the dominant contribution comes from $\sigma = 1$ and we are going to drop the other one (this corresponds to a boost of positive momentum $\alpha / \ell$). 
Moreover, one can verify that the case $ n = 0$ gives an exponentially small contribution in $\alpha$ hence we will focus only on the terms $n > 0$ (in this case one also neglects $\varepsilon$). 
Combining these results with the $\alpha \gg 1$ limit gives the simplified expression
\begin{align}
	c_n 
	&\simeq
	\sqrt{2 \pi}\alpha ^{3/2}
	\tilde b \,
	\frac{\mathcal{E}_0}{\hbar \omega}
	\frac{1}{2 \sqrt{2^n n!}}
	\frac{\left(i \alpha d /\ell\right)^n}{(i  n)^{2+ i \alpha }}
	e^{-\pi \alpha /2}
	e^{-\alpha ^2 d^2/(4 \ell^2)}
	e^{i \varphi}
	\\
	&=
	-
	\sqrt{2 \pi}\alpha ^{3/2}
	\tilde b
	\,
	\frac{\mathcal{E}_0}{\hbar \omega}
	\frac{\left(i \alpha d /\ell\right)^n}{2\sqrt{2^n n!}}
	\frac{1}{n^{2+i \alpha}}
	e^{-\alpha ^2 d^2/(4 \ell^2)}
	e^{i \varphi}
	\;,
	\label{eq:cn_approx}
\end{align}
where we have defined $\varphi \equiv - \alpha + \alpha \log \alpha + \pi/4$.

Before studying the wavefunction at late times, we can gain some intuition for the final result by analysing the transition probability $P_{0 \to n} \equiv \lvert \braket{n|\Psi(t = 0)}\rvert^2  = \rvert c_n\lvert^2$, assuming $n\neq 0$. 
We are going to show that $P_{0 \to n}$ is dominated by transitions to small $n$'s. 
The probability is 
\begin{equation}
	P_{0 \to n}
	\simeq 
	\frac{\pi \alpha ^3 \tilde b^2}{2n^{4}n!}
	\frac{\mathcal{E}_0^2}{(\hbar \omega)^2}
	\left(
	\frac{\alpha^2 d^2}{2 \ell^2}
	\right)^n
	e^{-\alpha ^2 d^2/(2 \ell^2)}
	\;.
	\label{eq:Pn_QM}
\end{equation}
Since $\alpha d / \ell \ll 1$, this expression decreases with increasing $n$ and the maximum is attained at $n=1$.
Therefore, the interaction populates states with large $n$ with a tiny probability. However, as we are going to see momentarily, excited states are very relevant when focussing on the tail of the distribution. The probability of transition to excited states is small (the factors of $\hbar$ at the denominator cancel) and  this suggests that the perturbative expansion in $\tilde b$ is reliable for this type of question.

\begin{figure*}[t!]
 \includegraphics[width=0.49\textwidth]{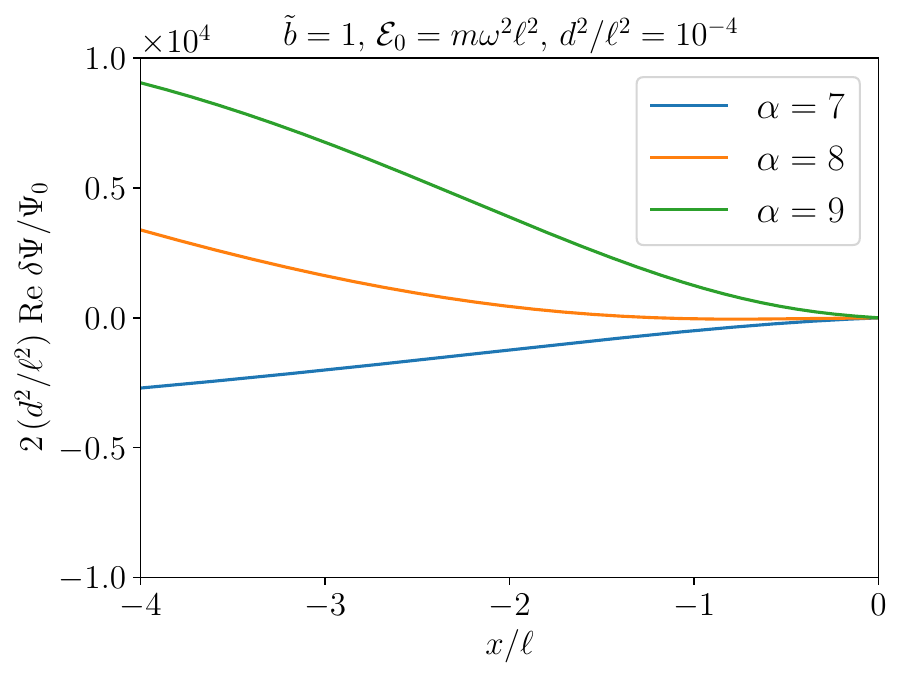}
 \includegraphics[width=0.49\textwidth]{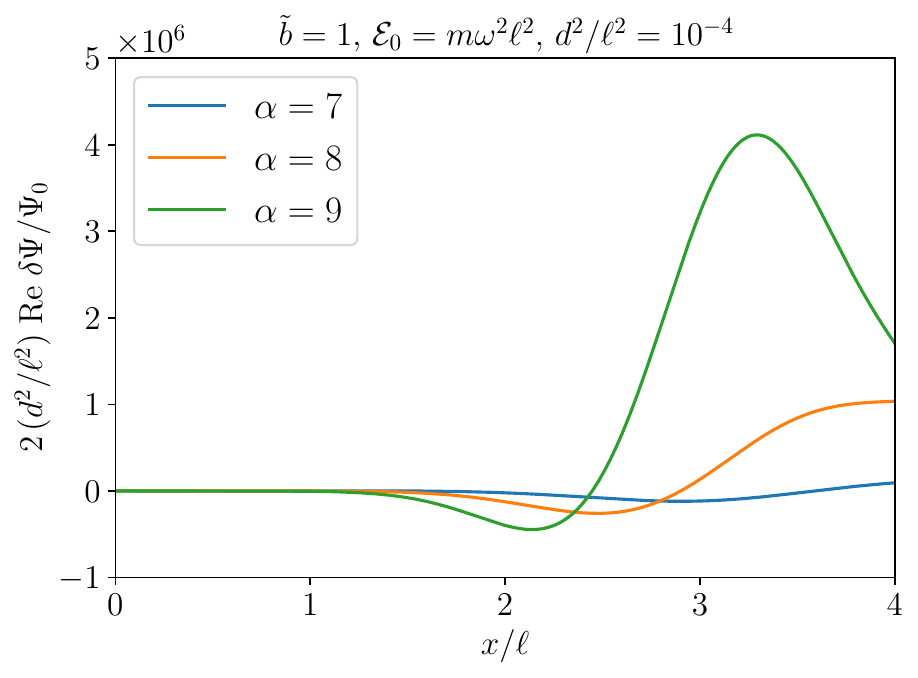}
 \caption{~Relative correction to the real part of the wavefunction as a function of $x / \ell$ for $\alpha = 7$, $8$ and $9$ for $\tilde b = 1$. We use units where $d^2 / \ell^2 = 10^{-4}$ and $\varepsilon = 10^{-2}$. With these values, the ranges of $x / \ell$ in the plot belong to the semiclassical limit. The wavefunction is multiplied by $d^2 / \ell^2$. Notice the similarities with the results in the main text, notably Fig.~\ref{fig:SFM-intermediate}. 
 \label{fig:qm_plot}
 \textbf{Left:} case of negative $x$. \textbf{Right:} case of positive $x$. 
 }
\end{figure*}

We can then look at the wavefunction in position space at $t = 0$, $\Psi(x) \equiv \braket{x|\Psi(t = 0)}$. 
Here we will need the expression for the harmonic oscillator eigenstates:
\begin{equation}
	\psi_n(x) 
	=
	\braket{x|n}
	= 
	\frac{\sqrt{d}}{\pi^{1/4}\sqrt{2^n n !} } 
	H_n\left(x / d\right)
	e^{- x^2/(2 d^2)}\;,
\end{equation}
where $H_n(x/d)$ are the Hermite polynomials. We are going to study the wavefunction in the semiclassical limit, $\hbar \to 0$, which is appropriate on the tail of the distribution. In this limit one must keep the highest degree term in $H_n(x/d)$ (i.e.~$H_n(x/d) \simeq (2 x / d)^n$) since $\hbar$ appears at the denominator here. The factor $d^{-n}$ will then cancel with $d^n$ entering Eq.~\eqref{eq:cn_QM}. 
Also, the factor $e^{-\alpha ^2 \mathit{d}^2/(4 \ell^2)}$ contains $\hbar$ in the exponent. This term, in the inflationary case, would read $e^{-\alpha^2 P_{\zeta} / 4}$ and $\alpha^2 P_{\zeta}$ being the loop-counting parameter, which can be set to zero in the semiclassical limit.
Using this consideration and the simplified expression for $c_n$ in Eq.~\eqref{eq:cn_approx} and working at order $\tilde b$, we write the correction to the probability distribution as $|\Psi|^2 \simeq |\Psi_0|^2 \left( 1 + 2 {\rm Re} \, \delta \Psi / \Psi_0\right)$, where
\begin{align}
	2 {\rm Re} \,  \frac{\delta \Psi(x)}{\Psi_0(x)} 
	&=
	2 {\rm Re} \sum_{n = 1}^{\infty} c_n \frac{\psi_n(x)}{\psi_0(x)}
	\nn \\
	&\simeq 
	2 \sum_{n = 1}^{\infty} {\rm Re} \, c_n
	\frac{(2 x / d)^{n}}{\sqrt{2^n n!}}
	\nn \\
	&\simeq 
	- \sqrt{2 \pi}\alpha ^{3/2}
	\tilde b
	\,
	\frac{\Cc E_0}{\hbar \omega}
	\sum_{n=1}^\infty
	\frac{\left(\alpha x /\ell\right)^n}{n^2 n! }
	\cos \left( \varphi - \alpha \log n + n \pi / 2 \right)
	\;.
	\label{eq:Psi_QM}
\end{align} 
The left-hand side of Eq.~\eqref{eq:Psi_QM} is the exact analogue of the correction to the wavefunction we studied in the inflationary case. 
It is useful to compare Eq.~\eqref{eq:Psi_QM} with the probability in Eq.~\eqref{eq:Pn_QM}: even though the probability to jump to the state $n$ is low, when we look at the tails with $|x| / d $ large, high $n$'s start to dominate (the parameter raised to the power $n$ is $\alpha x / \ell$ as opposed to $\alpha d / \ell$).
Taking into account the $1/n!$, one expects values of $n$ around $n_{\rm max} \sim \alpha x / \ell$ to dominate the sum in Eq.~\eqref{eq:Psi_QM}. Their contribution then scales as $\sim \exp(\alpha x / \ell)$, making the correction to the wavefunction potentially exponentially large, in analogy with the inflationary results. This explanation misses possible cancellations among the series coefficients, which can take either sign. Equation \eqref{eq:Psi_QM} can be evaluated numerically, as shown in Fig.~\ref{fig:qm_plot}: one can notice that, like in the case of inflation, this correction is highly asymmetric~\footnote{
The asymmetry of the wavefunction can be analysed by applying the Euler--Maclaurin formula to the series \eqref{eq:Psi_QM}. Such a formula allows one to represent the series as an integral over $n$ plus corrections that depend on the derivatives of the integrand evaluated at the end points of the integration. We check that those corrections are negligible in our case and therefore the series \eqref{eq:Psi_QM} can be represented by the integral over $n$. 
In the large $|x|$ and $\alpha$ limit, the integral can be done using the saddle-point approximation over $n$. We find that for $x > 0$ the dominant saddle point is approximately $n_s \sim i \alpha x/\ell$, which gives rise to an exponential factor $e^{\pi \alpha/2}$ in the wavefunction. It is important to note that such an exponential factor is due to the term $\alpha \log n$ in the cosine of Eq.~\eqref{eq:Psi_QM}.
On the other hand, for $x < 0$ we observe that the saddle point is close to zero and is not purely imaginary, so that the exponential factor on this side is smaller. 
Therefore, the wavefunction is highly asymmetric between positive and negative $x$, as shown in Fig.~\ref{fig:qm_plot}.}, featuring oscillations at frequency $\alpha$ and very large for positive $x$.
In this quantum mechanical example, the origin of the large asymmetry is more evident. Indeed, the dominant term in the perturbation $\Delta {\rm H}(t)$, selected by the resonance ($\sigma = 1$), corresponds with a boost with a positive momentum. Therefore, one expects larger effects on the wavefunction for positive $x$. Moreover, this operator is responsible for the phase $n^{-i\alpha}$ in the coefficients $c_n$ and in the series \eqref{eq:Psi_QM} (see Eq.~\eqref{eq:cn_QM}). Such a phase then leads to different behaviours of the series for different signs of $x$.

Alternatively, instead of starting from the general time-dependent perturbation theory, one can focus on the semiclassical limit. This can be obtained using the path integral representation of the wavefunction.
Alternatively, we can re-discover the semiclassical approximation in the following way. In $\delta \Psi(x)$ we keep the coefficients $c_n$ of Eq.~\eqref{eq:cn_integr_dt}, without performing the time integration. We then sum over the states $n$ and use the $\hbar \to 0$ limit of the Hermite polynomials. Inside this sum, one can notice that the phase $e^{-i(E_0 - E_n)t / \hbar} = e^{i \omega n t}$ combines with the coordinate $x$ in such a way that the final result, before the time integration, is a function of $x_{\rm cl}(t) = e^{i \omega t} x$: this is indeed the semiclassical solution for the harmonic oscillator at zeroth order in $\tilde b$. 
As in the inflationary case, at first order in $\tilde{b}$, it is enough to use the free solution to compute the wavefunction in the semiclassical limit. 
After performing this procedure, one finds the following expression for the correction to the wavefunction
\begin{align}
	\frac{\delta \Psi(x)}{\Psi_0(x)} 
	=
	-\frac{i \tilde b \Cc E_0}{\hbar}
	\int_{-\infty}^{0}\de t\, 
	W(t)
	\,
	\cos\left(\alpha \log(-t/t_0) + \alpha x_{\rm cl}(t) / \ell \right) 
	\;.
\end{align}
This is indeed the expression for $i \Delta S / \hbar$, with $\Delta S$ being the correction to the action, evaluated on the free classical solution $x_{\rm cl}(t)$. The resulting time integral can be solved in saddle-point for large $\alpha$ and large $x$. Similarly to the cases studied in inflation, this will be an oscillatory function in $\alpha x$. (In Fig.~\ref{fig:qm_plot} there is no appreciable difference between the exact correction, first line of Eq.~\eqref{eq:Psi_QM}, and the semiclassical approximation.)

\section{Conclusions and future directions}\label{sec:conclusions}

In this work, we studied the wavefunction of the universe in a simple single-field model with a resonant feature. 
We discovered a striking behaviour which is completely unexpected from perturbation theory: 
a feature with a tiny amplitude, which results in tiny deviations from a Gaussian wavefunction for typical fluctuations, 
have exponentially large effects for rare events in the tail of the distribution. Moreover the effect is large only for rare peaks of $\zeta$, while one has a small effects on troughs.

The skeptic may wonder whether our conclusion is an artifact of working at first order in the amplitude of the feature, but we stress that this is not the case. Going beyond first order is necessary to have a quantitative understanding of what is going on in the tail, but this does not change the fact that one has sizeable deviations from the Gaussian statistics, even for features with very small amplitudes. In this respect, we stress that, at least conceptually, there is no obstacle in applying our semi-classical method in the large $\zeta$ limit of the WFU beyond first order in the amplitude, by (numerically) solving the full non-linear equation of motion \eqref{eq:full-eom}, and evaluating the corresponding on-shell action. We leave this to future work: with the full WFU at all orders in $\tilde b$ one could investigate the impact of the new tail on eternal inflation and on the rate of formation of primordial black holes.

Besides, given the exponential sensitivity of the tail to tiny features, it is reasonable to also expect a large sensitivity of the WFU to details of the expansion history $H(t)$, which would be worthwhile to investigate. We have also pointed out the existence of a particularly striking regime where a non-perturbative description of the WFU is required even for typical fluctuations.
What are the observational consequences of such a regime, and potential links with recent developments in particle physics beyond the Standard Model, see e.g.~\cite{Falkowski:2019tft,Chang:2019vez,Hook:2023pba}, are interesting questions to which we plan to come back.

Our work can be developed in many directions, which are interesting both observationally and on purely theoretical grounds.
As we have explained, our formalism can be readily applied to study other types of small features, like localized ones. Moreover, for any type of feature, the unitary cutoff is pushed to infinity as its amplitude goes to zero and the theory becomes free. This is suggestive that the regime requiring a non-perturbative description of typical fluctuations may not be limited to oscillating features. For generic features in the expansion history, the function $h(t)$ in \eqref{eq:DS_1_general} may exhibit singularities in the complex plane which preclude the rotation to Euclidean time: whether this leads to specific features for the WFU is worth exploring. 
The UV completion of the type of models studied in this work contains additional states: it would be interesting to understand when and how they modify the predictions for the tail of the wavefunction. Eventually, it would be valuable to relate our non-perturbative computations of the wavefunction of the universe, which fully characterizes the state of primordial fluctuations, to the ongoing developments about how to best extract information from cosmological data, see for example \cite{deSanti:2023rsw} and references therein.

\section*{Acknowledgements}
It is a pleasure to thank D.~Cruces, M.~Mirbabayi, E.~Pajer, G.~A.~Palma, M.~Peloso, C.~D.~Pueyo, L.~Senatore, M.~Serone, S.~Sypsas, V.~Vennin and G.~Villadoro for illuminating discussions. S.R.P. is supported by the European Research Council under the European Union's Horizon 2020 research and innovation programme (grant agreement No.
758792, Starting Grant project GEODESI). V.Y. is supported by the WPI Research Center Initiative, MEXT, Japan. The work of V.Y. was supported by JSPS KAKENHI Grant No. JP22K20367.

\appendix

\section{Derivative versus polynomial form of the action}\label{App:action_check}
In this appendix we explicitly show that our action (\ref{eq:action_pi_simple}) indeed agrees with the action (24) of \cite{Behbahani:2011it}, and we demonstrate the practical advantage of using our form, even for the computation of $n$-point correlation functions. In doing so, it is important to remember that temporal boundary terms in the action contribute in general to correlation functions \cite{Arroja:2011yj,Burrage:2011hd,Rigopoulos:2011eq,Garcia-Saenz:2019njm}. In particular, total (temporal) derivative terms depending on $\dot{\pi}$ (by contrast to the ones involving $\pi$ only) contribute to correlation functions of $\pi$, which are our object of interest. Hence, we will only keep such boundary terms in the following discussion.

\subsection{Action of $\pi$ in two different forms}

Our starting point here is the action (\ref{eq:action_pi_simple}):
\begin{align}
S = \int \de^4x\,a^3 \MP^2 \dot{H}(t + \pi) (\partial_\mu \pi)^2 \;. 
\end{align}
Expanding $\dot{H}(t + \pi)$ as a power series gives
\begin{align}
S = \MP^2 \sum_{n = 0}^\infty \int \de^4x\, \frac{a^3}{n!} H^{(n + 1)}(t)~\pi^n \partial_\mu \pi \partial^\mu \pi  \;,
\end{align} 
where $H^{(n)}(t)$ denotes the $n^{\rm th}$ derivative with respect to time acting on $H(t)$. Here we are interested in the terms with $n \geq 1$. We rewrite the term $\pi^n \partial_\mu \pi$ as $\partial_\mu(\pi^{n + 1})/(n + 1)$, so that the interacting part of the action above becomes 
\begin{align}
S_{\textrm{int}} &= \MP^2 \sum_{n \geq 1}^\infty\int \de^4x\, \frac{a^3}{(n + 1)!} H^{(n + 1)}(t)\,\partial_\mu \pi^{n + 1} \partial^\mu \pi  \;, \nonumber \\
&= -\MP^2 \sum_{n \geq 1}^\infty \int \de^4x\, \frac{1}{(n + 1)!}  \partial_\mu \bigg[a^3 H^{(n + 1)}(t)\,\partial^\mu \pi  \bigg] \pi^{n + 1} + S_{\textrm{boundary}} 
\;,
\label{eq:app_pi_2}
\end{align} 
where we have performed an integration by parts in the second line, and we kept the relevant boundary term
\begin{align}
	S_{\textrm{boundary}}
	=
	-\MP^2 \sum_{n \geq 1}^\infty \int \de^4x\, 
	\partial_t \left[
	\frac{a^3}{(n + 1)!} H^{(n + 1)}(t)\, \pi^{n + 1} \dot{\pi} 
	\right]
	\,.
\end{align} 
We define $\Box \pi \equiv \partial_\mu(a^3 \partial^\mu \pi)/a^3$. The action (\ref{eq:app_pi_2}) then becomes 
\begin{align}
S_{\textrm{int}}  = -\MP^2 \sum_{n \geq 1}^\infty \int \de^4x\, \frac{1}{(n + 1)!}  \bigg[- a^3 H^{(n + 2)}(t)\,\pi^{n + 1}\dot{\pi} + a^3 H^{(n + 1)}(t)\,\pi^{n + 1}\Box \pi   \bigg] +S_{\textrm{boundary}} \;.
\end{align}
Performing an integration by parts on the first term on the RHS of the action above, we obtain
\begin{align}
S_{\textrm{int}}  = -\MP^2 \sum_{n > 2}^\infty \int \de^4x\, a^3 \bigg\{ \frac{1}{n!}  \bigg[ 3 H(t) H^{(n)}(t) + H^{(n + 1)}(t) \bigg] \pi^n + \frac{1}{(n - 1)!} H^{(n - 1)}(t)\,\pi^{n - 1}\Box \pi  \bigg\}+S_{\textrm{boundary}} \;. \label{step2}
\end{align}
where we have changed $n \rightarrow n - 2$. Additionally, note that the equation of motion derived from the quadratic Lagrangian for $\pi$ reads
\begin{align}
\frac{\delta \mathcal{L}_2}{\delta \pi}=-2 \MP^2 a^3 \dot{H}(t) \Box \pi+2 \MP^2 a^3 \ddot{H}(t) \dot{\pi}\,. \label{linear-eom}
\end{align}
Hence, to work out the action at first order in $b$, one can neglect the last term in \eqref{linear-eom} when inserted back in \eqref{step2}, which gives
\begin{align}
S_{\textrm{int}} =  \sum_{n > 2}^\infty \int \de^4x\, \bigg\{ -\frac{a^3 \MP^2}{n!}  \bigg[ 3 H(t) H^{(n)}(t) + H^{(n + 1)}(t) \bigg] \pi^n + f_n(\pi) \frac{\delta \mathcal{L}_2}{\delta \pi}  \bigg\} + \mathcal{O}(b^2)+S_{\textrm{boundary}}  \;, \label{eq:app_self_pi}
\end{align}
where we have defined 
\begin{align}
f_n(\pi) \equiv \frac{1}{2(n - 1)!} \frac{H^{(n - 1)}(t)}{\dot{H}(t)}\,\pi^{n - 1} \;.
\end{align}
We see that the action (\ref{eq:app_self_pi}) is the same as the action (24) of \cite{Behbahani:2011it}~\footnote{One can straightforwardly show that the action \eqref{eq:app_self_pi} is also equivalent to the $n^{\rm th}$-order interacting action obtained in \cite{Leblond:2010yq}, as shown in \cite{Behbahani:2011it}.}, although the authors here did not write the boundary terms, despite their role in the computation of $n$-point functions, as we will see in (\ref{bispectrum}). Therefore, our non-perturbative form of the action (\ref{eq:action_pi_simple}) is equivalent to their action.

Although what ultimately matters, in perturbative computations, are correlation functions of $\pi$, it is interesting to study the structure of the equation of motion of $\pi$. In particular, the fact that the action (\ref{eq:app_self_pi}) involves self-interactions of $\pi$ without derivatives acting on it seems to suggest that a constant $\pi$ is not a solution to the equation of motion, implying that $\pi$ is not conserved on super-horizon scales. However, this is not the case due to the contributions coming from the term in $\delta \mathcal{L}_2/\delta \pi$ to the equation of motion of $\pi$: one can verify that at each order in $\pi$ the equation of motion derived from (\ref{eq:app_self_pi}) admits a constant solution. In contrast, this property is manifestly valid, and non-perturbatively, from our form (\ref{eq:action_pi_simple}).

\subsection{Time-independence of the bispectrum}
\label{bispectrum}

It is instructive to see explicitly the late-time constancy of the 3-point function of $\pi$ computed from the form \eqref{eq:app_self_pi}, and to contrast it with the computation starting from our form (\ref{eq:action_pi_simple}) of the action. We will see that the constancy of the bispectrum is immediate in the latter case, whereas in the former, it requires taking into subleading terms in the expansion in large $\alpha$ as well as cancellations between the contributions from the bulk and the boundary terms. This provides a non-trivial consistency check of our computations, and showcases the usefulness of our form of the action.

\paragraph{{\bf Polynomial form of the action.}} There are three types of terms in the Lagrangian \eqref{eq:app_self_pi}: the bulk terms that are polynomial in $\pi$, the terms proportional to the linear equation of motion, and boundary terms. When computing correlation functions in perturbation theory, terms proportional to the linear equation of motion never contribute, as the interaction-picture fields are by definition free fields obeying that equation. We treat the two other types of terms in turn.

The contribution to the 3-point function of $\pi$ at time $\eta$ from the bulk terms in \eqref{eq:app_self_pi}, reads
\begin{align}
	\langle \pi(\boldsymbol{k}_1,\eta)\pi(\boldsymbol{k}_2,\eta) \pi(\boldsymbol{k}_3,\eta)) \rangle'_{\textrm{bulk}}
	&=
	\frac{1}{(4 \epsilon_\star  \MP^2)^2 2 \epsilon_\star H^4 \prod_i k_i^3 } 
	\times \nonumber \\
	& \hspace{0.5cm}
	\times \textrm{Im} \left[ 
	\int_{-\infty(1-i \epsilon)}^\eta \frac{\de \eta'}{\eta'^{4}} \left( H^{(4)}+3H H^{(3)} \right)(\eta') f(\eta')f^*(\eta) 
	\right] \;,
	\label{bulk-con}
\end{align}
with $f(\eta)=e^{i k_{\rm t} \eta} \prod_i (1-i k_i \eta)$ and $k_{\rm t}=\sum_i k_i$. The prime means that we drop the factor $(2\pi)^3 \delta^{(3)}(\vect k_1 + \vect k_2 + \vect k_3)$. All the integrals here can be done analytically, but in terms of incomplete Gamma functions which are not very illuminating for our purposes. Instead, remember that our goal is simply to show that the bispectrum of $\pi$ goes to a constant at late times. Hence, we will only keep track of the relevant contributions at the asymptotic future, which are simply oscillations in $\eta^{\pm i \alpha}$, and we check that they cancel in the final result. 
For that purpose, it is sufficient to expand $f(\eta)$ at late times:
\begin{align}
	f(\eta)
	=
	1
	+\frac{\eta^2}{2} \sum_i k_i^2
	+\frac{i\eta^3 }{3} \sum_i k_i^3
	+{\cal O}(\eta^4) 
	\;,
\end{align}
where we have to go up to third order as one needs
\begin{align}\label{bulk-contribution_im}
	\textrm{Im} 
	\left[ 
	f(\eta')f^*(\eta)
	\right]
	=
	\frac{1}{3}
	(\eta'^{3}- \eta^3) 
	\sum_i k_i^3
	+ \textrm{subleading} \;.
\end{align}
Here, we use the $i \epsilon$ prescription that is relevant only in the asymptotic past, so that, for our purposes, taking the imaginary part in \eqref{bulk-contribution_im} effectively applies only to $f(\eta')f^*(\eta)$. As for the terms denoted as subleading, they do not lead to contributions that survive at late times. At first order in $b$, one can consider $H$ constant in the term $H H^{(3)}$, and with 
\begin{align}
	\dot{H}(\eta)
	=
	-\epsilon_\star H_\star^2 
	\left[
	1-\frac{6 b}{\alpha} 
	\cos\left(\alpha \log\left(\eta/\eta_\star\right) - \tilde{\delta} 
	\right) 
	\right]
	\,,
\end{align}
one obtains that the terms that survive at late times are
\begin{align}
	\langle \pi(\boldsymbol{k}_1,\eta)&\pi(\boldsymbol{k}_2,\eta) \pi(\boldsymbol{k}_3,\eta)) \rangle'_{\textrm{bulk}}
	\supset 
	- \frac{H_\star b\, \alpha^2}{(4 \epsilon_\star  \MP^2)^2 } \frac{\sum_i k_i^3}{\prod_i k_i^3} \times \nonumber \\
	&  
	\times \int^\eta_{-\infty(1-i \epsilon)}  \frac{\de \eta'}{\eta'^{4}} \left(\eta'^{3}- \eta^3 \right) \left[ \sin \left( \alpha \log\left(\eta'/\eta_\star\right) - \tilde{\delta} \right)+\frac{3}{\alpha} \cos \left( \alpha \log\left(\eta'/\eta_\star\right)  - \tilde{\delta} \right)  \right] \nonumber \\
	&= 
	-\frac{3 H_\star b}{(4 \epsilon_\star  \MP^2)^2 } \frac{\sum_i k_i^3}{\prod_i k_i^3} \times  \sin \left( \alpha \log\left(\eta/\eta_\star\right)  - \tilde{\delta}
	\right)
	\,.
\label{contribution-bulk}
\end{align}
Note that cancellations between the various contributions---the sine and the cosine terms, as well as the ones with and without $\eta$-dependence in the integrand---lead to the amplitude of the result being independent of $\alpha$. In particular, it was important to take into account the cosine term, despite the fact that its contribution to the integrand is subdominant in the limit of large $\alpha$.

Let us now consider the boundary term 
\begin{equation} 
	S^{(3)}_{\textrm{boundary}}
	=
	-\frac{\MP^2}{2} 
	\int \de t \de ^3  x
	\, 
	\partial_t 
	\left( a^3 \ddot{H}(t) \pi^2 \dot{\pi} \right)
	\,.
\end{equation}
The contribution from any such term can be worked out from first principles using the commutation relation (see e.g.~Sec.~3.3 of \cite{Garcia-Saenz:2019njm}). Using that the (linear) conjugate momentum to $\pi$ is $\partial {\cal L}_2/ \partial \dot{\pi}=-2 a^3 \MP^2 \dot{H} \dot{\pi}$, this gives here:
\begin{align}
	\langle \pi(\vect{k}_1,\eta)&\pi(\vect{k}_2,\eta) \pi(\vect{k}_3,\eta)) \rangle'_{\textrm{boundary}}
	=
	-\left(\frac{\ddot{H}}{2 \dot{H}}\right)(\eta) \left( P_\pi(k_1,\eta) P_\pi(k_2,\eta)
	+ \textrm{ 2 perms.} 
	\right) \;,
\end{align}
where, at first order in $b$, one can use the standard power spectrum for $\pi$, $P_\pi(k,\eta)=(1+k^2 \eta^2)/(4 \epsilon_\star \MP^2 k^3)$. Keeping the terms that survive at late times, this gives
\begin{align}
	\langle 
	\pi(\vect{k}_1,\eta) & \pi(\vect{k}_2,\eta) \pi(\vect{k}_3,\eta)) 
	\rangle'_{\textrm{boundary}}
	=
	\frac{3 H_\star b}{(4 \epsilon_\star  \MP^2)^2 } \frac{\sum_i k_i^3}{\prod_i k_i^3} \times  
	\sin \left(\alpha \log\left(\eta/\eta_\star\right)  - \tilde{\delta} \right)
	\,,
\end{align}
which precisely cancel with \eqref{contribution-bulk}. Therefore, we have explicitly shown that the $3$-point function of $\pi$ goes to a constant at late times, using the polynomial form of the action accompanied with the necessary boundary terms.

\paragraph{{\bf Derivative form of the action:}} When using the derivative form of the action (\ref{eq:action_pi_simple}), the cubic part reads
\begin{equation}
	S^{(3)}
	=
	\int \de^4 x\, a^3 \MP^2 
	\ddot{H} \pi (-\dot{\pi}^2+(\partial_i \pi)^2/a^2)
	\,,
\end{equation}
which gives for the time-dependent $3$-point function
\begin{align}
	\langle \pi(\vect{k}_1,\eta)&\pi(\vect{k}_2,\eta) \pi(\vect{k}_3,\eta)) \rangle'
	=
	\frac{1}{(4 \epsilon_\star  \MP^2)^2  \epsilon_\star H^2 \prod_i k_i^3 }  
	\times \nonumber \\
	& 
	\times \textrm{Im} \left[ \int_{-\infty(1-i \epsilon)}^\eta \frac{\de \eta'}{\eta'^{2}} \ddot{H}(\eta') \left(\vect{k}_2 \cdot \vect{k}_3  f^*(\eta) f(\eta')+k_2 k_3 f^*(\eta) g_1(\eta') \right) \right]
	+\textrm{2 perms.} \;,
\label{bulk-contribution}
\end{align}
with $f(\eta)$ as above, and $g_1(\eta)=e^{i k_{\rm t} \eta}(1-i k_1 \eta) k_2 k_3 \eta^2$. Contrary to the polynomial form, here all the integrals are manifestly convergent, and the late-time constancy of the $3$-point function is immediate.

Let us also compute the bispectrum starting from \eqref{bulk-contribution}. Taking $\eta \to 0$ and writing the sine function as a sum of two exponentials, this gives, with $\beta= - \tilde{\delta} -\alpha \log \eta_\star$:
\begin{align}
	B(k_1,k_2,k_3)
	=
	\ &\frac{ 3 H_\star b}{(4 \epsilon_\star  \MP^2)^2 \prod_i k_i^3 }  
	\times 
	\nonumber \\
	&\times \textrm{Im} 
	\left[
	\sum_{n=0}^3 
	\sum_{\sigma=\pm1} 
	\Gamma(i \sigma \alpha+n-1) 
	\frac{a_n}{k_{\rm t}^{i \sigma \alpha+n-1}} \
	e^{ i \sigma \beta } 
	e^{-\sigma \alpha \pi/2} 
	\right] \;,
\label{bispectrum}
\end{align}
with 
\begin{equation}
\begin{aligned}
	a_0 &=-\frac12 \sum_i k_i^2 \;, \qquad \hspace{2.32cm} a_1 =-\frac{1}{2} k_{\rm t} (\sum_i k_i^2)\,,  
	\\
	a_2 &=\frac{k_{\rm t}}{2} \bigg(-\sum_{i \neq j} k_i^2 k_j +k_1 k_2 k_3 \bigg) \;, \quad 
	a_3 =- k_1 k_2 k_3\bigg(\sum_i k_i^2/2+ \sum_{i< j} k_i k_j\bigg) \,.
\end{aligned}
\end{equation}
Simplifying, one finds
\begin{equation}
	B(k_1,k_2,k_3)
	=
	\frac{ 3 H_\star b \,  \alpha^3\,  \cosh(\alpha \pi/2) }{(4 \epsilon_\star  \MP^2)^2 \prod_i k_i^3 }
	\, 
	\textrm{Im} \, z 
	\;,
\end{equation}
with 
\begin{equation}
	z
	=
	\left[ 
	\prod_i k_i-\frac{i}{\alpha}
	\left( \prod_i k_i-\sum_{i \neq j} k_i^2 k_j \right)
	-\frac{\sum_i k_i^3}{\alpha^2}  
	\right] 
	\Gamma(-1-i \alpha) e^{i(\alpha \log k_{\rm t}-\beta+\pi/2)} 
	\;.
\end{equation}
In the limit of large $\alpha$, with $\cosh(\pi \alpha/2) \to e^{\pi \alpha/2}/2$ and $\Gamma(-1-i \alpha) \to \sqrt{2 \pi}\,e^{-\pi\alpha/2}\alpha^{-3/2} e^{i(\alpha-\alpha \log \alpha+3 \pi/4)}[1+13 i/(12\alpha)]$ taking into account next-to-leading order (NLO) terms, this reduces to 
\begin{equation}
	B(k_1,k_2,k_3)
	=
	\frac{ 3 H_\star b \,  \alpha^{3/2} \sqrt{2 \pi} }{2(4 \epsilon_\star  \MP^2)^2 \prod_i k_i^2 } 
	\left[ 
	\sin(\alpha \log k_{\rm t}+\varphi)
	+\frac{1}{\alpha} \cos( \alpha \log k_{\rm t}+\varphi)
	\left(\sum_{i \neq j} \frac{k_i}{k_j}+\frac{1}{12}\right) 
	\right] 
	\;,
\end{equation}
with $\varphi=\alpha-\alpha \log \alpha+5 \pi/4-\beta$, and where we have kept the first subleading term, see \cite{DuasoPueyo:2023viy} for similar result.

\section{Beyond the decoupling limit: mixing with gravity}\label{App:mixing}

In the main text, we have derived the full nonlinear action \eqref{eq:action_pi_simple}, at all orders in $\pi$ and zeroth-order in the decoupling limit. This action scales like ${\cal O}(\epsilon)$. Here, we derive first-order corrections to this action, i.e. up to ${\cal O}(\epsilon^2)$, also keeping the full nonlinear structure in $\pi$. From this, we deduce an upper bound on $\bar{\zeta}$ for the mixing with gravity to be negligible.

We perform the space-time dependent time diffeomorphism $t \rightarrow t + \pi(t,\vect{x})$, starting from the unitary gauge action \eqref{eq:EFT_action}. We use the ADM parametrization of the metric:
\begin{equation}
\de s^2=-N^2 \de t^2+\hat g_{ij}(\de x^i+N^i \de t)(\de x^j+N^j \de t)\,,
\ee
we neglect tensor modes, and we choose the spatially flat gauge $\hat g_{ij}=a^2\delta_{ij}$. One thus obtains the action
\begin{equation}
\begin{aligned}
\label{eq:action-contraint}
S=\int \de^4x\;a^3\MP^2 & \left\{\frac{1}{2N}(E_{ij}^2-E^2) -N(3 H^2(t+\pi)+\dot{H}(t+\pi))\right.\\ \nonumber
&\qquad+\left. \dot{H}(t+\pi)\left[-N^{-1}(1+\dot\pi)^2+2N^{-1}(1+\dot\pi)N^i\partial_{i}\pi+N\left(\partial_{i}\pi\right)^2-N^{-1}(N^i\partial_{i}\pi)^2\right]  \right\}\,,
\end{aligned}
\end{equation}
where
\begin{equation}
	E_{ij}
	=
	\frac{1}{2}\dot{\hat g}_{ij} 
	- N_{(i;j)}
	=
	a^2 H\delta_{ij}
	- N_{(i;j)}\ ,
	\quad 
	N_{(i;j)}
	=
	{1\over2}(N_{i;j}+N_{j;i})\,,
\end{equation}
and $;$ stands for the covariant derivative with respect to the spatial metric $\hat g_{ij}$, which in this gauge is simply an ordinary derivative.  

Performing similar manipulations as in the main text, this reads
\begin{equation}
\begin{aligned}
	S 
	= 
	\int \de^4x\, a(t)^3 \MP^2 
	&
	\left\{ 
	-\dot{H}(t + \pi) (\dot{\pi}^2-(\partial_i \pi)^2) - 3 (H(t+\pi)-H(t))^2   
	\right.\\ \nonumber
	\qquad & 
	+ \left.  
	\delta N \left( 3 N^{-1} H^2(t)-3H^2(t+\pi)-2 N^{-1} H  N_i^{\,;i}\right)
	+\frac12 N^{-1} (N_{(i;j)}^2-(N_i^{\,;i})^2 )  
	\right.\\ \nonumber
	\qquad & 
	+ \left. 
	\dot{H}(t+\pi) \left( \delta N( N^{-1}(1+\dot \pi)^2-1 +(\partial_i \pi)^2)+2(1+\dot \pi) N^{-1} N^i \partial_i \pi -N^{-1}(N^i \partial_i \pi)^2   
	\right) \right\} \;,
\label{full-action}
\end{aligned}
\end{equation}
where contractions are made with the spatial metric $\hat g_{ij}=a^2\delta_{ij}$, we wrote $N=1+\delta N$, and no approximation has been made so far. Varying the action with respect to $\delta N$ and $N_i$, we obtain, respectively, the lapse constraint:
\begin{equation}
\begin{aligned}
	&&3(H^2(t)-H^2(t+\pi))-3 H^2(t+\pi) \delta N(2+\delta N)-2 H(t) \partial^i N_i -\frac12 (N_{(i;j)}^2-(\partial^i N_i)^2 )\quad \\
	&&+\dot{H}(t+\pi)[ 2 \dot \pi +\dot \pi^2 + (N^i\partial_{i}\pi)^2 -2(1+\dot\pi)N^i\partial_{i}\pi+\left(\partial_{i}\pi\right)^2 (1+\delta N)^2-\delta N(2+\delta N) ]=0\,,
\label{lapse-constraint}
\end{aligned}
\end{equation}
and shift constraint:
\begin{equation}
\begin{aligned}
	2 N^{-1}\dot{H}(t+\pi)[1+\dot\pi -N^j\partial_{j}\pi] \partial_{i}\pi=\partial_j  \biggl[ \frac12 N^{-1}(N_{i}^{\,;j}+N^j_{\,;i})-\delta_i^j ( 2 H(t) \delta N/N+N_k^{\,\,;k}) \bigg] \,.
\label{shift-constraint}
\end{aligned}
\end{equation}
In order to derive the first correction to the decoupling limit action, we see that it is enough to work out $\delta N$ and $N_i$ at first order in $\epsilon$, which we denote with the superscript $^{(1)}$.  From Eqs.~\eqref{lapse-constraint}--\eqref{shift-constraint}, and using the Helmoltz decomposition of the shift, $N_i=\partial_i \psi + \tilde{N}_i$, with $\partial^i \tilde{N}_i=0$, one obtains the compact expressions
\begin{equation}
\delta N^{(1)}=\frac{1}{H(t)} \partial^{-2} \partial_i X_i\, \quad \textrm{and} \quad \tilde{N}^{(1)}_i= 4 a^2 \partial^{-2} \left(\partial_i \partial^{-2} \partial_j X_j -X_i\right)\,, \quad \textrm{with} \quad X_i \equiv-\dot{H}(t+\pi) (1+\dot \pi)\partial_i \pi \,, 
\end{equation}
and 
\begin{equation}
2 H(t) \frac{\partial^2 \psi^{(1)}}{a^2}
 =3 \left(H^2(t)-H^2(t+\pi) \right)-6 H^2(t + \pi) \delta N^{(1)}+\dot{H}(t+\pi)[ 2 \dot{\pi}+\dot{\pi}^2+(\partial_i \pi)^2]\,,
\end{equation}
where indices are simply contracted with $\delta_{ij}$ here. One then obtains the final expression of the action including first-order $\epsilon$ corrections:
\begin{equation}
\begin{aligned}
S = \int \de^4x\, a(t)^3 \MP^2 & \left\{ 
	 -\dot{H}(t + \pi) (\dot{\pi}^2-(\partial_i \pi)^2) - 3 (H(t+\pi)-H(t))^2 - 3 (H^2(t+\pi)-H^2(t)) \delta N^{(1)}  \right.\\ 
\qquad & + \left.   3 H^2(t) (\delta N^{(1)})^2 +\dot{H}(t+\pi)(2 \dot{\pi}+\dot{\pi}^2+(\partial_i \pi)^2) \delta N^{(1)} +\frac12 \tilde{N}^{(1)}_{(i;j)} \tilde{N}^{(1)(i;j)} -2 X_i \tilde{N}^{(1) i}  \right\}\,,
\label{result-mixing}
\end{aligned}
\end{equation}
where we note that the explicit expression of $\psi^{(1)}$ is actually not needed for this result because of structural cancellations in the computation.\\

Let us now use the action \eqref{result-mixing} to identify the regime of validity of the decoupling limit analysis. For this, in the same spirit as in Sec.~\ref{sec:ODE_analysis}, we consider a boundary profile $\bar{\zeta}$ characterized by a typical momentum scale $k$. This way, the effects of spatial derivatives simply read $\partial_i \sim k_i$ and $\partial^{-2} \sim k^{-2}$. We also use the behaviour of the free mode function \eqref{eq:zeta_free_fourier} to deduce the estimates $\pi \sim \bzeta/H(1+k \eta)$ and $\dot{\pi} \sim \bzeta (k \eta)^2$. We thus find $X_i \sim \epsilon H \bzeta (1+\bzeta (k \eta)^2) (1+k \eta) k_i$, $\tilde{N}^{(1)}_i \sim a^2/k^2 X_i$ and in particular
\begin{equation}
\delta N^{(1)} \sim  \epsilon \bzeta\,[1+\bzeta (k \eta)^2] (1+k \eta)\,.
\label{N1}
\end{equation}
The validity of the decoupling limit necessitates $\delta N^{(1)} \ll 1$. The right-hand side of \eqref{N1} grows with $k \eta$, but keep in mind that the corresponding interactions are shut off deep inside the horizon, due to the $i \epsilon$ prescription projecting onto the interacting vacuum. However, we should demand that $\delta N^{(1)} \ll 1$ at the resonance, such that $|k \eta| \sim \alpha$. This imposes the bound
\begin{equation}
|\bar{\zeta}| \ll \frac{1}{\sqrt{\epsilon \alpha^3}}\,.
\label{eq:bound-zeta}
\end{equation}
This turns out to be the most stringent bound on $\bzeta$ when requiring the ${\cal O}(\epsilon^2)$ interactions to be negligible compared to the leading ${\cal O}(\epsilon)$ part of the action. Checking this is straightforward, except for the second term in \eqref{result-mixing}. The corresponding polynomial interactions $(H(t+\pi)-H(t))^2 \sim \epsilon^2 H^4 \pi^2$ do not decay outside the Hubble radius, contrary to the leading two-derivative part of the action. This is simply a manifestation of the fact that $\pi$ acquires a mass beyond the decoupling limit. Requiring that these interactions are subdominant at the late-time saddle point \eqref{eq:late_saddle_ode} gives the bound $\epsilon |\bzeta| \ll 1$, which is indeed less stringent than \eqref{eq:bound-zeta}.\\

Importantly, let us highlight that the regime $\alpha^2 |\bzeta| \sim 1$, in which full non-perturbative results are needed and we found qualitative deviations from perturbation theory, always lies within the regime of validity \eqref{eq:bound-zeta} of the decoupling limit analysis. Notice as well that in order to get the range of validity for $\bzeta$ of the decoupling limit it is crucial to use our non-perturbative form of the action; it is not enough to stick to the perturbative analysis discussed in \cite{Behbahani:2011it}.

\section{Correlators from the WFU at one loop}\label{App:bndry_loops}

In this section we are going to investigate how Witten diagrams are related to the correlators at loop level. 
In the WFU approach, equal-time correlators are obtained by performing a path integral over field configurations at that given time. At one-loop level, on top of Witten-diagram loops (needed to obtain the WFU), there are in principle additional loops originating from `averaging' tree-level Witten diagrams over boundary (late-times) field configurations.
In turn, this step can lead to additional divergences that are not manifest in the WFU. When estimating the size of loops, we therefore need to take into account both contributions. 

On the other hand, correlators can also be evaluated using different methods that do not involve the WFU, as for instance the `in-in formalism' \cite{Maldacena:2002vr}. In this case, one does not make a distinction between WFU and boundary loops. Notice also that the propagators running inside loops used in the two methods are different: the in-in uses Wightman functions, while the WFU uses bulk-to-bulk propagators.
As it is perhaps expected, on the WFU side we will obtain that the two sets of loop diagrams combine to yield back the in-in result: the effect of boundary loops is to change the boundary conditions of the propagator.~\footnote{For a similar discussion on the relation between the in-in and WFU approaches to loop diagrams, see \cite{Lee:2023jby}.}

For concreteness, we consider $\lambda \phi^4$ in dS, although our conclusion appears general. The action is taken to be
\begin{equation}
	S
	=
	\int \de^4 x \,
	a^4(\eta)
	\left[
	- \frac{1}{2} (\partial_\mu \phi)^2
	- \frac{\lambda}{4!}\phi^4
	\right] 
	\;.
\end{equation}
We furthermore focus on the equal time two-point function $\braket{\phi(\eta, \vect x) \phi(\eta, \vect y)}$ at linear order in $\lambda$.
At this order, the WFU at time $\etaf$, $\Psi[\bar \phi; \etaf]$, contains a tree-level correction to the four-point coefficient $\psi_4$ and a one-loop correction to the two-point coefficient $\psi_2$. Both are needed to obtain the one-loop correlator at $\mathcal{O}(\lambda)$.

In the following calculations we will need the bulk-to-bulk and the bulk-to-boundary propagators for massless fields in dS.
First, let us define the wave-modes in $k$-space that solve the free equation of motion for $\phi$
\begin{equation}
\label{eq:wavemodes_dS}
	\varphi_\pm(k, \eta)
	\equiv
	(1 \mp i k \eta) e^{\pm i k \eta}
	\;.
\end{equation}
Then, the bulk-to-bulk propagator is obtained as
\begin{align}
\label{eq:bulk_to_bulk}
	G(\eta, \eta'; \vect k)
	&= 
	-\frac{i H^2}{2 k^3}
	\left[
		\theta(\eta-\eta') \varphi_+(k, \eta') \varphi_-(k, \eta)
		+
		\theta(\eta'-\eta) \varphi_+(k, \eta) \varphi_-(k, \eta')
	\right.
	\nonumber
	\\
	&
	\;
	\qquad
	\qquad
	\left.
		-
		\frac{\bar \varphi_-}{\bar \varphi_+} \varphi_+(k, \eta) \varphi_+(k, \eta')
	\right]
	\;,
\end{align}
where $\bar \varphi_\pm(k) \equiv \varphi_\pm (k, \etaf)$ and $\theta$ is the Heaviside theta function. By construction, the function $G(\eta, \eta'; \vect k)$ vanishes at early times (when $\eta$, $\eta' \to -\infty(1-i \epsilon)$) and at late times (when $\eta$, $\eta' \to \etaf$).
Requiring that $G$ satisfies the free equations of motion with a $\delta(\eta - \eta')$ source then fixes the normalization in Eq.~\eqref{eq:bulk_to_bulk} (i.e.~we are working with canonically-normalized fields). Note that this normalization changes if we work with $\zeta$ (in this case the factor $H^2$ gets replaced by $P_\zeta$).
On the other hand, the bulk-to-boundary propagator is 
\begin{equation}
	K_{\vect k}(\eta)
	\equiv
	\frac{\varphi_+(k, \eta)}{\bar \varphi_+(k)}
	\;.
\end{equation}
This function is obtained as the free solution of the equations of motion satisfying $K_{\vect k}(\etaf) = 1$ and vanishing for $\eta \to -\infty(1-i \epsilon)$.

After introducing these quantities, we are ready to evaluate the Witten diagrams contributing to the WFU.
The tree-level contribution is obtained by evaluating the interaction part of the action times $i$ on the free modes for $\phi$ (given by Eq.~\eqref{eq:zeta_free_fourier}, but for $\phi$). Therefore, we obtain
\begin{align}
	\log \Psi_4[\bar \phi; \etaf]
	&=
	-\frac{i \lambda}{4!}
	\int_{-\infty}^{\etaf} \frac{\de \eta}{(H \eta)^4}
	\int 
	\prod_{i = 1}^{4}
	\left[
	\frac{\de^3 \vect k_i}{(2\pi)^3} \, \phi(\eta, \vect k_i)
	\right]
	(2 \pi)^3 \, \delta^{(3)}(\vect k_{\rm t})
	\nn \\
	& = 
	\frac{1}{4!}
	\int 
	\prod_{i = 1}^{4}
	\frac{\de^3 \vect k_i}{(2\pi)^3} \, 
	(2 \pi)^3 \, \delta^{(3)}(\vect k_{\rm t})
	\, 
	\psi_4(\vect k_1, \ldots, \vect k_4; \etaf) 
	\,
	\bar \phi(\vect k_1) \ldots \bar \phi(\vect k_i)
	\;,
\end{align}
where $\log \Psi_n[\bar \phi; \etaf]$ is the correction to the exponent of the WFU with $n$ fields and $\vect k_{\rm t} \equiv \vect k_1 + \vect k_2 + \vect k_3 + \vect k_4$. Thus, written in terms of bulk-to-boundary propagators $K_{\vect k_i}(\eta)$, the coefficient $\psi_4$ reads
\begin{equation}
	\label{eq:psi_4_loops}
	\psi_4(\vect k_1, \vect k_2, \vect k_3, \vect k_4; \etaf)  
	=
	-i \lambda
	\int_{-\infty}^{\etaf} 
	\frac{\de \eta}{(H \eta)^4} 
	\prod_{i = 1}^{4}
	K_{\vect k_i}(\eta)
	\;.
\end{equation}

The one-loop contribution to $\psi_2$ can instead be obtained using the standard Witten rules (see \cite{Anninos:2014lwa} for a derivation in $\rm AdS$) 
\begin{align}
	\log \Psi_2[\bar \phi; \etaf]
	&=
	\frac{\lambda}{4}
	\int_{-\infty}^{\etaf} \frac{\de \eta}{(H \eta)^4}
	\int 
	\frac{\de^3 \vect k}{(2\pi)^3}
	K_{\vect k}^2(\eta) 
	\,
	\bar \phi(\vect k) \bar \phi(-\vect k)
	\int\frac{\de^3 \vect p}{(2 \pi)^3} G(\eta, \eta; \vect p)
	\nn \\
	& \equiv 
	\frac{1}{2!}
	\int 
	\frac{\de^3 \vect k}{(2\pi)^3}
	\, 
	\delta \psi_2(\vect k; \etaf) 
	\,
	\bar \phi(\vect k) \bar \phi(-\vect k)
	\;,
\end{align}
where we used the total momentum-conserving delta function to remove one integral.
Therefore, the correction to $\psi_2$ is given by
\begin{equation}
\label{eq:psi_2_loops}
	\delta \psi_2(\vect k; \etaf)
	= 
	\frac{\lambda}{2}
	\int_{-\infty}^{\etaf} \frac{\de \eta}{(H \eta)^4}
	K_{\vect k}^2(\eta)
	\int\frac{\de^3 \vect p}{(2 \pi)^3} G(\eta, \eta; \vect p)
	\;.
\end{equation}

Explicitly, the WFU at order $\lambda$ can be expanded as follows
\begin{equation}
	\Psi[\bar \phi; \etaf] 
	= 
	\Cc N 
	\,
	\Bigl(
		1 + \log \Psi_2[\bar \phi; \etaf] + \log \Psi_4[\bar \phi; \etaf]
	\Bigr)
	\exp
	\left[
		\frac{1}{2!}
		\int \frac{\de^3 \vect k}{(2 \pi)^3}
		\psi_2(\vect k; \etaf)
		\,
		\bar \phi(\vect k) \bar \phi(-\vect k)
	\right] 
	\,, \label{eq:WFU_app}
\end{equation}
where $\Cc N^{-1} \equiv \int \Cc D \bar \phi \, |\Psi[\bar \phi; \etaf]|^2$ is a normalization constant.
The two-point correlator in momentum space is then obtained using the standard Born rule of quantum mechanics:
\begin{equation}\label{eq:2pt_path_int}
	\braket{\phi(\etaf, \vect p)\phi(\etaf, \vect p')}
	= 
	\int \Cc D \bar \phi \;
	\bar \phi(\vect p) \bar \phi(\vect p')
	\,
	\bigl|\Psi[\bar \phi; \etaf]\bigr|^2
	\;.
\end{equation}

\begin{figure}
\centering
  \includestandalone[width=0.8\textwidth]{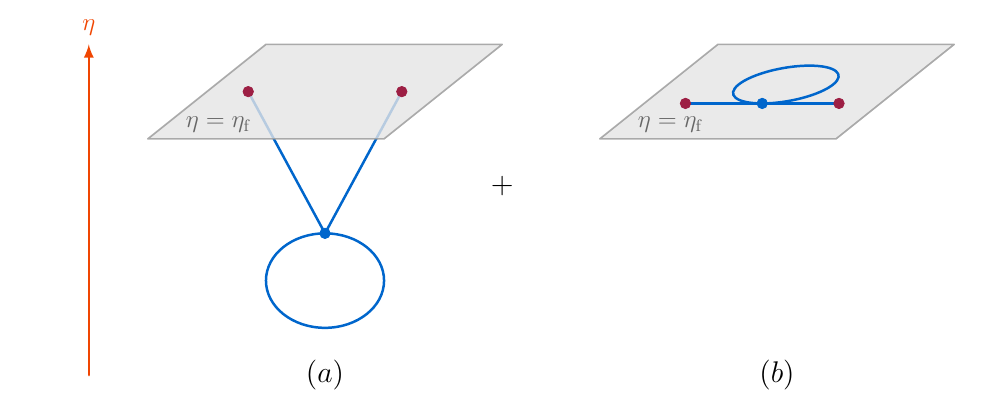}
  \caption{
  \quad 
  The two one-loop contributions to $\braket{\phi(\etaf, \vect p) \phi(\etaf, \vect p')}$. 
  The grey surface represents the boundary at $\eta = \etaf$.
  On the left, $(a)$, we have the one-loop diagram on the bulk, $\log \Psi_2$, that contributes directly to the WFU. On the right, $(b)$, we have the boundary diagram where we average over two boundary fields in $\log \Psi_4$.
  }
  \label{fig:bndry_loops} 
\end{figure}

In performing this path integral at order $\lambda$ we encounter various types of contributions. There are, for instance, bubble diagrams, that are however cancelled once the normalization $\Cc N$ is taken into account. In principle, there could be disconnected diagrams, but at this order they do not appear.
Finally, there are two types of connected diagrams. First, from $\delta \psi_2$, a diagram where the external $\bar \phi(\vect p)$, $\bar \phi(\vect p')$ connect to the internal $\bar \phi(\vect k) \bar \phi(-\vect k)$. This can be interpreted as a loop contribution from the bulk (given the origin of $\delta \psi_2$). 
Second, from $\psi_4$, we have a loop diagram where two internal $\bar \phi(\vect k_i)$'s are contracted among each other. Therefore, this corresponds to a boundary loop.
These two one-loop diagrams are represented respectively in Fig.~\ref{fig:bndry_loops} (a) and (b).

One can straightforwardly check that
\begin{equation}
	{\rm Re}\, \psi_2(\vect k; \etaf) 
	=
	-\frac{k^3}{H^2 \bar \varphi_+(k) \bar \varphi_-(k)}
	\;.
\end{equation}
At this point, we are ready to compute the order $\lambda$ correction in Eq.~\eqref{eq:2pt_path_int} (the integral can be performed, for instance, by first introducing a generating functional and by taking its functional derivatives). We obtain
\begin{align}
	&\braket{\phi(\etaf, \vect p)\phi(\etaf, \vect p')}'_{(1)}
	=
	\frac{1}{2 \left( {\rm Re \,} \psi_2(\vect p; \etaf)\right)^2}
	\left[
		{\rm Re}\, \delta \psi_2(\vect p; \etaf)
		-
		\frac{1}{4}
		\int \frac{\de^3 \vect k }{(2\pi)^3}
		\frac{{\rm Re \, }\psi_4(\vect p, \vect p, \vect k, \vect k; \etaf)}{{\rm Re \, \psi_2(\vect k; \etaf)}}
	\right]
	\nn \\
	&=
	\frac{\lambda}{4 \left( {\rm Re \,} \psi_2(\vect p; \etaf)\right)^2}
	\, {\rm Re}\,
	\int_{-\infty}^{\etaf} \frac{\de \eta}{(H \eta)^4}
	K_{\vect p}(\eta)^2
	\int \frac{\de^3 \vect k }{(2\pi)^3}
	\left[
		G(\eta, \eta; \vect k)
		+
		\frac{i K_{\vect k}^2(\eta)}{2 \, {\rm Re \,} \psi_2(\vect k; \etaf)}
	\right]
	\;. \label{eq:appB:2pt_loop}
\end{align}
In the correlator, the subscript $(1)$ stands for the first correction in $\lambda$, while the prime means we remove $(2\pi)^3 \delta^{(3)}(\vect p + \vect p')$. In going to the second line we used Eqs.~\eqref{eq:psi_4_loops} and \eqref{eq:psi_2_loops}. Notice that, although the two contributions of the first line come with different symmetry factors, when expressed in terms of bulk objects this difference cancels. Indeed, as it is clear from Fig.~\ref{fig:bndry_loops}, both diagrams have the same combinatorics from a bulk perspective.
In the second line on the RHS of (\ref{eq:appB:2pt_loop}), the two terms in the square bracket are given by
\begin{align}
	\label{eq:G_bb_loops}
	G(\eta, \eta; \vect k)
	&=
	-
	\frac{i H^2}{2 k^3}
	\left[
		\varphi_+(k, \eta) \varphi_-(k, \eta) - \frac{\bar \varphi_-(k)}{\bar \varphi_+(k)} \varphi_+(k, \eta) \varphi_+(k, \eta)
	\right]
	\;,
	\\
	\label{eq:KK_loops}
	\frac{i K_{\vect k}^2(\eta)}{2 \, {\rm Re \,} \psi_2(\vect k; \etaf)}
	&=
	-\frac{i H^2}{2 k^3} 
	\bar \varphi_+(k) \bar \varphi_-(k)
	\left(\frac{\varphi_+(k, \eta)}{\bar \varphi_+(k, \eta)}\right)^2
	\;.
\end{align}
By combining these two pieces, the term that imposes the boundary conditions at $\etaf$ in $G(\eta, \eta, \vect k)$ (the last in Eq.~\eqref{eq:G_bb_loops}) is cancelled by the boundary loop \eqref{eq:KK_loops}.
Therefore, we arrive at the overall one-loop contribution to the correlator
\begin{equation}
	\label{eq:phiphi_loop}
	\braket{\phi(\etaf, \vect p)\phi(\etaf, \vect p')}'_{(1)}
	= 
	\frac{\lambda H^2}{8 \left( {\rm Re \,} \psi_2(\vect p; \etaf)\right)^2}
	\, {\rm Im}\,
	\int_{-\infty}^{\etaf} \frac{\de \eta}{(H \eta)^4}
	K_{\vect p}(\eta)^2
	\int \frac{\de^3 \vect k }{(2\pi)^3}
	\frac{1}{k^3}
	\varphi_+(k, \eta) \varphi_-(k, \eta)
	\;, 
\end{equation}
where indeed we recognize the propagator of the in-in relations inside the loop.
Notice also that the WFU loop, $\delta \psi_2$ in Eq.~\eqref{eq:psi_2_loops}, is convergent in the IR, while the loop for the correlator \eqref{eq:phiphi_loop} is divergent \cite{Gorbenko:2019rza}.
Moreover, given that the loops are typically divergent, in order to have finite correlators after renormalization we arrive at the conclusion that the WFU is necessarily a divergent object on itself. The divergences from the additional path integral in Eq.~\eqref{eq:2pt_path_int} will then make the final result UV finite.
One can also check that Eq.~\eqref{eq:phiphi_loop} agrees with the in-in computation \cite{Senatore:2009cf}. The cancellation between bulk and boundary terms is completely independent on the spacetime, the theory and the evaluation time $\etaf$, as it is based on the properties of the in-in formalism. Therefore, the conclusion also applies to the model of resonant non-Gaussianities.  

\section{Estimate of one-loop Witten diagrams at $\mathcal{O}(\tilde{b})$}\label{app:1-loop_estimate}

In this appendix we want to show that the one-loop Witten diagrams with $n$ external legs at first order in $\tilde{b}$ (see Fig.~\ref{fig:Witten_diags}) scale as $\sqrt{\alpha}\, \tilde{b}\, \bzeta^2 (\alpha^2 \bzeta)^{n - 2} (\alpha^2 P_\zeta) / P_\zeta$. In comparison with the tree-level diagrams with the same number of external legs,  these loop corrections are suppressed by $\alpha^2 P_\zeta \lesssim 1$. Note that this condition is  weaker than the one of perturbative unitarity, $\alpha^4 P_\zeta \lesssim 1$, see the discussion around Eq.~\eqref{unitarity}.

For illustrative purposes, let us focus on the one-loop Witten diagram with two external legs (second graph in the second line of Fig.~\ref{fig:Witten_diags}). The ingredients for computing such a diagram are two bulk-to-boundary propagators, one bulk-to-bulk propagator and the four-point vertex which is given by the quartic interaction:
\begin{align}\label{eq:vertex_4}
	\mathcal{L}_4 
	= 
	\frac{\tilde{b} \alpha^2 \zeta^2}{4 \eta^2 P_\zeta} 
	\bigg[
	\zeta'^2 - (\partial_i \zeta)^2
	\bigg] 
	\cos
	\left(
	\alpha\log(\eta/\eta_\star) - \tilde{\delta}
	\right) 
	\;.
\end{align}
The coupling above can be straightforwardly obtained from expanding the action (\ref{eq:action_Zeta}) for small $\zeta$ up to quartic order.~\footnote{In general, the non-linear self-coupling of $\zeta$'s at first order in $\tilde{b}$ reads
\begin{align}\label{eq:vertex_n+2}
	\mathcal{L}_{n + 2} 
	= 
	-\frac{\tilde{b} \alpha^n \zeta^n}{2 \eta^2 P_\zeta n!} 
	\bigg[
	\zeta'^2 - (\partial_i \zeta)^2
	\bigg] 
	\left\{
	\begin{matrix}
		(-1)^{n/2}\cos\left(\alpha\log(\eta/\eta_\star) - \tilde{\delta}\right) \;, & n + 2 \in {\rm even} \\
		(-1)^{(n+1)/2}\sin\left(\alpha\log(\eta/\eta_\star) - \tilde{\delta}\right) \;, & n + 2 \in {\rm odd} 
	\end{matrix}
	\right. 
	\;.
\end{align}
}

Derivatives in the action can act on internal or external legs. Let us focus first on the case where the two derivatives act on the two bulk-to-boundary propagators: this is the leading contribution (highest power of $\alpha$) as we will verify later. Using the vertex \eqref{eq:vertex_4}, the one-loop coefficient of the WFU with two external legs is given by~\footnote{We define the wavefunction coefficients as in Eq.~(\ref{eq:WFU_app}).} 
\begin{align}\label{eq:M_1-loop_1}
	\psi^{\rm 1-loop}_{2}(\vect{k}_1, \vect{k}_2) 
	\supset
	\frac{\tilde{b} \alpha^2 }{4 P_\zeta} 
	\int^{\etaf}_{-\infty} 
	\frac{\de \eta}{\eta^2} 
	K'_{\vect{k}_1}(\eta) K'_{\vect{k}_2}(\eta) 
	\cos(\alpha\log(\eta/\eta_\star) - \tilde{\delta}) 
	\int \frac{\de^3 \vect k}{(2\pi)^3} 
	G(\eta,\eta; \vect{k}) 
	\;,
\end{align}
where the subscript denotes the number of external legs, $\vect{k}_1$ and $\vect{k}_2$ are the external momenta, and $\vect{k}$ is the internal momentum. 
Note that in the expression (\ref{eq:M_1-loop_1}) we considered two time derivatives acting on $K_{\vect{k}}(\eta)$. One gets the same scaling in $\alpha$ also for spatial derivatives: this is straightforward to verify using the saddle-point approximation ($\alpha \gg 1$) to the $\eta$-integral in the early-time limit.~\footnote{Subleading corrections in the limit $|k\eta| \gg 1$ contain less power of $\eta$: on the saddle point of the $\eta$ integral, their contributions have fewer power of $\alpha$ compared to the leading contributions.} Using the formula \eqref{eq:G_bb_loops} and the fact that $K'_{\vect{k}}(\eta) = \eta k^2 e^{ik\eta}/\bar{\varphi}_+(k)$, Eq.~\eqref{eq:M_1-loop_1} becomes
\begin{align}
	\psi^{\rm 1-loop}_{2}(\vect{k}_1, \vect{k}_2) 
	&\supset 
	-\frac{i\tilde{b} \alpha^2 H^2}{8}  
	\frac{k_1^2 k_2^2}{\bar{\varphi}_+(k_1) \bar{\varphi}_+(k_2)}  
	\int^{\etaf}_{-\infty} \de \eta
	\,
	e^{i k_{\rm t} \eta}  
	\cos\big(\alpha\log(\eta/\eta_\star) - \tilde{\delta}\big) 
	\times	
	\nonumber \\
	& \hspace{0.45cm}
	\times
	\int \frac{\de^3 \vect k}{(2\pi)^3} 
	\frac{1}{k^3}
	\left[
	\varphi_+(k, \eta) \varphi_-(k, \eta) 
	- \frac{\bar \varphi_-(k)}{\bar \varphi_+(k)} \varphi_+(k, \eta) \varphi_+(k, \eta)
	\right] 
	\;,
	\label{eq:M_1-loop_3}
\end{align}
where $k_{\rm t} \equiv k_1 + k_2$.
Setting $\eta_{\rm f} = 0$ (\footnote{Since one gets a non-zero result for $\eta_{\rm f} = 0$, it is safe to neglect subleading terms. As we will discuss, it is useful to keep $\eta_{\rm f} \neq 0$ to study the Minkowski limit of the result.}) we have
\begin{align}
	\psi^{\rm 1-loop}_{2}(\vect{k}_1, \vect{k}_2) 
	&\supset 
	-\frac{i\tilde{b} \alpha^2 H^2}{8}  
	k_1^2 k_2^2
	\int_{-\infty}^{0} \de\eta
	~ ~e^{i k_{\rm t} \eta}  
	\cos\big(\alpha\log(\eta/\eta_\star) - \tilde{\delta}\big) 
	\times	
	\nonumber \\
	& \hspace{0.45cm} 
	\times 
	\frac{4 \pi}{(2\pi)^3} 
	\int_0^{\Lambda a(\eta)} 
	\frac{\de k}{k}  
	\left[(1 + k^2 \eta^2) - (1 - i k \eta)^2 e^{2 i k \eta}
	\right]
\;,
\end{align}
where $\Lambda$ is a fixed physical cutoff. Notice that this physical cutoff  $\Lambda$ 
appears together with the scale factor at time $\eta$, cutting off the integral in comoving momentum $k$ (see a more detailed discussion below Eq.~(26) of \cite{Senatore:2009cf}). Then, we change the variable using $k_{\rm p} \equiv k/a(\eta) = -\eta H k $, where $k_{\rm p}$ denotes a physical momentum. Thus, the integral above becomes 
\begin{align}\label{eq:M_1-loop_2}
	\psi^{\rm 1-loop}_{2}(\vect{k}_1, \vect{k}_2) 
	&\supset 
	-\frac{i\tilde{b} \alpha^2 H^2 }{16 \pi^2}  
	k_1^2 k_2^2 
	\int_{-\infty}^{0} \de\eta
	~e^{i k_{\rm t} \eta}  
	\cos\big(\alpha\log(\eta/\eta_\star) - \tilde{\delta}\big) 
	\times	\nonumber \\
	& \hspace{0.45cm} 
	\times  
	\int_0^{\Lambda } 
	\frac{\de k_{\rm p}}{k_{\rm p}} 
	\left[
	\bigg(
	1 + \frac{k_{\rm p}^2}{H^2}\bigg) - \bigg(1 + \frac{ik_{\rm p}}{H} 
	\bigg)^2 
	e^{-2 i k_{\rm p}/H}
	\right]
	\;.
\end{align} 
We now see that the integral over $k_{\rm p}$ in the second line does not depend on $\eta$ and in fact it potentially leads to quadratic and logarithmic divergences. 
More precisely, we obtain
\begin{align}\label{eq:Loop_cutoff}
\log \Psi_2^{\rm 1-loop}  
&\supset
\sqrt{\alpha}~\tilde{b}\,\bzeta^2 \alpha^2 
	\frac{k_1^2 k_2^2 }{k_{\rm t}}  
	\,
	e^{-i [\alpha \log(X_\star) + \tilde{\delta}]} \bigg[\log\bigg(\frac{\Lambda}{H}\bigg) + \frac{\Lambda^2}{2 H^2} + \, {\rm finite \, terms} \bigg]\;,
\end{align}
where we have used $X_\star = -k_{\rm t} \eta_\star$, and we have evaluated the $\eta$ integral on the perturbative saddle point, i.e.~$\eta_s = -\alpha/k_{\rm t}$. Note that the expression above is an estimate, assuming that $\bar \zeta(\vect k) \simeq \bar \zeta$  and it is peaked at some momentum. 
From Eq.~\eqref{eq:Loop_cutoff}, we see that the quadratic divergence, as expected, can be removed by adding a local counter-term to the WFU, resulting in, for example, renormalization of the mass term. On the other hand, the logarithmic divergence of the form $\log(\Lambda/H)$, as argued in \cite{Senatore:2009cf}, cannot be removed by a local counter-term since for the modes inside the horizon such a divergence becomes non-local, $\log((a(t)\Lambda)/k)$. With this reasoning, this logarithmic divergence is physical and we can read off the dependence on $\alpha$ for $\psi^{\rm 1-loop}_{2}$. 

Alternatively, one could draw the same conclusion using the dimensional regularization (dim. reg.) to compute $\psi^{\rm 1-loop}_{2}$. In dim.~reg.~one usually performs the $k$ integral in $d = 3 + \varepsilon$ dimensions. We use $d$ to denote the number of spatial dimensions. Additionally, in $(d + 1)$-dimensions the free mode function becomes the Hankel function $(-H\eta)^{d/2} H^{(1)}_{d/2}(- k\eta)$. Since the integral involving the Hankel function is very complicated, one can then use a trick, proposed in \cite{Melville:2021lst}, to obtain a simple form of the mode function in $d = 3 + \varepsilon$ dimensions. One considers a massive scalar field in $(d+1)$ dimensional de Sitter space, and analytically continues both the number of spatial dimensions and the mass of the field in such a way that the index of the Hankel function remains $3/2$. By doing so, the mode function in $d = 3 + \varepsilon$ dimensions takes a simple form, $(-H\eta)^{\varepsilon/2} (1 - i k \eta) e^{i k \eta}$. We see that this mode function is different from the one in $d = 3$ dimensions by the overall normalization factor $(-H \eta)^{\varepsilon/2}$. Using such a simple form of the mode function it is then straightforward to perform the integral in $k$ analytically in the limit $\eta_{\rm f} = 0$.
At this point, one can see that there are several terms appearing after the integration: a finite term, a term going as $1/\varepsilon$ and a term that contains $\log(H/\mu)$ where $\mu$ is a renormalization scale. 
Note that the finite term and the term with $1/\varepsilon$ can be altogether removed in some renormalization scheme.
After that, similar to the calculation in \eqref{eq:Loop_cutoff}, one can apply the saddle-point approximation to perform the $\eta$ integral. 
Finally, we obtain the same $\alpha$ dependence of $\psi^{\rm 1-loop}_{2}$ together with the logarithmic divergence of the form $\log(H/\mu)$ which, as explained earlier, cannot be removed by adding local counter-terms.~\footnote{
	We can convince ourselves that $\log(H/\mu)$ originates from a non-analytic term in $k$ by performing the calculation in dim.~reg. in the early-times limit $|\etaf| \gg 1$, where all the modes are inside the horizon and the mode functions are approximated by the Minkowski ones (by time-translational invariance in this limit, we can set the final time $t_{\rm f}= 0$). The time and momentum integrals in $\psi_2^{1-{\rm loop}}$ in this limit take the following form 
	\begin{equation}
	\sum_{\sigma} 
	\int_{-\infty}^{0} \de t 
	\,
	e^{-i t(k_{\rm t p} -\sigma \omega)} 
	\int \frac{\de^d \vect k_{\rm p}}{(2 \pi)^d} \frac{\mu^\delta }{k_{\rm p}}
	\left[
	1 - e^{-2i k_{\rm p} t}
	\right]
	\propto
	\frac{k_{\rm t p}^2 + \omega^2}{4 \delta}
	+
	\frac{1}{8}
	\left[
	(k_{\rm t p}^2 + \omega^2) \log\frac{4 \mu^2}{k_{\rm t p}^2 - \omega^2}
	+ k_{\rm t p} \omega \log \frac{k_{\rm t p} - \omega}{k_{\rm t p} + \omega}
	\right]
	\;,
	\end{equation}
	where in the first step we expanded the cosine in exponentials ($\sigma = \pm 1$) and in the second step we performed the summation over $\sigma$. The subscript $\rm p$ stands for physical momentum. We first performed the time integral and then the one over $k_{\rm p}$.
	Notice that the boundary term $e^{-2i k_{\rm p} t}$ mixes the $t$ and $k_{\rm p}$ dependences and does not vanish in dim.~reg.~. Note the non-analytic dependence of the finite terms in $\delta$.
}
 
The result \eqref{eq:Loop_cutoff} indicates that this one-loop correction is suppressed compared to the tree-level one by $\alpha^2 P_\zeta \ll1$. This confirms the estimates of Sec.~\ref{sec:finite-wfu}. 
In addition, following the same method we used, it is straightforward to generalise the scaling of $\log \Psi_2^{\rm 1-loop}$ to $\log \Psi_n^{\rm 1-loop}$ where $n$ is the number of external legs:
\begin{align}
	\log \Psi^{\rm 1-loop}_{n} 
	\sim 
	\tilde{b} \sqrt{\alpha}\,\frac{ \bzeta^2 }{P_\zeta}\,
	(\alpha^2 \bzeta)^{n - 2} (\alpha^2 P_\zeta) 
	\;.
\end{align}  
The suppression with respect to tree-level is always $\alpha^2 P_\zeta \ll1$.

Let us discuss what happens when derivatives do not act on the external legs: contributions such as $K_{\vect{k}_1}(\eta) K'_{\vect{k}_1}(\eta) G'(\eta, \eta;\vect{k})$ and $K_{\vect{k}_1}(\eta) K_{\vect{k}_1}(\eta) G''(\eta, \eta; \vect{k})$ to the one-loop coefficient $\psi_2^{\rm 1-loop}$. Following the same procedure as above, one can straightforwardly show that these contributions are suppressed by $\alpha$, compared to Eq.~\eqref{eq:Loop_cutoff}. 
More explicitly, we have 
\begin{align}
\label{eq:1-loop_other_deri}
	\log \Psi_2^{\rm 1-loop}(KK'G') 
	\sim 
	\frac{\tilde{b}\sqrt{\alpha}}{\alpha}
	\,
	\frac{ \bzeta^2}{P_\zeta}\, (\alpha^2 P_\zeta) 
	\;, 
	\quad 
	\log \Psi_2^{\rm 1-loop}(KKG'') 
	\sim 
	\frac{\tilde{b}\sqrt{\alpha}}{\alpha^2}
	\,
	\frac{ \bzeta^2}{P_\zeta}\, (\alpha^2 P_\zeta) 
	\;.
\end{align}
The reason for this $\alpha$ suppression is the following. In this case, there is at least one time derivative acting on the bulk-to-bulk propagator, implying that it will generate a factor of internal momentum $k$. Then, changing the variables to the physical momentum ($k_{\rm p} = -\eta H k$) leads to an additional factor of $1/\eta$, so that the integrand in $\eta$ contains fewer powers of $\eta$ compared with the  $K'_{\vect{k}_1}(\eta) K'_{\vect{k}_1}(\eta) G(\eta, \eta;\vect{k})$ contribution. Therefore, evaluating the $\eta$ integral on the saddle point ($k_{\rm t}\eta_s \sim -\alpha$) we find that in comparison with Eq.~\eqref{eq:Loop_cutoff} these contributions are suppressed in $\alpha$ as shown in Eq.~\eqref{eq:1-loop_other_deri}.

Applying the same technique to the higher-loop corrections, one can deduce that the $\ell$-loop wavefunction coefficient at $\mathcal{O}(\tilde{b})$ with $n$ external legs scales as
\begin{align}
\log \Psi^{\rm \ell-Loop}_{n} \sim \tilde{b}\sqrt{\alpha} \, \frac{\ \bzeta^2}{P_\zeta} \, (\alpha^2 \bzeta)^{n - 2}\, (\alpha^2 P_\zeta)^{\ell} \;.
\end{align}

\section{Numerical methods}\label{app:num_method}

In this appendix we explain the numerical methods used to compute the on-shell action in Sec.~\ref{sec:pde_num}. 
The first step towards numerically integrating the action is to obtain an accurate value for the free solution $\zeta(\tau, r)$ with the late-time boundary condition.
We then provide two methods to obtain such a solution.
The first method makes use of the analytical expression for the Fourier transform $\zeta(\tau, k)$, i.e.~the Euclidean version of Eq.~\eqref{eq:zeta_free_fourier}. 
Then, the real-space solution is obtained by performing an inverse-Fourier transform at each time step, which is implemented numerically via a fast Fourier transform (FFT).
Since our system possesses spherical symmetry, the solution then only depends on $k$ and the inverse-Fourier transform simply becomes one dimensional.
After that, we plug the real-space solution into the action and evaluate the integral numerically, although there are some technical points to be careful about, as we will explain below.
The second method, on the other hand, relies on solving the linear PDE for $\zeta(\tau, r)$ with prescribed boundary conditions. 
Then, we can straightforwardly compute the action on such a numerical solution. 
This method in fact is similar to what was implemented in \cite{Celoria:2021vjw} (in this reference the equation for $\zeta(\tau, r)$ was however non-linear).
We will refer to these two methods as FFT and PDE methods, respectively.

Before entering into the details of the numerical integration, let us comment on the Gaussian profile at late times, see Eq.~(\ref{eq:gaussian_prof}). In fact, in this specific case the solution $\zeta(\tau, r)$ can be obtained analytically in terms of exponential-integral functions, using the inverse-Fourier transform of the (Euclidean-rotated) $\zeta(\tau, k)$ in (\ref{eq:zeta_free_fourier}), giving
\begin{equation}
	\zeta(\tau, r) / \bar \zeta
	=
	(1-4 \tau ^2) 
	\,
	{\rm Re}\, W(z)
	+\frac{2\tau}{r}
	(\tau ^2 - r^2) 
	\,
	{\rm Im}\, W(z)
	-\frac{2 \tau }{\sqrt{\pi }} 
	\;,
\label{eq:zeta_analytic}
\end{equation}
where $W(z) \equiv e^{-z^2} (1 - i \, {\rm erfi}(z))$, ${\rm erfi}(z) \equiv - i {\rm erf}(i z)$ is the imaginary error function and $z \equiv r + i \tau$. 
It should be noted that dealing with the above expression is not always straightforward. Indeed, an accurate evaluation of Eq.~\eqref{eq:zeta_analytic} requires the implementation of arbitrary-precision numerics (otherwise large numerical errors appear when evaluating $W(z)$ for complex argument in the regions $|\tau| \gg 1$, $r \gg 1$), resulting in very long evaluation times. 
Actually, we employ this analytical expression (\ref{eq:zeta_analytic}) only for negative $\bar \zeta$ since high accuracy/precision is needed, as discussed in Sec.~\ref{sec:pde_num}. However, this procedure of finding an analytical expression for $\zeta(\tau, r)$ cannot be generally applied to an arbitrary late-time configuration.

Here we explain in detail how to perform the numerical integration over $\tau$ and $r$. 
First, note that $\tau \in (-\infty, 0)$ and $r \in (0, +\infty)$.
In order to obtain a better accuracy/precision, we divide our integral (\ref{eq:DeltaS1_E_PDE2}) into three pieces,
\begin{align}\label{app:three_pieces}
\Delta S_{\rm E, 1} = \Delta S_{1, {\rm early}} + \Delta S_{1, \rm{grid}} + \Delta S_{1, \rm{late}} \;, 
\end{align} 
where $\Delta S_{1, {\rm early}}$ is the integration over $\tau \in (-\infty, \tau_{\rm min})$, $\Delta S_{1, \rm{grid}}$ the one over $\tau \in (\tau_{\rm min}, \tau_{\rm max})$, and $\Delta S_{1, \rm{late}}$ the one over $\tau \in (\tau_{\rm max}, 0)$.
The reason for this separation, as we will see below, is essentially the fact that the integral over $\tau$ in both $\Delta S_{1, {\rm early}}$ and $\Delta S_{1, {\rm late}}$ can be performed analytically, which indeed improves the matching with the saddle-point approximation. Therefore, we are left with only the numerical integration in $r$ for $\Delta S_{1, {\rm early}}$ and $\Delta S_{1, {\rm late}}$. Of course, we still have to do the numerical integrations over $\tau$ and $r$ in $\Delta S_{1, \rm{grid}}$,~\footnote{In fact, when performing the numerical integration it is more useful to change coordinates $\{\tau, r\}$ to $\{\tilde{t}, \tilde{r}\}$, defined as 
\begin{equation}
	\tilde{t} 	 \equiv - \log (-\tau H)\;, \quad 
	 \tilde{r}   \equiv   \log(H r)
	\;.
\end{equation}
Notice that both $\tilde{t} $ and $ \tilde{r}$ run from $-\infty$ to $+\infty$.
We see that the oscillation of the on-shell action is periodic in $\tilde{t}$ with frequency $\alpha$, whereas
its frequency increases as a function of $\tau$ at late times. Therefore, this suggests that performing the numerical integral over the variable $\tilde{t}$ is more robust (the sampling can be done on a uniform grid), compared to the integral over $\tau$.} but this is less subtle since both $\tau_{\rm min}$ and $\tau_{\rm min}$ are finite. Below, we provide a detailed analysis of both $\Delta S_{1, {\rm early}}$ and $\Delta S_{1, {\rm late}}$.

Let us analyse the late-time contribution $\Delta S_{1, {\rm late}}$. 
In this regime, expanding the solution $\zeta(\tau, r)$ for small $|\tau|$, see Eq.~\eqref{eq:zeta_late0}, gives rise to the late-time limit action up to the leading contributions,
\begin{align}
	\Delta S_{1, {\rm late}} 
	\simeq
	\frac{2 \pi}{P_{\zeta}}
	\int_{0}^{+\infty}\de r\, r^2 
	\int_{\tau_{\rm max}}^{0} \de \tau
	&
	\bigg[
	\left(
	\nabla^2 \bar \zeta(r) + \partial_r \bar \zeta(r) \, \partial_r \nabla^2 \bar \zeta(r)
	\right)
	\cos q(\tau, r)
	\nn \\
	&
	\qquad \qquad \qquad
	- \frac{\alpha}{2} \nabla^2 \bar \zeta(r) \, (\partial_r \bar \zeta(r))^2
	\sin q(\tau, r)
	\bigg]
	\;, \label{eq:app_S1_late}
\end{align}
where we have used the fact that the late-time profile is spherically symmetric and we have defined 
$q(\tau, r) \equiv \alpha \left(    \log(\tau/\eta_\star)+ \bar{\zeta}(r)-i \pi/2 \right)-\td$. It is important to note that the limit $\tau \to 0^{-}$ is regular by construction. Performing the integral over $\tau$ in (\ref{eq:app_S1_late}) analytically, we therefore obtain 
\begin{align}
	\Delta S_{1, {\rm late}} 
	\simeq
	&
	-
	\frac{2 \pi \tau_{\rm min}}{(1 + \alpha^2)P_{\zeta}}
	\int_{0}^{+\infty}\de r\, r^2 
	\bigg[
	\left(
	\nabla^2 \bar \zeta(r) + \partial_r \bar \zeta(r) \, \partial_r \nabla^2 \bar \zeta(r)
	\right)
	( \cos q_{\rm max}(r) + \alpha \sin q_{\rm max}(r))
	\nn \\
	&
	\hspace{5cm}
	- \frac{\alpha}{2} \nabla^2 \bar \zeta(r) \, (\partial_r \bar \zeta(r))^2
	( \sin q_{\rm max}(r) - \alpha \cos q_{\rm max}(r))
	\bigg] \label{Delta_S_1late}
	\;,
\end{align}
where $q_{\rm max/min}(r) \equiv q(\tau_{\rm max/min}, r)$. 
Notice that in the formula shown above we have kept the terms up to $\mathcal{O}(\tau_{\rm max})$. 
Actually, in our numerical implementation we include up to order $\mathcal{O}(\tau_{\rm max}^{12})$ and when choosing $\tau_{\rm max}$ we check that additional corrections are negligible.
The integration over $r$ in (\ref{Delta_S_1late}) is then performed numerically.

Finally, let us consider the early-time contribution $\Delta S_{1, {\rm early}}$.
Following the same procedure as for the late-time limit action, we expand the solution $\zeta(\tau, r)$ for large $|\tau|$ to obtain 
\begin{align}
	\Delta S_{1, \rm{early}}
	\simeq
	- \frac{2 \pi}{P_{\zeta}} 
	\int_{0}^{+\infty} \de r \, r^2 
	(\partial_r \bar \zeta(r))^2 
	\int_{-\infty}^{\tau_{\rm min}} \frac{\de \tau}{\tau^2} 
	\cos(q(\tau, r)) \;.
\end{align}
Notice that in the expression above the fact that $\zeta(\tau, r)$ quickly decays for large $|\tau|$ implies that the dominant contribution in the action at early times comes from the counter term in Eq.~\eqref{eq:DeltaS1_E}. 
Evaluating the integral over $\tau$ analytically, we thus obtain
\begin{align}
	\Delta S_{1, \rm{early}}
	= 
	\frac{2 \pi}{\tau_{\rm min}(1 + \alpha^2) P_{\zeta}} 
	\int_{0}^{+\infty} \de r \, r^2 
	(\partial_r \bar \zeta(r))^2 
	\left(
	\cos q_{\rm min}(r) - \alpha \sin q_{\rm min}(r)
	\right) \label{Delta_S_1early}
	\;.
\end{align}
As before, we are left with the radial integration which can be done numerically. Notice that this contribution decays very slowly at early times, as $\sim 1 / \tau_{\rm min}$. 
This suggests that including $\Delta S_{1, \rm{early}}$ allows us to choose moderately large values for $\tau_{\rm min}$, without considering an extremely large grid.

\bibliographystyle{utphys}
\bibliography{biblio}

\end{document}